\documentclass[twocolumn,prc,aps,showpacs,superscriptaddress,nofootinbib]{revtex4}
 
\usepackage{amssymb}
\usepackage{amsmath}
\usepackage{graphicx} 
\usepackage{epsfig}   
\usepackage{dcolumn}  
\usepackage{rotating} 
\usepackage[dvips]{color}

\begin{document}

\title{Semi-Inclusive Charged-Pion Electroproduction off Protons and Deuterons:
Cross Sections, Ratios and Access to the Quark-Parton Model at Low Energies}

\newcommand*{\YERPHY}{Yerevan Physics Institute, Yerevan 0036, Armenia}
\newcommand*{\JLAB}{Thomas Jefferson National 
Accelerator Facility, Newport News, Virginia 23606, USA}
\newcommand*{\HAMPTON}{Hampton University, Hampton, Virginia 23668, USA }
\newcommand*{\ARGONNE}{Physics Division, Argonne National Laboratory, 
Argonne, Illinois 60439, USA }
\newcommand*{\BUCH}{Bucharest University, Bucharest, Romania }
\newcommand*{\CALTECH}{California Institute of Technology, 
Pasadena, California 91125, USA }
\newcommand*{\DUKE}{Triangle Universities Nuclear Laboratory 
and Duke University, 
Durham, North Carolina 27708, USA }
\newcommand*{\FIU}{Florida International University, 
University Park, Florida 33199, USA }
\newcommand*{\GETTY}{Gettysburg College, Gettysburg, 
Pennsylvania 18103, USA}
\newcommand*{\GWU}{The George Washington University, 
Washington, D.C. 20052, USA }
\newcommand*{\HOU}{University of Houston, Houston, TX 77204, USA }
\newcommand*{\JMU}{James Madison University, Harrisonburg, 
Virginia 22807, USA }
\newcommand*{\MARYLAND}{University of Maryland, College Park, 
Maryland 20742, USA }
\newcommand*{\MSS}{Mississippi State University, Mississippi 
State, Mississippi 39762, USA }
\newcommand*{\NCSU}{North Carolina A \& T State University, 
Greensboro, North Carolina 27411, USA}
\newcommand*{\OHIO}{Ohio University, Athens, Ohio 45071, USA }
\newcommand*{\REGINA}{University of Regina, Regina, Saskatchewan, 
S4S 0A2, Canada }
\newcommand*{\RPI}{Rensselaer Polytechnic Institute, 
Troy, New York 12180, USA}
\newcommand*{\RUTGERS}{Rutgers, The State University of New Jersey, 
Piscataway, New Jersey, 08855, USA }
\newcommand*{\UCONN}{University of Connecticut, Storrs, 
Connecticut 06269, USA }
\newcommand*{\UMASS}{University of Massachusetts Amherst, 
Amherst, Massachusetts 01003, USA}
\newcommand*{\UNC}{University of North Carolina Wilmington, 
Wilmington, North Carolina 28403, USA }
\newcommand*{\UVA}{University of Virginia, Charlottesville, 
Virginia 22901, USA }
\newcommand*{\VASS}{Vassar College, Poughkeepsie, 
New York 12604, USA }
\newcommand*{\VU}{VU-University, 1081 HV Amsterdam, The Netherlands }
\newcommand*{\JOHAN}{University of Johannesburg, 
Johannesburg, South Africa }
\newcommand*{\WITS}{University of the Witwatersrand, 
Johannesburg, South Africa }

\author{R.~Asaturyan} 
\affiliation{\YERPHY}
\author{R.~Ent} 
\affiliation{\JLAB}
\affiliation{\HAMPTON}
\author{H.~Mkrtchyan}  
\affiliation{\YERPHY}
\author{T.~Navasardyan}
\affiliation{\YERPHY}
\author{V.~Tadevosyan} 
\affiliation{\YERPHY}
\author{G.S.~Adams} 
\affiliation{\RPI}
\author{A.~Ahmidouch}
\affiliation{\NCSU}
\author{T.~Angelescu}
\affiliation{\BUCH}
\author{J.~Arrington} 
\affiliation{\ARGONNE}
\author{A.~Asaturyan}
\affiliation{\YERPHY}
\author{O.K.~Baker}
\affiliation{\JLAB}
\affiliation{\HAMPTON}
\author{N.~Benmouna}
\affiliation{\GWU}
\author{C.~Bertoncini}
\affiliation{\VASS}
\author{H.P.~Blok}
\affiliation{\VU}
\author{W.U.~Boeglin} 
\affiliation{\FIU}
\author{P.E.~Bosted} 
\affiliation{\JLAB}
\affiliation{\UMASS}
\author{H.~Breuer}
\affiliation{\MARYLAND}
\author{M.E.~Christy} 
\affiliation{\HAMPTON}
\author{S.H.~Connell} 
\affiliation{\JOHAN}
\author{Y.~Cui}
\affiliation{\HOU}
\author{M.M.~Dalton} 
\affiliation{\WITS}
\author{S.~Danagoulian}
\affiliation{\NCSU}
\author{D.~Day}
\affiliation{\UVA}
\author{J.A.~Dunne} 
\affiliation{\MSS}
\author{D.~Dutta}
\affiliation{\DUKE}
\author{N.~El~Khayari} 
\affiliation{\HOU}
\author{H.C.~Fenker}
\affiliation{\JLAB}
\author{V.V.~Frolov} 
\affiliation{\CALTECH}
\author{L.~Gan} 
\affiliation{\UNC}
\author{D.~Gaskell}
\affiliation{\JLAB}
\author{K.~Hafidi} 
\affiliation{\ARGONNE}
\author{W.~Hinton} 
\affiliation{\HAMPTON}
\author{R.J.~Holt}
\affiliation{\ARGONNE}
\author{T.~Horn} 
\affiliation{\JLAB}
\author{G.~M.~Huber} 
\affiliation{\REGINA}
\author{E.~Hungerford}
\affiliation{\HOU}
\author{X.~Jiang} 
\affiliation{\RUTGERS}
\author{M.~Jones} 
\affiliation{\JLAB}
\author{K.~Joo}
\affiliation{\UCONN}
\author{N.~Kalantarians} 
\affiliation{\HOU}
\author{J.J.~Kelly} 
\affiliation{\MARYLAND}
\author{C.E.~Keppel} 
\affiliation{\JLAB}
\affiliation{\HAMPTON}
\author{V.~Kubarovsky} 
\affiliation{\JLAB}
\author{Y.~Li}
\affiliation{\HOU}
\author{Y.~Liang} 
\affiliation{\OHIO}
\author{D.~Mack}
\affiliation{\JLAB}
\author{S.~P.~Malace}
\affiliation{\HAMPTON}
\author{P.~Markowitz} 
\affiliation{\FIU}
\author{E.~McGrath} 
\affiliation{\JMU}
\author{P.~McKee}
\affiliation{\UVA}
\author{D.G.~Meekins} 
\affiliation{\JLAB}
\author{A.~Mkrtchyan}
\affiliation{\YERPHY}
\author{B.~Moziak} 
\affiliation{\RPI}
\author{G.~Niculescu} 
\affiliation{\JMU}
\author{I.~Niculescu} 
\affiliation{\JMU}
\author{A.K.~Opper}
\affiliation{\OHIO}
\author{T.~Ostapenko} 
\affiliation{\GETTY}
\author{P.E.~Reimer} 
\affiliation{\ARGONNE}
\author{J.~Reinhold}
\affiliation{\FIU}
\author{J.~Roche} 
\affiliation{\JLAB}
\author{S.E.~Rock} 
\affiliation{\UMASS}
\author{E.~Schulte}
\affiliation{\ARGONNE}
\author{E.~Segbefia} 
\affiliation{\HAMPTON }
\author{C.~Smith} 
\affiliation{\UVA}
\author{G.R.~Smith}
\affiliation{\JLAB}
\author{P.~Stoler} 
\affiliation{\RPI}
\author{L.~Tang}
\affiliation{\JLAB}
\affiliation{\HAMPTON}
\author{M.~Ungaro} 
\affiliation{\RPI}
\author{A.~Uzzle} 
\affiliation{\HAMPTON}
\author{S.~Vidakovic}
\affiliation{\REGINA}
\author{A.~Villano} 
\affiliation{\RPI}
\author{W.F.~Vulcan} 
\affiliation{\JLAB}
\author{M.~Wang}
\affiliation{\UMASS}
\author{G.~Warren} 
\affiliation{\JLAB}
\author{F.~R.~Wesselmann} 
\affiliation{\UVA}
\author{B.~Wojtsekhowski}
\affiliation{\JLAB}
\author{S.A.~Wood} 
\affiliation{\JLAB}
\author{C.~Xu} 
\affiliation{\REGINA}
\author{L.~Yuan} 
\affiliation{\HAMPTON}
\author{X.~Zheng} 
\affiliation{\ARGONNE}
 
\newpage
\date{\today}

\begin{abstract}
A large set of cross sections for semi-inclusive electroproduction of charged pions ($\pi^{\pm}$) 
from both proton and deuteron targets was measured. The data are in the deep-inelastic scattering 
region with invariant mass squared $W^2 >$ 4 GeV$^2$ (up to $\approx$ 7 GeV$^2$)
and range in four-momentum transfer squared 
$2<Q^2<4$ (GeV/$c$)$^2$, and cover a range in the Bjorken scaling variable $0.2<x<0.6$. 
The fractional energy of the pions spans a range $0.3<z<1$, with small transverse momenta with 
respect to the virtual-photon direction, $P_t^2<0.2$ (GeV/$c$)$^2$. The invariant mass that goes 
undetected, $M_x$ or $W^\prime$, is in the nucleon resonance region, $W^\prime<2$ GeV. The new 
data conclusively show the onset of quark-hadron duality in this process, and the relation of this 
phenomenon to the high-energy factorization ansatz of electron-quark scattering and subsequent 
quark $\rightarrow$ pion production mechanisms. The $x$, $z$ and $P^2_t$ dependences of several 
ratios (the ratios of favored-unfavored fragmentation functions, charged pion ratios, 
deuteron-hydrogen and aluminum-deuteron ratios for $\pi^+$ and $\pi^-$) have been studied. The 
ratios are found to be in good agreement with expectations based upon a high-energy quark-parton 
model description. We find the azimuthal dependences to be small, as compared to exclusive pion 
electroproduction, and consistent with theoretical expectations based on tree-level factorization 
in terms of transverse-momentum-dependent parton distribution and fragmentation functions. In the 
context of  a simple model, the initial transverse momenta of $d$ quarks are found to be slightly 
smaller than for $u$ quarks, while the transverse momentum width of the favored fragmentation 
function is about the same as for the unfavored one, and both fragmentation widths are larger than 
the quark widths. 
\end{abstract}

\pacs{13.60.Le, 13.87.Fh}
\maketitle

\section{INTRODUCTION}

There has been a growing realization that understanding of the resonance region in inelastic 
scattering, and the interplay between resonance behavior and a high-energy scaling phenomenon in 
particular, represents a critical gap that must be filled if one is to fathom fully the nature of 
the quark--hadron transition in Quantum ChromoDynamics (QCD). The decade or so preceding the 
development of QCD saw tremendous effort devoted to describing hadronic interactions in terms of 
$S$-matrix theory and self-consistency relations. One of the profound discoveries of that era was 
the remarkable relationship between low-energy hadronic cross sections and their high-energy 
behavior, in which the former on average appears to mimic certain features of the latter.

At low energies, one expects the hadronic scattering amplitude to be dominated by just a few 
resonance poles. As the energy increases, the density of resonances in each partial wave, as well 
as the number of partial waves itself, grows, making it harder to identify contributions from 
individual resonances, and more useful to describe the scattering amplitude in terms of a sum of 
$t$-channel Regge poles and cuts. Progress towards synthesizing the two descriptions came with the 
development of finite-energy sum rules, relating dispersion integrals over the amplitudes at low 
energies to high-energy parameters.

The observation \cite{bg70-71} of such a nontrivial relationship between inclusive 
electron--nucleon scattering cross sections at low energy, in the region dominated by the nucleon 
resonances, and that in the deep-inelastic scattering (DIS) regime at high energy, similarly 
predated QCD. Initial interpretations of this duality naturally used the theoretical tools 
available at the time, finite-energy sum rules or consistency relations between hadronic 
amplitudes inspired by the developments in Regge theory that occurred in the 1960s 
\cite{col77,ddln02}. With the advent of QCD, De R\'{u}jula, Georgi, and Politzer offered a 
qualitative explanation of Bloom-Gilman duality~\cite{DeR77} in terms of either small or 
on-average-canceling higher-twist contributions, but a quantitative understanding of the origin of 
the duality phenomenon in QCD remains elusive, although some insight has been obtained through 
phenomenological model calculations \cite{DKDS71}.

Even if it remains counterintuitive that there should be a strong relationship between the 
resonance region, in which the lepton scatters from a target hadron traditionally treated as a 
bound system of massive constituent quarks, and the deep-inelastic region, where the lepton 
essentially scatters from a single free quark, it is essential to provide precise data on the 
onset of this phenomenon. In regions where single-quark scattering is well established, a rich 
plethora of nucleon structure information can be gathered from such reactions through the 
quark-parton model \cite{CGK72,CG72,COT74}.

In this article, we will concentrate on the largely unexplored low-energy domain of semi-inclusive 
electron scattering, $e N \rightarrow e \pi^\pm X$, in which a charged pion $\pi^\pm$ is detected 
in the final state in coincidence with the scattered electron. The process of semi-inclusive 
deep-inelastic scattering (SIDIS) has been shown to factorize~\cite{Ji04}, in the high energy 
limit, into lepton-quark scattering followed by quark hadronization. Our focus will be on the 
process where a quark fragments into a pion, such that the electroproduced pion carries away a 
large fraction, but not all, of the exchanged virtual photon's energy. 

The quark-hadron duality phenomenon has been predicted~\cite{Car98,Afa00,CI01}, and subsequently 
verified~\cite{Nav07} for high-energy meson electroproduction. The relation of the duality 
phenomenon with the onset of factorization in electron-quark scattering and subsequent quark 
hadronization was also postulated and shown to hold in {\sl e.g.} the SU(6) quark 
model~\cite{CGK72,CG72,COT74,CI01,Cl-Mel03}. If so, one obtains access to the virtue of 
semi-inclusive meson production, that lies in the ability to identify, in a partonic basis, 
individual quark species in the nucleon by tagging specific mesons in the final state, thereby 
enabling both the flavor and spin of quarks and antiquarks to be systematically determined. 
Ideally, one could even directly measure the quark transverse momentum dependence  of the quark 
distribution functions $q(x,k_t)$ by detecting all particles produced in the hadronization process 
of the struck quark. A large set of pion electroproduction data from both hydrogen and deuterium 
targets has been obtained in experiment E00-108 spanning the nucleon resonance region. Cross 
sections for semi-inclusive electroproduction of charged pions ($\pi^{\pm}$) from both proton and 
deuteron targets were measured for $0.2<x<0.6$,~$2<Q^2<4$ (GeV/$c$)$^2$,~$0.3<z<1$, and 
$P_t^2<0.2$ (GeV/$c$)$^2$. The results from this experiment permit a first study of a possible 
low-energy access to the quark-parton model, either directly through cross section measurements or 
indirectly through their ratios, possibly lowering the energy threshold to access the quark-parton 
model if higher-twist contributions would  fully cancel.

In Section 2, we will describe in detail the relation between kinematical cuts to separate current 
and target region fragmentation events as optimally as possible, quark-hadron duality, and a 
low-energy onset of a factorized (or precociously factorized) description in terms of 
electron-quark scattering and subsequent hadronization of the struck (current) quark. 
In Section 3, we will relate the findings of earlier low-energy experiments to a quark-parton 
model description, and extensions thereof beyond the infinite momentum frame including 
azimuthal-angle and transverse-momentum dependences. Sections 4, 5 and 6 will cover the 
experimental details, the data analysis procedures, and the systematic uncertainties, 
respectively. Finally, Sections 7 and 8 will describe the experimental results in terms of 
dependences of ratios and cross sections on various kinematic variables, including some nuclear 
dependences, followed by the conclusions.

\section{Towards a High-Energy Description of Semi-Inclusive Pion 
Electroproduction}

\subsection{Semi-Inclusive Deep Inelastic Scattering}

In semi-inclusive deep inelastic scattering (SIDIS), a hadron $h$ (in our case a charged pion 
$\pi^\pm$) is detected in coincidence with a scattered electron, with a sufficient amount of 
energy and momentum transferred in the scattering process. Under the latter conditions, the 
reaction can be seen as knockout of a quark and subsequent (independent) hadronization.

\begin{figure}
\begin{center}
\epsfxsize=3.40in
\epsfysize=2.00in
\epsffile{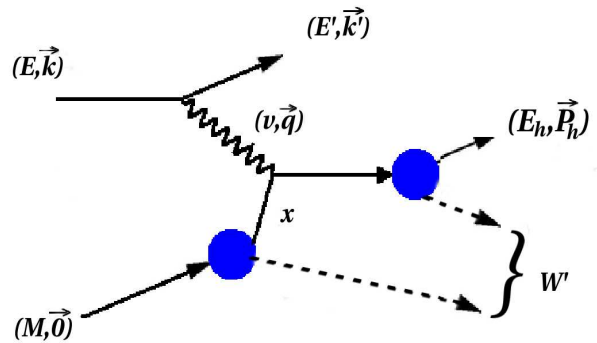}
\caption{\label{fig:mesonproddiag}
Schematic diagram of meson electroproduction.}
\end{center}
\end{figure}

Fig.~\ref{fig:mesonproddiag} gives a schematic picture of this process, including the kinematics. 
An electron with four-momentum $(E,\vec{k})$ scatters from a nucleon with mass $M$ (taken to be 
the proton mass $M_p$ at rest), resulting in a scattered electron with four-momentum  
$(E^\prime,\vec{k^\prime})$, thereby exchanging a virtual photon with four-momentum 
$q=(\nu,\vec{q})$ with a quark. A meson with four momentum $m=(E_h,\vec{P}_h)$ is produced, with 
the residual hadronic system characterized by an invariant mass $W^\prime$. As usual, the 
four-momentum transfer squared is defined as $Q^2= -q^2$ and the Bjorken variable as 
$x=Q^2/2M\nu$. The latter can be interpreted as the fraction of the light-cone momentum of the 
target nucleon carried by the struck quark. Furthermore, \( z \) is defined as 
\( z=(p\cdot m)/(p\cdot q ) \). In the target rest (lab) frame, this becomes \( z=E_{h}/\nu  \), 
the fraction of the virtual photon energy taken away by the meson. In the elastic limit, 
\( z=1 \), and the meson carries away all of the photon's energy. Finally, we define \( P_t \) to 
be the transverse momentum of the meson in the virtual photon-nucleon system.

At high values of $Q^2$ and $\nu$, the cross section (at leading order in the strong coupling 
constant $\alpha_s$) for the reaction $N(e,e^\prime \pi)X$ can be written in the following way 
(see Ref.~\cite{Dakin73}),
\begin{eqnarray}
\label{eq:semi-parton}
\left.
\begin{array}{lll}
{\frac {\frac {d\sigma} {d\Omega_e dE_e dz dP_t^2 d\phi} }
{\frac {d\sigma} {d\Omega_e dE_e}}} = {\frac {dN} {dz}} b e^{-bP_t^2}
{\frac {1 + A cos(\phi) + B cos(2\phi)} {2\pi}}, ~\\
&  & \\
\frac {dN}  {dz} \sim \sum_i e_i^2\ q_i(x,Q^2)\ D_{q_i \to \pi}(z,Q^2) ,~
\end{array}
\right.
\end{eqnarray}
where \( i \) denotes the quark flavor and \( e_{i} \) is the quark charge, and the 
fragmentation function $D_{q_i \to \pi}(z,Q^2)$ gives the probability for a quark to evolve into a 
pion $\pi$ with a fraction $z$ of the quark (or virtual photon) energy, $z=E_{\pi}/\nu$. The first 
part of this formula expresses that the cross section factorizes into the product of the virtual 
photon--quark interaction and the subsequent quark hadronization. A consequence of factorization 
is that the fragmentation function is independent of $x$, and the parton distribution function 
$q_i(x,Q^2)$ independent of $z$. Both parton distribution and fragmentation functions, however, 
depend on $Q^2$ through logarithmic $Q^2$ evolution~\cite{strat97}. The second part describes the 
dependence on the transverse momentum $P_t$, assumed to be Gaussian, and the general 
dependence~\cite{Donnelly} of the cross section in the unpolarized case on the angle $\phi$, the 
angle between the electron scattering plane and the pion production plane, with $A$ and $B$, 
reflecting the interference terms ${\sigma}_{LT}$ and ${\sigma}_{TT}$, respectively, being 
functions of $x,Q^2,z,P_t$.
An important variable for the analysis is the missing mass \( M_{x} \), which is the invariant 
mass of the undetected residual system. Here we refer to this quantity as \( W' \) ($M_x$) to 
highlight the fact that \emph{it could play a role analogous to \( W \) for duality in the 
inclusive case}~\cite{Afa00}. If we neglect the pion mass, \( W'^{2} \) is given by
\begin{eqnarray}
\label{eq:wpr_sq1}
{W^\prime}^2 = W^2 - 2z\nu \, (M+\nu -\left| \vec{q} \right| \, 
\cos \theta _{qm})\, ,
\end{eqnarray}
where \( \nu =E-E' \) and \( \theta _{qm} \) is the lab angle between the virtual photon momentum 
\( \left| \vec{q} \right|  \) and the outgoing meson momentum \( \left| \vec{p} \right|  \). As in 
the usual inclusive scattering case, the square of the (inclusive) invariant mass \( W \) is given 
by
\begin{eqnarray}
\label{eq:w_sq}
W^2 = M^2 + Q^2\, (\frac{1}{x}-1)\, .
\end{eqnarray}
If we further limit the outgoing meson to be collinear with the virtual photon momentum and 
require that \( Q^{2}/\nu ^{2}\ll 1 \), we can express \( W'^{2} \) in terms of \( z \), \( x \), 
and \( Q^{2} \) as
\begin{eqnarray}
\label{eq:wpr_sq2}
{W^\prime}^2 = M^2 + Q^2\, (1-z)\, (\frac{1}{x}-1)\, .
\end{eqnarray}
The quantitative differences between Eqs.~(\ref{eq:wpr_sq1}) and (\ref{eq:wpr_sq2}) are small for 
the described experimental results, and not visible on any of the figures in the remainder of this 
article.

For the remainder of this article, we will equate the ``nucleon resonance region'' to the 
condition that $W^\prime < 2$ GeV, even if the invariant mass $W$ will be beyond the usually 
defined resonance region, $W > 2$ GeV. As can be easily read from Eq.~(\ref{eq:wpr_sq2}), the 
larger $z$ the fewer hadronic states will be involved in the semi-inclusive pion electroproduction 
process, with $z = 1$ the (deep) exclusive limit.

\subsection{Factorization}
If one neglects the dependence of the cross section on the pion transverse momentum $P_t$ and the 
angle $\phi$, the SIDIS cross section as given in Eq.~(\ref{eq:semi-parton}) can be written as 
\begin{equation}
\label{eq:simplefactorization}
\sigma \propto \sum_i q_i(x,Q^2)\ D_{q_i \to \pi}(z,Q^2).
\end{equation}
(At higher orders one has to worry about gluon fragmentation functions, but this can be neglected 
for the energy and momentum transfers under consideration  here \cite{BKK95}). The question is how 
well this factorization into independent functions of $x$ and $z$ is fulfilled in practice.

Initial investigations of the hadronization process were made in electron--positron annihilation 
and in deep inelastic scattering. By now a wealth of data has been accumulated to parameterize the 
fragmentation functions as function of $z$ and $Q^2$. It is well known that for the case of SIDIS 
one has to worry about separating pions directly produced by the struck quark (termed ``current 
fragmentation'') from those originating from the spectator quark system 
(``target fragmentation''). This has been historically done for high-energy SIDIS by using 
separation in rapidity, $\eta$, with the latter defined in terms of the produced pion energy and 
the longitudinal component of the momentum (along the $\vec q$ direction),
\begin{eqnarray}
\label{eq:rapidity}
\eta &=& {\frac 1 2}
         \ln\left( \frac { E_\pi - p^z_\pi} {E_\pi + p^z_\pi }
            \right)\ .
\end{eqnarray}
Early data from CERN \cite{BERG75,BERG87} suggest that a difference in rapidities, $\Delta\eta$, 
between pions produced in the current and target fragmentation regions (``rapidity gap'') of at 
least $\Delta\eta \approx 2$ is needed to kinematically separate the two regions.

\begin{figure}
\begin{center}
\epsfxsize=3.40in
\epsfysize=3.40in
\epsffile{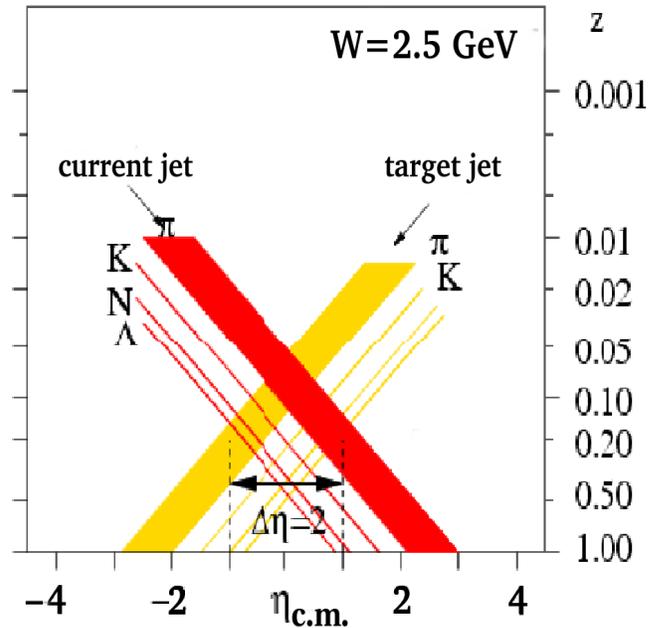}
\caption{\label{fig:mulders} (Color online) 
Relation between elasticity $z$ and center of mass rapidity $\eta_{CM}$ in semi-inclusive 
electroproduction of various hadrons for $W=2.5~GeV$, assuming null transverse momentum. The band 
for pions reflects the influence of transverse momentum.}
\end{center}
\end{figure}

It has been argued that such kinematic separation is even possible at lower energies, or low 
$W^2$, if one considers only electroproduced pions with large elasticity $z$, {\em i.e.}, with 
energies close to the maximum energy transfer~\cite{BERG87,MULD00}. Figure~\ref{fig:mulders} shows 
a plot of rapidity versus $z$ for $W = 2.5$~GeV. At $W = 2.5$~GeV, a rapidity gap of 
$\Delta \eta \geq 2$ would be obtained with $z > 0.4$ for pion electroproduction. For larger $W$, 
such a rapidity gap could already be attained at a lower value of $z$ (see 
Ref.~\cite{MULD00,MEK05}). For instance, one would anticipate a reasonable kinematic separation 
between the current and target fragmentation processes for $z > 0.2$ at $W = 5$~GeV.
The other issue is at which energy scales we can make the assumption of independence of the hard 
scattering process from the hadronization process. At low energies we would normally view the 
nucleon as a collection of constituent quarks, and the factorization ansatz could break down due 
to effects of final state interactions, resonant nucleon excitations and higher-twist 
contributions~\cite{Melnit01}, even if a sufficient rapidity gap would be established. For this 
article, we will simply assume that factorization in terms of a hard scattering and subsequent 
hadronization (called kinematical factorization in the remainder of this article) holds 
{\it provided} kinematical separation between current and target fragmentation is possible, 
{\it and} one is beyond the nucleon resonance region, $W^{\prime} > 2$~GeV.

To give credence to the latter assumption, we observe that in the annihilation process 
$e^+e^- \rightarrow h X$, experimental data \cite{SIEG,HANS} beyond $z \approx 0.5$ at 
$W = 3$~GeV ($W^\prime = 1.94$~GeV) were historically described in terms of fragmentation 
functions. The region extends to $z \geq 0.2$ for $W = 4.8$~GeV ($W^\prime = 2.84$~GeV) and to 
$z \geq 0.1$ for $W = 7.4$~GeV ($W^\prime = 4.14$~GeV). For $z > 0.3$, fragmentation functions 
have also been obtained from data \cite{DREWS} on $ep \rightarrow e^\prime \pi^\pm X$ at an 
incident energy $E = 11.5$~GeV, with $3 < W < 4$~GeV. All of these data are beyond the 
($W^\prime >$ 2 GeV) nucleon resonance region as defined above, and seem indeed reasonably well 
understood in terms of a simple fragmentation description.

\subsection{Quark-Hadron Duality and Precocious Factorization}
At energies $W^\prime < 2$ GeV, it is not obvious that the pion electroproduction process 
factorizes in the same manner as in Eq.~(\ref{eq:semi-parton}). At energies where hadronic 
phenomena dominate, the pion electroproduction process may rather be described through the 
excitation of nucleon resonances, $N^*$, and their subsequent decays into mesons and lower lying 
resonances, $N'^*$ \cite{Melnit01, CM09}. It has been argued that a factorization similar to the 
one at high-energy may appear to hold at low energies due to the quark-hadron duality 
phenomenon~\cite{Car98,Afa00,CI01}. For that phenomenon to occur, non-trivial cancellations of the 
angular distributions from various decay channels~\cite{IJMV01,CI01,CM09} would be required to 
produce the fast-forward moving pion at the high-energy limit.

\begin{figure}
\begin{center}
\epsfxsize=3.50in
\epsfysize=2.60in
\epsffile{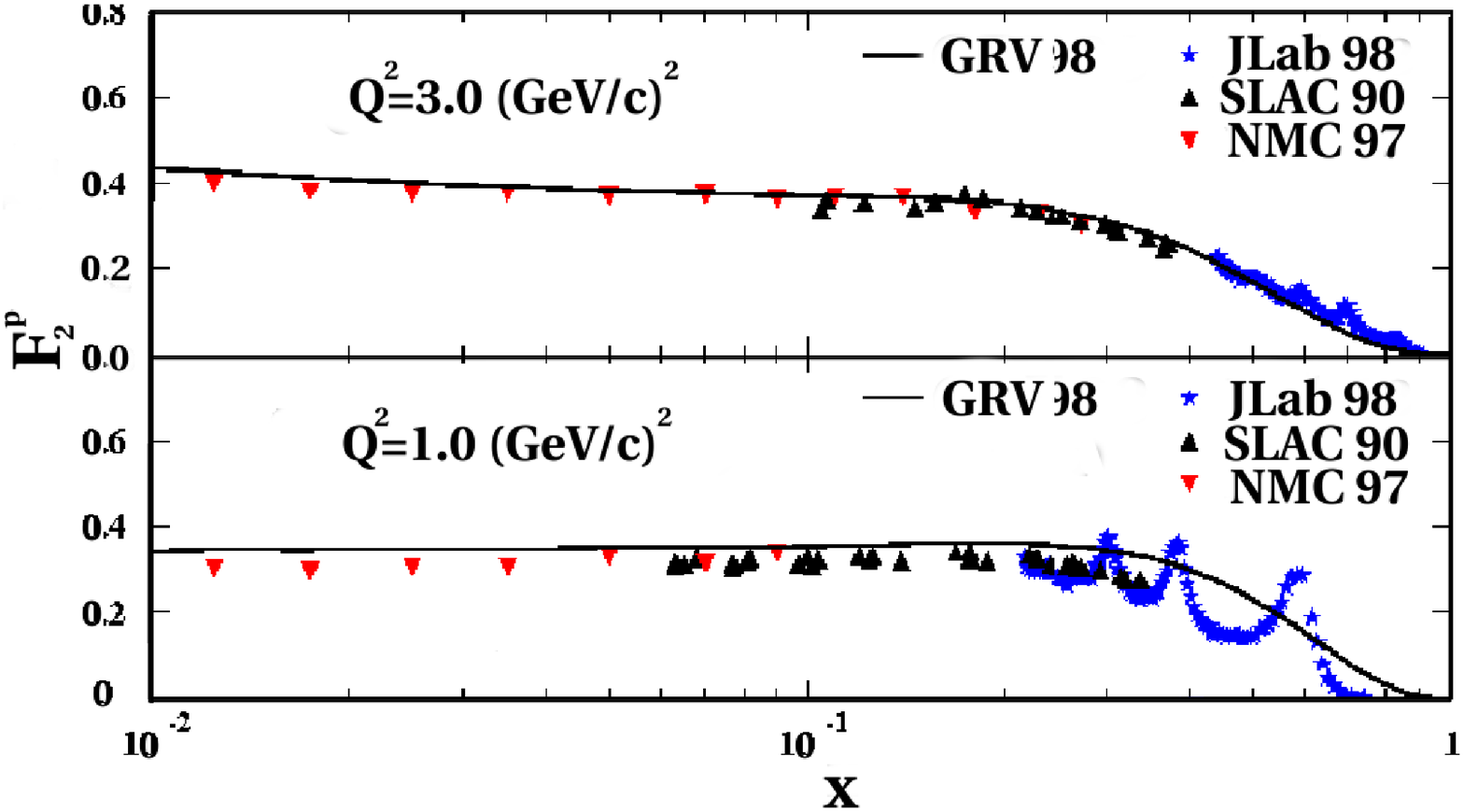}
\caption{\label{fig:domingoplot} (Color online)
The structure function $F_{2}$ versus $x$ for resonance data from Jefferson Lab and 
SLAC~\cite{Nic99,Nic00b}, and SLAC and NMC deep-inelastic data, at two different values of 
$Q^{2}$. The solid curves show the GRV parameterization~\cite{GRV} at $Q^{2}=3$~(GeV/$c$)$^{2}$ 
and $Q^{2}=1$~(GeV/$c$)$^{2}$.} 
\end{center}
\end{figure}

In the early 1970s, Bloom and Gilman made the phenomenological observation that there exists a 
duality between inelastic electron-proton scattering in the resonance region and in the 
Deep-Inelastic Scattering (DIS) regime~\cite{bg70-71}. More detailed studies over the last decade 
have shown that quark-hadron duality is exhibited over a broader kinematic range, and with greater 
precision, than was previously known~\cite{Nic99,Nic00a}. Duality was found to also work quite 
well locally, with various resonance regions averaging to DIS scaling expectations to good 
approximation ($<10\%$), even down to low momentum transfer values ($Q^2 \approx 0.5$ 
(GeV/$c$)$^2$. Alternatively, the individual resonance scans average to \emph{some global} curve 
even down to $Q^2 \approx 0.1$ (GeV/$c$)$^2$~\cite{Nic99,Nic00b}. 
This global curve then coincides with the DIS scaling expectations at larger $Q^2$ (or $x$). 
This is illustrated in Fig.~\ref{fig:domingoplot}.
The observation of duality tells us that higher twist terms mostly cancel, or are small, even at 
these low values of $Q^2$, when averaging over a sufficient (but relatively small) amount of 
resonances and the underlying non-resonant background contributions. This implies that 
single-quark scattering remains the dominant process, even though one visually can see the effect 
of the quark-quark interactions by the resonance peak enhancements. The quark-quark interactions 
modify the measured spectrum, bound to create confined quarks, but do so only locally.

Duality studies in inclusive scattering have been extended to spin-structure functions, which was 
predicted from both perturbative~\cite{Carl98} and nonperturbative QCD 
arguments~\cite{CI01,Cl-Gi}. The first experiment accessing the spin-dependent asymmetries was 
SLAC experiment E143~\cite{Baum80,Abe97,Abe98}. Spin asymmetry data reported by the HERMES 
Collaboration~\cite{Air03} and JLab (CLAS~\cite{Fat03} and Hall A~\cite{E01-012}) in the nucleon 
resonance region were also found to be in reasonable agreement with those measured in the 
deep-inelastic region~\cite{Abe98,Air98,Adeva,Anth}, with possible exceptions in the $N-\Delta$ 
resonance region. More recently, both CLAS and Hall A Collaborations have accumulated a set of 
precision data to study the onset of quark-hadron duality in polarized inclusive electron-nucleon 
scattering as function of $Q^2$~\cite{Yun,E01-012}, with good agreement found at larger $Q^2$ 
($>$ 2 (GeV/$c$)$^2$).

While the phenomenon of duality in inclusive electron scattering is thus well-established, duality 
in the related case of semi-inclusive meson electroproduction was not experimentally tested before 
this experiment. To experimentally investigate the existence of quark-hadron duality in 
semi-inclusive pion electroproduction processes, and how this may be related to a precocious 
(low-energy) factorization and partonic description, was one of the main goals of the E00-108 
experiment.
 
Carlson~\cite{Car98} suggested several phenomena one could look for to explore any possible dual 
behavior between electroproduction of mesons in the resonance region and the high-energy scaling 
expectations by using `meson tagging' in the final state, in close analogy to the original 
inclusive case findings by Bloom and Gilman~\cite{bg70-71}:
\begin{itemize}
\item Do we observe scaling behavior as $Q^2$ increases?
\item Do the resonances tend to fall along the DIS scaling curve?
\item Does the ratio of resonant to non-resonant strength remain roughly 
constant with increase of $Q^2$, as was one of Bloom-Gilman's original 
observations?
\end{itemize}
Existing experimental charged pion electroproduction data show hardly any nucleon resonance 
structure at $W^\prime > 1.4$~GeV, and seem to scale~\cite{CDW95}, hinting at partial answers to 
some of these questions. In addition, an initial investigation of duality in semi-inclusive pion 
production was made in Ref.~\cite{CI01}, where the factorization between parton distribution and 
fragmentation functions was found to hold when summing over the N* resonances in the SU(6) quark 
model.

The existence of low-energy kinematical factorization \cite{CI01,EHK} in combination with the 
quark-hadron duality phenomenon may very well lead to a precocious description of the SIDIS 
process at low energy in terms of the quark-parton model. Applying that to the case discussed 
here, one could anticipate that factorization and a quark-parton model description work reasonably 
well for $z >$ 0.4 and at relatively low $W^\prime$ scales (below 2 GeV).

In this discussion we have neglected the dependence of measured pion yields, as in 
Eq.~(\ref{eq:simplefactorization}), on the pion transverse momentum, $P_t$. 
At high energies the dependence on $P_t$ has historically been described with a Gaussian 
dependence as $\exp(-b P_t^2$), where $b^{-1}$ represents the average transverse momentum squared 
of the struck quark. At lower energies, the measured $P_t$ dependence must reflect to some extent 
the decay angular distributions of the electroproduced resonances in regions where these 
resonances dominate. One would therefore expect the $P_t$ dependence to vary with $W^\prime$ at 
low $W^\prime$.

\section{SEMI-INCLUSIVE DEEP INELASTIC SCATTERING AND THE QUARK-PARTON MODEL}

In this Section we will first revisit experimental information at relatively low energies, in 
order to see if that exhibits characteristics of a factorized description as portrayed by 
Eq.~(\ref{eq:simplefactorization}). Theoretically, such a factorized description is only valid at 
leading order in $\alpha_s$, and after integration over the transverse momentum $P_t$ and the
azimuthal angle $\phi$. Then, we will investigate what can be learned from a phenomenological 
description of measurements of such transverse momenta and azimuthal angles, where the factorized 
description breaks down even at leading order, and a whole series of further assumptions must be 
made to relate these data to a quark-parton description.

\subsection{Low-Energy $x-z$ Factorization and the Quark-Parton Model}
Several pieces of evidence suggest that factorization may hold at low energies in meson-tagged 
reactions. Initially, skepticism existed about the applicability of the quark-parton model at 
energies below those historically used at, {\sl e.g.}, the Electron Muon Collaboration 
experiment~\cite{Slo88}, because of the possibility of overlapping current and target 
fragmentation regions. Interest grew with the findings of the HERMES experiment at DESY, where an 
intriguing similarity was found between results from semi-inclusive deep inelastic scattering at 
moderate energies~\cite{Ack98} and a Drell-Yan experiment at far higher energies~\cite{Haw98}. 
This similarity suggested that factorization, and a quark-parton model description may after all 
be valid at energies where it is not necessarily expected to work. 

The HERMES experiment measured semi-inclusive pion electroproduction 
(\( \gamma ^{*}N\rightarrow \pi ^{\pm }X \)) in the DIS regime, over the ranges 
\( 13<\nu <19 \)~GeV and \( 21<W^{2}<35 \)~(GeV)\( ^{2} \), with an average four-momentum transfer 
\( \left\langle Q^{2}\right\rangle =2.3 \)~(GeV/\( c)^{2} \). The HERMES analysis explicitly 
assumed factorization in order to extract the sea asymmetry \( \overline{d}-\overline{u} \). 
In particular, it was assumed that the charged pion yield \( N^{\pi ^{\pm }} \) factorized into 
quark density distributions \( q_{i}(x) \) and fragmentation functions 
\( D_{q_{i}}^{\pi ^{\pm }}(z) \):
\begin{equation}
\label{eq:factorization}
N^{\pi ^{\pm }}(x,z)\, \propto \, \sum _{i}e_{i}^{2}\left[ \: q_{i}(x)\: 
D_{q_{i}}^{\pi ^{\pm }}(z)\: +\: \overline{q}_{i}(x)\: 
D_{\overline{q}_{i}}^{\pi ^{\pm }}(z)\right] \, 
\end{equation}
Indeed, agreement was found between the extracted flavor asymmetry of the nucleon quarks sea 
results of HERMES, and the FermiLab Drell-Yan experiment E866 that first reported this flavor 
asymmetry (at dramatically higher energies). Revisiting these HERMES data, at an average 
$W >$ 5 GeV, and constrained to fractions of the virtual photon energy, $z$, of larger than 0.2, 
it is perhaps not so surprising that these data support that the factorization assumption used in 
the HERMES analysis appears to be valid for the nucleon sea, even at relatively low energy loss. 
A rapidity gap $\eta >$ 2, rendering potential sufficient separation between the current and 
target fragmentation regions and thus kinematical factorization, can already be attained at values 
of $z$ of 0.2 (see Ref.~\cite{MULD00,MEK05}) for $W = 5$ GeV. Given that in the HERMES kinematics 
also $W^\prime$ remains larger than 2 GeV, both requirements we assumed to be needed for a valid 
high-energy factorized description are fulfilled.

At even lower energies, with kinematics close to the experiment reported upon here, a series of 
measurements of semi-inclusive pion electroproduction was carried out at Cornell in the mid 1970s, 
with both hydrogen and deuterium targets~\cite{Beb75,Beb76,Beb77a}. These Cornell measurements 
covered a region in \( Q^{2} \) (\( 1 < Q^{2} < 4 \)~(GeV/$c$)$^2$) and 
\( \nu \) (\( 2.5 < \nu < 6 \)~GeV). The results of these measurements were analyzed in terms of 
an invariant structure function (comparable to \(N^{\pi ^{\pm }}(x,z) \) of 
Eq.~\ref{eq:factorization}), written in terms of the sum of products of parton distribution 
functions and parton fragmentation functions. The authors concluded that this invariant structure 
function shows no \( Q^{2} \) dependence, and a weak dependence on \( W^{2} \), within their 
region of kinematics, which would be consistent with a factorized quark-parton model description.

This is the more striking if one realizes that the Cornell kinematics cover a region in 
\( W^{2} \) between 4 and 10~GeV$^2$, and in \( z \) between 0.1 and 1. In fact, if one would 
calculate $W^\prime$, these results are for an appreciable fraction of their kinematics in the 
region  \( 1 < W' < 2 \)~GeV, which is generally associated with the nucleon resonance region. 
Even more, the final pion momentum is often only 1 GeV/$c$, such that final pion-nucleon 
scattering effects, especially differences between both pion charge flavors, cannot be neglected. 
To complete this enumeration, \( P_t \), the average transverse momentum of the meson, was 
typically less than 0.1~GeV/$c$. Unfortunately, not enough statistics and information are 
available to warrant a careful check of duality or factorization in the Cornell data, even if the 
data are suggestive that quark-hadron duality in charged pion electroproduction, and a precocious 
low-energy factorization, may work in these kinematics.

\subsection{Transverse momenta and azimuthal angles}
\label{sec:pt_phi}
A central question in the understanding of nucleon structure is the orbital motion of partons. 
Much is known about the light-cone momentum fraction, $x$, and virtuality scale, $Q^2$, 
dependences of the up and down quark parton distribution functions (PDFs) in the nucleon. 
In contrast,  very little is presently known about the dependence of these functions on the 
transverse momentum $k_t$ of the parton. Simply based on the size of the nucleon in which the 
quarks are confined, one would expect characteristic transverse momenta of order a few hundred 
MeV/c, with larger values at small Bjorken $x$, where the sea quarks dominate, and smaller values 
at high $x$, where all of the quark momentum is longitudinal in the limit $x=1$. Increasingly  
precise studies of the nucleon spin sum rule~\cite{EMC,E155,HERMES,RHIC} strongly suggest that the 
net spin carried by quarks and gluons is relatively small, and therefore the net orbital angular 
momentum must be significant. This in turn implies significant transverse momentum of quarks. 
Questions that naturally arise include: what is the flavor and helicity dependence of the 
transverse motion of quarks and gluons, and can these be modeled theoretically and measured 
experimentally?

In the E00-108 experiment, we detect only a single hadronization product: a charged pion carrying 
a (large) energy fraction $z$ of the available energy. The probability of producing a pion with a 
transverse momentum $P_t$ relative to the virtual photon ($\vec{q}$) direction is described by a 
convolution of the quark distribution functions  and $p_t$-dependent fragmentation functions 
$D^+(z,p_t)$ and $D^-(z,p_t)$, where $p_t$ is the transverse momentum of the pion relative to the 
quark direction, $k_t$ is the struck quark intrinsic transverse momentum, with the 
condition~\cite{Anselmino} $P_t = z k_t + p_t$. The ``favored'' and ``unfavored'' fragmentation 
functions  $D^+(z,p_t)$ and $D^-(z,p_t)$ refer to the cases where the produced pion contains the 
flavor of the struck quark or not. ``Soft'' non-perturbative processes are 
expected~\cite{Anselmino} to generate relatively small values of $p_t$ with an approximately 
Gaussian distribution in $p_t$. Hard QCD processes are expected to generate large non-Gaussian 
tails for $p_t>1$ GeV/$c$, but probably do not play a major role in the interpretation of the 
E00-108 experiment, for which the total transverse momentum $P_t<0.45$ GeV/$c$.

Because the average value of $\phi$ in the E00-108 experiment is correlated with $P_t$, 
(see Fig.~\ref{fig:phipt}), we first need to study the $\phi$ dependence. The cross sections for 
each target and pion flavor were parameterized in the form of Eq.~\ref{eq:semi-parton}. The 
assumed Gaussian $P_t$ dependence (with slopes $b$ for each case) is an effective parameterization 
that seems to describe the data adequately for use in making radiative and bin-centering 
corrections. Small values of $A$ and $B$ are expected from non-zero parton motion, as described by 
Cahn~\cite{Cahn} and Levelt-Mulders~\cite{Levelt-Mulders}. In general, any non-zero parton motion 
effects, be it kinematic or dynamic, are proportional to $P_t$ for $A$, and $P^2_t$ for $B$, 
respectively~\cite{Cahn,Levelt-Mulders,Berger,Oganes}.

The more recent treatment of Ref.~\cite{Anselmino} similarly gives results for $A$ and $B$ that 
are very close to zero (especially for $B$). Other possible higher twist contributions will also 
be proportional to powers of $P_t/\sqrt{Q^2}$~\cite{Metz,Bacchetta}, and therefore suppressed at 
our low average values of $P_t$. Specifically, the twist-2 Boer-Mulders~\cite{Boercalc} 
contribution to $B$ is essentially zero in the models of Ref.~\cite{Boercalc,Vincenzo}. For the 
kinematics of the E00-108 experiment, the value of $B$ for $\pi^+$ is expected to be positive and 
could change approximately linearly with $x$, $z$ and $P_t$ from $\sim$0.002 to 
$\sim$0.02 GeV/$c$, see Fig. 6 and Fig. 7 in Ref.~\cite{GaGo-Schlegel}. 
For $\pi^-$, it is expected to be negative and the dependences on $x,~z$ and $P_t$ to be much 
weaker. In contrast, the  values of $A$ and $B$ are much larger in exclusive pion production than 
those predicted  for SIDIS.

\section{EXPERIMENT} 

The experiment E00-108 ~\cite{E00-108} ran in the summer of 2003 in Hall C at Jefferson Lab. An 
electron beam with energy of 5.479 GeV and currents ranging between 20 and 60 $\mu A$ was provided 
by the CEBAF accelerator. Incident electrons were scattered from 4-cm-long liquid hydrogen or 
deuterium targets and detected in the Short Orbit Spectrometer (SOS). The SOS central momentum 
remained constant throughout the experiment, with a value of 1.702 GeV/$c$. The electroproduced 
mesons (predominantly pions) were detected in the High Momentum Spectrometer (HMS), with momenta 
ranging from 1.3 to 4.1 GeV/$c$. A detailed description of the spectrometers and set-up can be 
found in Ref.~\cite{HBlok08}. The experiment consisted of three parts: i) at a fixed electron 
kinematics of ($x,Q^2$) = (0.32, 2.30 (GeV/$c$)$^2$), $z$ was varied from 0.3 to 1 by changing the 
HMS momentum, with nearly uniform coverage in the pion azimuthal angle, $\phi$, around the virtual 
photon direction, but at a small average $P_t$ of 0.05 GeV/$c$ ($z$-scan); ii) for $z=0.55$, $x$ 
was varied from 0.2 to 0.6, with a corresponding variation in $Q^2$ from 1.5 to 4.6 (GeV/$c$)$^2$, 
by changing the SOS angle, keeping the pion centered on the virtual-photon direction (and again 
average $P_t$ of 0.05 GeV/$c$) ($x$-scan); iii) for ($x,Q^2$) = (0.32, 2.30 (GeV/$c$)$^2$), $z$ 
near  0.55, $P_t$ was scanned from 0 to 0.4 GeV/$c$ by increasing the HMS angle (with average 
$\phi$ near 180 degrees) ($P_t$-scan). The kinematic settings are listed in Table~\ref{kin-tab}. 

\begin{widetext} 
\begin{center}
\begin{table}
\caption{\label{kin-tab} Kinematic settings (z-scan, x-scan and $P_t$-scan) in experiment E00-108. 
The electron beam energy $E$ was 5.479 GeV. The scattered electrons were detected in SOS set at 
constant momentum 1.702 GeV/$c$ throughout the experiment.}
{\centering  \begin{tabular}{||c|cccccc|cccc||}
\hline
\hline
$\theta_e$ & $\nu$ &  $Q^2$      & $x$  &  $W^2$  & $|\vec{q}_{}|$ & $\theta_{q}$ & $\theta_m$ & $z$   & $p_m$ & ${W'}^2$\\
  deg      & GeV   & (GeV/c)$^2$ &      & GeV$^2$ & GeV/c          &     deg      &  deg       &       & GeV/c & GeV$^2$ \\   
\hline
\hline
  28.71    & 3.794 & 2.30        & 0.32 & 5.70    & 4.09           & 11.54        & 11.54      & 0.37  & 1.397 & 3.92    \\ 
           &       &             &      &         &                &              &            & 0.42  & 1.606 & 3.65    \\
           &       &             &      &         &                &              &            & 0.49  & 1.846 & 3.35    \\
           &       &             &      &         &                &              &            & 0.56  & 2.122 & 3.00    \\
           &       &             &      &         &                &              &            & 0.64  & 2.439 & 2.60    \\
           &       &             &      &         &                &              &            & 0.74  & 2.803 & 2.13    \\
           &       &             &      &         &                &              &            & 0.85  & 3.222 & 1.60    \\
           &       &             &      &         &                &              &            & 0.97  & 3.703 & 1.00    \\
\hline
\hline
25.70      & 3.794 & 1.85        & 0.26 & 6.16    & 4.03           & 10.55        & 10.55      & 0.55  & 2.082 & 3.25    \\
28.71      & 3.794 & 2.30        & 0.32 & 5.70    & 4.09           & 11.54        & 11.54      & 0.55  & 2.082 & 3.05    \\
31.75      & 3.794 & 2.80        & 0.39 & 5.20    & 4.15           & 12.47        & 12.47      & 0.55  & 2.082 & 2.82    \\
34.55      & 3.794 & 3.30        & 0.46 & 4.70    & 4.21           & 13.27        & 13.27      & 0.55  & 2.082 & 2.60    \\
37.17      & 3.794 & 3.80        & 0.53 & 4.20    & 4.27           & 13.95        & 13.95      & 0.55  & 2.082 & 2.37    \\
39.63      & 3.794 & 4.30        & 0.60 & 3.70    & 4.32           & 14.54        & 14.54      & 0.55  & 2.082 & 2.15    \\
\hline
\hline
28.71      & 3.794 & 2.30        & 0.32 & 5.70    & 4.09           & 11.54        & 11.54      & 0.55  & 2.082 & 3.29    \\
           &       &             &      &         &                & 11.54        & 13.54      & 0.55  & 2.082 & 3.29    \\
           &       &             &      &         &                & 11.54        & 15.54      & 0.55  & 2.082 & 3.29    \\
           &       &             &      &         &                & 11.54        & 17.54      & 0.55  & 2.082 & 3.29    \\
           &       &             &      &         &                & 11.54        & 19.54      & 0.55  & 2.082 & 3.29    \\
\hline
\hline
\end{tabular}\par}
\end{table}
\end{center}
\end{widetext}

\begin{figure}
\begin{center}
\epsfxsize=3.40in
\epsfysize=3.40in
\epsffile{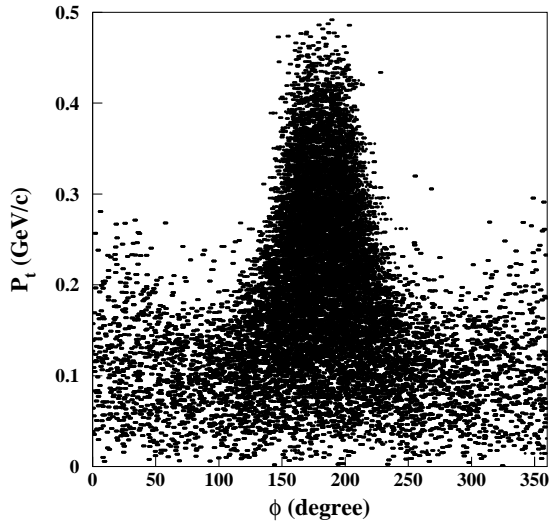}
\vspace*{-0.5in}
\caption{\label{fig:phipt} $P_t$ distribution of the data from this experiment as a function of 
azimuthal angle $\phi$.}
\end{center}
\end{figure}

The $\phi$ distribution as a function of $P_t$ is shown for all three data sets combined in 
Fig.~\ref{fig:phipt}. 
Except for the largest $x$-setting in the $x$-scan, the virtual-photon-nucleon invariant mass $W$ 
was always larger than 2.1 GeV (typically 2.4 GeV), in the traditional deep-inelastic region for 
inclusive scattering. In order to avoid complications from \( \pi N \) final-state interactions 
the momenta of the outgoing pions were kept greater than 2~GeV/\( c \) in most cases. 
All measurements were performed for both $\pi^+$ and $\pi^-$.


\section{DATA ANALYSIS}

The raw data collected by the data acquisition system were processed by the standard Hall C 
analysis engine (ENGINE), which decodes the data into physical quantities on an event by event 
basis. The main components of the data analysis include tracking, event reconstruction, 
determination of and correction for experimental and kinematic offsets, particle identification 
and event selection, background estimation and subtraction, correction for detector efficiencies, 
and electronic and computer dead times. Many steps of the analysis here are similar to the 
$F_\pi$ data analysis described in Ref.~\cite{HBlok08}. Below we will discuss some of the steps, 
and will emphasize details relevant to the E00-108 experiment.

{\bf Accidentals:}
Random coincidences occur between events from any two beam bursts within the coincidence timing 
gate. The resulting coincidence timing structure of random coincidences is peaked every 2 ns 
(defined by the beam microstructure of the CEBAF accelerator). The random events under the real 
coincidence peak cannot be identified but their contribution can be estimated. The data were 
corrected for these random contributions by selecting a number of random peaks and subtracting 
their average content from the content of the real coincidence peak. The accidentals were taken 
for an $\sim$80 ns interval (for 40 bursts far from the coincidence peak ($e^\prime\pi^-$ at 
negative polarity of the HMS spectrometer, and $e^\prime$p and $e^\prime\pi^+$ peaks at positive 
polarity)), and the average number of accidentals (within 2 ns) was defined as 
$N_{Accidentals}$( 2 ns) = [$N_{Accidentals}$ (80 ns)]/40.

{\bf Electron Identification:}
Electrons were identified in the SOS using a combination of the SOS gas \v{C}erenkov detector and 
calorimeter. The gas \v{C}erenkov detector was used as a threshold detector with a mean signal 
of 
$\sim$7 photoelectrons per electron. Good electron events were selected for a photoelectron (pe) 
cut $N_{pe}> 0.5$. This cut was chosen to ensure good efficiency over the full acceptance, even 
after accounting for the position dependence of the  pe yield. To determine the efficiency of 
the \v{C}erenkov detector, an electron sample was selected from data with the calorimeter cut 
$E_{cal}/P_{e^-}>0.8$, where $E_{cal}$ is a total energy deposited in the calorimeter, and 
$P_{e^-}$ is the momentum of the particle in the electron arm (SOS) defined by tracking. 
The \v{C}erenkov detector efficiency is then given by the ratio of events with and without the 
\v{C}erenkov detector cut. The efficiency was found to be $\gtrsim99.8\%$ for a photoelectron cut 
of $N_{pe}>0.5$.

Electrons deposit their entire energy in the calorimeter peaking $E_{cal}/P_{e^-}$ distribution 
at 1. Good electron events in the calorimeter were selected by applying the 
cut $E_{cal}/P_{e^-} > 0.7$. This cut removes most of the pions, while keeping high electron 
detection efficiency. The efficiency of the calorimeter was determined in a similar fashion as for 
the gas \v{C}erenkov detector. A particle identification cut was placed on the gas \v{C}erenkov 
detector, and the calorimeter efficiency was estimated as the ratio of events passing the 
calorimeter cut to the total number of events. The corresponding efficiency was estimated to be 
$\gtrsim99.5\pm0.1\%$. The pion rejection factor (the ratio of pions with and without cut on 
energy deposition in calorimeter) in this case was $\sim$ 1:20.

{\bf Pion Identification:} 
Pions in the HMS are selected with aerogel and gas \v{C}erenkov detectors. The gas \v{C}erenkov 
detector's function was to separate electrons from negatively-charged pions. The HMS calorimeter 
did not play a significant role and was used only for cross checks, and for the gas \v{C}erenkov 
detector efficiency determination.

Pions under the \v{C}erenkov radiation threshold do not in principle produce a signal in the 
detector. However, pions may produce $\delta$-electrons, which will result in a photoelectron 
number greater than zero. Applying a cut to reject electrons may then reject pion events as well. 
To determine the pion efficiency of the \v{C}erenkov detector cut positive polarity $\pi^+$ data 
were used with a calorimeter cut (to take out contribution from positrons). The ratio of events 
passing the \v{C}erenkov detector cut to all the events is then the \v{C}erenkov detector pion 
efficiency. The pion \v{C}erenkov detector efficiency is ${99.6\pm0.05\%}$ for a cut $N_{pe}< 2$.

The separation of pions from protons ( and partly from kaons) relies on the HMS aerogel 
detector~\cite{Asa05}. Whether or not a particle traversing the aerogel \v{C}erenkov detector 
produces a signal depends on the index of refraction of the aerogel material ($n_{ref}$) and the 
particle velocity ($\beta=v/c$). The mean number of photoelectrons (for aerogel material with 
$n_{ref}$=1.015) was $N_{pe}\sim$7-8 and slightly varied with a particle momentum (for $z$-scan 
data). The aerogel \v{C}erenkov detector efficiency is ${99.5\pm0.02\%}$ for a threshold cut of 
$N_{pe} > 1$.

Real electron-proton coincidences are eliminated via coincidence time cuts in the analysis. In all 
kinematic settings of the experiment  the electron-hadron coincidence time distribution is well 
described by Gaussian with $\sigma\leq$ 250 ps (in average). In the analysis a cut $\pm$1.2 ns is 
used on the e-$\pi$ coincidence time. At the highest momentum setting of the HMS 
($P_{HMS}$=4.1 GeV/$c$), which is the worst case, there is still about 3 ns separation between 
electron-proton and electron-pion coincidence peaks. Even in the absence of proton rejection from 
the aerogel \v{C}erenkov detector,  the protons (and partly kaons at low momenta) would be removed 
in the random subtraction.

\subsection{Background subtraction and corrections}

{\bf Contribution from Target Walls:}
Events from the aluminum walls of the cryogenic target cell were subtracted by performing empty 
target runs ({\it dummy runs}). The {\it dummy data} are analyzed in the same way as the regular 
data including the same method of random coincidence subtraction and applying the same analysis 
cuts. The effective charge-normalized yields are then subtracted  from the real data yields taking 
into account the difference in the wall thickness between the target cell (0.133 mm) and dummy 
target (1 mm). In most cases, the estimated contribution of the target can to the measured yield 
is quite small, about ${2-3\%}$. The uncertainty in the ratio of the thickness of the dummy 
relative to the target can $(2-3\%)$ contributes to a negligible uncertainty to the total yield.

{\bf Radiative Corrections:}
Essentially all of the events that ``radiate in'' to a given bin come from either: (i) incoming 
electrons with a lower actual energy than the nominal beam energy, because they have radiated a 
photon; or (ii) scattered electrons with higher energy than the one measured in the spectrometer, 
because they radiated a photon. In both cases, the value of $\nu$ at the vertex is lower than the 
reconstructed one, hence z is larger and $W^\prime$ is smaller than the nominal value.

The radiative tails within our semi-inclusive pion electroproduction data were estimated using the 
Monte Carlo package SIMC. The radiative correction formula coded is based on the work of Mo and 
Tsai~\cite{MoTsai69}, which originally was derived  for inclusive electron scattering, but was 
modified for use in coincidence experiments~\cite{Ent01}. Details of the implementation are 
described in Ref.~\cite{Koltenuk99}. The original formulation of the radiative correction 
procedure used in SIMC was for $(e,e^\prime p)$ reaction. The formula were extended to pion 
electroproduction by D. Gaskell~\cite{Gaskell01}.

As a cross-check, we also estimated radiative corrections using the code POLRAD. The standard 
FORTRAN code POLRAD-2.0 \cite{Aku97} was written for radiative correction (RC) calculations in 
inclusive and semi-inclusive deep-inelastic scattering of polarized leptons by polarized nucleons 
and nuclei. The program, which is based theoretically on the original approach proposed in the 
Ref.~\cite{Aku-Shu83}, was created to suit the demands of experiments with fixed polarized nuclear 
targets and at a collider. A new version of POLRAD~\cite{Aku-Shu94} was created to calculate the 
RC for semi-inclusive (polarized) experiments. In this case the cross section depends additionally 
on the variable $z$. 

The radiative corrections calculated with POLRAD-2.0 are in good agreement with SIMC. On average 
the RC's are on the level of $\sim 6-8\%$ for all our data sets at $z<0.7$ and reach $\sim 15\%$ 
at $z\gtrsim$~0.9. The relative values of radiative corrections at our kinematic settings are 
listed in Tables \ref{tab:radcorr_x}, \ref{tab:radcorr_z} and \ref{tab:radcorr_pt2}.

\begin{table}
\caption{\label{tab:radcorr_x} The values of radiative corrections for $x$-scan data.}
{\centering \begin{tabular}{||c|c|c|c|c|c|c||}
\hline
\hline
     $x$ &  $z$ & $Q^2$      &  $\pi_H^+$ & $\pi_H^-$ &  $\pi_D^+$ & $\pi_D^-$\\
         &      &(GeV/c)$^2$ &   ($\%$)   & ($\%$)    &   ($\%$)   & ($\%$)   \\
\hline
   0.22 & 0.55 & 1.59 & 4.2$\pm$0.4& 5.6$\pm$0.6& 4.6$\pm$0.5& 5.4$\pm$0.5    \\
   0.26 & 0.55 & 1.88 & 4.5$\pm$0.5& 5.8$\pm$0.6& 4.9$\pm$0.5& 5.6$\pm$0.6    \\
   0.30 & 0.55 & 2.17 & 4.9$\pm$0.5& 6.2$\pm$0.6& 5.2$\pm$0.5& 5.9$\pm$0.6    \\
   0.34 & 0.55 & 2.46 & 5.3$\pm$0.5& 6.5$\pm$0.7& 5.6$\pm$0.6& 6.3$\pm$0.6    \\
   0.38 & 0.55 & 2.75 & 5.8$\pm$0.6& 6.9$\pm$0.7& 6.1$\pm$0.6& 6.7$\pm$0.7    \\
   0.42 & 0.55 & 3.04 & 6.2$\pm$0.6& 7.4$\pm$0.7& 6.5$\pm$0.7& 7.1$\pm$0.7    \\
   0.46 & 0.55 & 3.32 & 6.8$\pm$0.7& 7.8$\pm$0.8& 7.0$\pm$0.7& 7.5$\pm$0.8    \\
   0.50 & 0.55 & 3.61 & 7.3$\pm$0.7& 8.3$\pm$0.8& 7.5$\pm$0.8& 8.0$\pm$0.8    \\
   0.54 & 0.55 & 3.90 & 7.9$\pm$0.8& 8.9$\pm$0.9& 8.1$\pm$0.8& 8.5$\pm$0.9    \\
   0.58 & 0.55 & 4.19 & 8.5$\pm$0.9& 9.4$\pm$0.9& 8.7$\pm$0.9& 9.1$\pm$0.9    \\
\hline
\hline
\end{tabular}\par}
\end{table}

\begin{table}
\caption{\label{tab:radcorr_z} The values of radiative corrections for $z$-scan data.}
{\centering \begin{tabular}{||c|c|c|c|c|c|c||}
\hline
\hline
    $x$  &  $z$ & $Q^2$      &  $\pi_H^+$ & $\pi_H^-$ &  $\pi_D^+$ & $\pi_D^-$\\
         &      &(GeV/c)$^2$ & ($\%$)     & ($\%$)    & ($\%$)     & ($\%$)   \\
\hline
   0.32 & 0.37 & 2.31 & 1.6$\pm$0.2& 3.3$\pm$0.3& 2.1$\pm$0.2& 3.2$\pm$0.3    \\
   0.32 & 0.42 & 2.31 & 2.4$\pm$0.2& 4.1$\pm$0.4& 2.8$\pm$0.3& 3.9$\pm$0.4    \\
   0.32 & 0.49 & 2.31 & 3.4$\pm$0.3& 5.1$\pm$0.5& 3.8$\pm$0.4& 4.8$\pm$0.5    \\
   0.32 & 0.55 & 2.31 & 4.5$\pm$0.5& 6.2$\pm$0.6& 4.9$\pm$0.5& 5.8$\pm$0.6    \\
   0.32 & 0.64 & 2.31 & 5.9$\pm$0.6& 7.5$\pm$0.8& 6.2$\pm$0.6& 7.0$\pm$0.7    \\
   0.32 & 0.74 & 2.31 & 7.8$\pm$0.8& 9.3$\pm$0.9& 8.1$\pm$0.8& 8.8$\pm$0.9    \\
   0.32 & 0.85 & 2.31 &10.8$\pm$1.1&11.9$\pm$1.2&11.0$\pm$1.1&11.5$\pm$1.2    \\
   0.32 & 0.97 & 2.31 &18.3$\pm$1.3&18.5$\pm$1.9&18.3$\pm$1.8&18.3$\pm$1.8    \\
\hline
\hline
\end{tabular}\par}
\end{table}

\begin{table}
\caption{\label{tab:radcorr_pt2} The values of radiative corrections for $P_t$-scan data. Note, 
for this set of measurements the scattered electron kinematic was fixed at $Q^2$=2.31~$(GeV/c)^2$}
{\centering \begin{tabular}{||c|c|c|c|c|c|c||}
\hline
\hline
  $x$  & $z$ & $P_t^2$     & $\pi_H^+$& $\pi_H^-$& $\pi_D^+$& $\pi_D^-$\\
       &     &(GeV/c)$^2$  &($\%$)    & ($\%$)   & ($\%$)   & ($\%$)   \\
\hline
 0.32 & 0.55& 0.01 & 6.3$\pm$0.5& 7.9$\pm$0.7& 6.4$\pm$0.5& 7.4$\pm$0.5\\
 0.32 & 0.55& 0.03 & 5.4$\pm$0.7& 6.7$\pm$0.7& 5.4$\pm$0.7& 6.1$\pm$0.7\\
 0.32 & 0.55& 0.05 & 4.7$\pm$0.7& 5.9$\pm$0.7& 4.3$\pm$0.7& 5.0$\pm$0.7\\
 0.32 & 0.55& 0.07 & 3.9$\pm$0.6& 4.9$\pm$0.6& 4.8$\pm$0.6& 5.5$\pm$0.6\\
 0.32 & 0.55& 0.09 & 1.4$\pm$0.7& 2.2$\pm$0.7& 3.3$\pm$0.7& 3.9$\pm$0.7\\
 0.32 & 0.55& 0.11 & 1.4$\pm$0.7& 2.2$\pm$0.8& 2.5$\pm$0.7& 3.1$\pm$0.8\\
 0.32 & 0.55& 0.13 &-1.6$\pm$0.9&-1.0$\pm$0.9&-0.8$\pm$0.9&-0.1$\pm$0.9\\
 0.32 & 0.55& 0.15 &-1.4$\pm$1.2&-0.7$\pm$1.2&-1.5$\pm$1.2&-0.8$\pm$1.2\\
 0.32 & 0.55& 0.17 &-3.9$\pm$1.8&-3.4$\pm$1.9&-1.1$\pm$1.8&-0.5$\pm$1.8\\
 0.32 & 0.55& 0.19 &-8.3$\pm$3.2&-7.9$\pm$3.3&-5.7$\pm$3.2&-5.3$\pm$3.2\\
\hline
\hline
\end{tabular}\par}
\end{table}

{\bf Exclusive Pions:}
In addition, we subtracted radiative events coming from the exclusive reactions
$e + p \rightarrow e^\prime + \pi^+ + n$ and
$e + n \rightarrow e^\prime + \pi^- + p$. 
This required a model for the cross section of exclusive pion electroproduction that is valid for 
a large range of $W$ (from the resonance region to $W \approx 2.5$~GeV) at relatively large $Q^2$. 
The model used in this analysis started with the parameterization of exclusive $\pi^+$ and $\pi^-$ 
production cross section data from~\cite{Bra79} at $W\approx2.2$~GeV and $Q^2=0.7$ and 
$1.35$~(GeV/$c$)$^2$. This parameterization describes the more recent data taken at Jefferson Lab 
as part of the Charged Pion Form Factor program~\cite{HBlok08,Tadev07,Horn06} ($W=1.95$~GeV, 
$Q^2$=0.6$-$1.6~(GeV/$c$)$^2$ and $W=2.2$~GeV, $Q^2$=1.6,~2.45~(GeV/$c$)$^2$) reasonably well.

While the starting parameterization is appropriate for describing exclusive pion production above 
the resonance region, it does rather poorly for values of $W$ significantly smaller than 2 GeV. 
Since no existing model or parameterization describes exclusive pion production both in the 
resonance region and at large $W$, we chose to adjust our starting model by-hand to give good 
agreement with the MAID model~\cite{MAID} of pion electroproduction in the resonance region. This 
by-hand adjustment began with the assumption that the longitudinal contribution was well described 
by the starting model, even at relatively low $W$. Discrepancies between the starting fit and the 
MAID calculation were attributed to the transverse cross section and were removed by assuming a 
more modest $W$ dependence therein. We further simplified the model by assuming that the $TT$ and 
$LT$ interference terms mostly averaged to zero over our experimental acceptance so that they 
contributed negligibly to the radiative events.  

We ran SIMC with this modified model for exclusive $\pi^+$ electroproduction on the proton and for 
$\pi^+$ and $\pi^-$ production on the deuteron for all our kinematic settings ($z$-scan, $x$-scan 
and $P_t$-scan).

Contributions from exclusive pions were subtracted on a bin by bin basis. On average, the 
contribution from the exclusive tail was estimated to be 4-5$\%$ for the $x$-scan data, 5-15$\%$ 
for the $z$-scan data at $z<0.8$, and 8-10$\%$ for the $P_t$-scan results 
(see Tables~\ref{tab:exclusive_x},~\ref{tab:exclusive_z}, and \ref{tab:exclusive_pt2}). 

\begin{table}
\caption{\label{tab:exclusive_x} The relative contribution of radiative exclusive tail for 
$x$-scan data.}
{\centering \begin{tabular}{||c|c|c|c|c|c|c||}
\hline
\hline
     $x$ &  $z$ & $Q^2$      &  $\pi_H^+$  & $\pi_H^-$ &  $\pi_D^+$  & $\pi_D^-$   \\
         &      &(GeV/c)$^2$ &   ($\%$)    & ($\%$)    &   ($\%$)    & ($\%$)      \\
\hline
   0.22 & 0.55  & 1.59       & 6.1$\pm$0.2 &   -       & 3.7$\pm$0.1 & 5.1$\pm$0.2 \\
   0.26 & 0.55  & 1.88       & 5.2$\pm$0.1 &   -       & 3.5$\pm$0.1 & 5.1$\pm$0.1 \\
   0.30 & 0.55  & 2.17       & 4.6$\pm$0.1 &   -       & 3.4$\pm$0.1 & 5.3$\pm$0.1 \\
   0.34 & 0.55  & 2.46       & 4.6$\pm$0.1 &   -       & 3.3$\pm$0.1 & 5.1$\pm$0.1 \\
   0.38 & 0.55  & 2.75       & 4.2$\pm$0.1 &   -       & 2.9$\pm$0.1 & 4.8$\pm$0.1 \\
   0.42 & 0.55  & 3.04       & 3.8$\pm$0.1 &   -       & 2.7$\pm$0.1 & 4.9$\pm$0.1 \\
   0.46 & 0.55  & 3.32       & 3.7$\pm$0.1 &   -       & 2.6$\pm$0.1 & 4.2$\pm$0.1 \\
   0.50 & 0.55  & 3.61       & 3.1$\pm$0.1 &   -       & 2.3$\pm$0.1 & 3.6$\pm$0.1 \\
   0.54 & 0.55  & 3.90       & 3.2$\pm$0.1 &   -       & 1.9$\pm$0.1 & 3.1$\pm$0.1 \\
   0.58 & 0.55  & 4.19       & 2.5$\pm$0.1 &   -       & 1.5$\pm$0.1 & 2.5$\pm$0.1 \\
\hline
\hline
\end{tabular}\par}
\end{table}

\begin{table}
\caption{\label{tab:exclusive_z} The relative contribution of radiative exclusive tail for 
$z$-scan data.}
{\centering \begin{tabular}{||c|c|c|c|c|c|c||}
\hline
\hline
   $x$  &  $z$ & $Q^2$       &   $\pi_H^+$  & $\pi_H^-$ &  $\pi_D^+$   & $\pi_D^-$    \\
        &      & (GeV/c)$^2$ & ($\%$)       & ($\%$)    & ($\%$)       & ($\%$)       \\
\hline
   0.32 & 0.33 & 2.31        &  3.6$\pm$0.2 &     -     &  2.6$\pm$0.1 &  3.6$\pm$0.2 \\
   0.32 & 0.38 & 2.31        &  3.9$\pm$0.1 &     -     &  3.1$\pm$0.1 &  4.6$\pm$0.1 \\
   0.32 & 0.44 & 2.31        &  4.3$\pm$0.1 &     -     &  3.4$\pm$0.1 &  4.7$\pm$0.1 \\
   0.32 & 0.50 & 2.31        &  4.1$\pm$0.1 &     -     &  2.8$\pm$0.1 &  5.0$\pm$0.1 \\
   0.32 & 0.55 & 2.31        &  5.9$\pm$0.1 &     -     &  4.4$\pm$0.1 &  7.6$\pm$0.1 \\
   0.32 & 0.61 & 2.31        &  7.5$\pm$0.1 &     -     &  5.8$\pm$0.1 &  8.7$\pm$0.2 \\
   0.32 & 0.66 & 2.31        &  8.8$\pm$0.1 &     -     &  6.4$\pm$0.1 & 10.3$\pm$0.2 \\
   0.32 & 0.72 & 2.31        & 11.0$\pm$0.2 &     -     &  7.7$\pm$0.1 & 12.2$\pm$0.2 \\
   0.32 & 0.78 & 2.31        & 13.8$\pm$0.2 &     -     &  8.7$\pm$0.2 & 15.1$\pm$0.3 \\
   0.32 & 0.83 & 2.31        & 15.7$\pm$0.3 &     -     &  9.5$\pm$0.2 & 18.0$\pm$0.4 \\
   0.32 & 0.89 & 2.31        & 21.8$\pm$0.4 &     -     & 15.0$\pm$0.3 & 30.3$\pm$0.6 \\
   0.32 & 0.94 & 2.31        & $\gtrsim$90  &      -    & $\gtrsim$90  & $\gtrsim$90  \\
\hline
\hline
\end{tabular}\par}
\end{table}

\begin{table}
\caption{\label{tab:exclusive_pt2} The relative contribution of radiative exclusive tail for 
$P_t$-scan data. Note, for these measurements the value of four-momentum transfer square was kept 
at $Q^2$=2.31~(GeV/c)$^2$.}
{\centering \begin{tabular}{||c|c|c|c|c|c|c||}
\hline
\hline
 $x$  & $z$  & $P_t^2$     &  $\pi_H^+$   & $\pi_H^-$ & $\pi_D^+$   & $\pi_D^-$     \\
      &      & (GeV/c)$^2$ & ($\%$)       & ($\%$)    & ($\%$)      & ($\%$)        \\
\hline
 0.32 & 0.55 & 0.01        &  5.8$\pm$0.1 &     -     &  4.0$\pm$0.1 &  3.1$\pm$0.1 \\
 0.32 & 0.55 & 0.03        &  6.9$\pm$0.1 &     -     &  4.6$\pm$0.1 &  3.7$\pm$0.1 \\
 0.32 & 0.55 & 0.05        &  7.4$\pm$0.1 &     -     &  5.3$\pm$0.1 &  7.3$\pm$0.2 \\
 0.32 & 0.55 & 0.07        &  8.7$\pm$0.2 &     -     &  5.5$\pm$0.1 &  9.2$\pm$0.3 \\
 0.32 & 0.55 & 0.09        &  9.0$\pm$0.2 &     -     &  6.0$\pm$0.2 &  8.4$\pm$0.3 \\
 0.32 & 0.55 & 0.11        &  9.6$\pm$0.3 &     -     &  6.4$\pm$0.3 &  9.6$\pm$0.5 \\
 0.32 & 0.55 & 0.13        & 11.1$\pm$0.5 &     -     &  6.8$\pm$0.4 & 10.3$\pm$0.6 \\
 0.32 & 0.55 & 0.15        & 10.8$\pm$0.6 &     -     &  9.8$\pm$0.8 & 11.7$\pm$0.9 \\
 0.32 & 0.55 & 0.17        & 16.8$\pm$1.6 &     -     & 11.0$\pm$1.8 & 15.6$\pm$2.2 \\
 0.32 & 0.55 & 0.19        & 15.9$\pm$3.3 &     -     & 22.0$\pm$5.5 & 22.1$\pm$5.5 \\
\hline
\hline
\end{tabular}\par}
\end{table}

The radiative tail from exclusive events is the dominant correction for our data at $z>0.8$. For 
$z\gtrsim0.9$ the contributions from exclusive pions become more than 50$\%$.

We also performed an alternative analysis using the code HAPRAD~\cite{Aku99}. The two results 
agree to within ${\pm 10-15 \%}$ in the relative contribution of the radiative exclusive tail. 
Thus, the resulting uncertainty is only at the 1$\%$ level or less.

{\bf Diffractive $\rho$:} Some of the detected events may originate from the decay of diffractive 
vector meson production. The underlying physics of this process, which can be described as that 
the virtual photon fluctuates into a vector meson, which subsequently can interact with the 
nucleon through multiple gluon (Pomeron) exchange, is distinctively different from the interaction 
of a virtual photon with a single current quark. Again, we used SIMC to evaluate such a 
diffractive $\rho$ meson contribution.

The p(e,e$^\prime\rho^\circ$)p cross section calculation was based on the PYTHIA \cite{pythia} 
generator, adopting similar modifications as implemented by the HERMES collaboration to describe 
lower-energy processes~\cite{Liebing}. Additional modifications were implemented to improve
agreement with $\rho^0$ cross section data from CLAS in Hall B at 
Jefferson Lab~\cite{Hadjidakis}.

The p(e,e$^\prime\rho^\circ$)p cross section can be written as
\begin{equation}
\sigma^{ep \rightarrow \rho p}(\nu,Q^2) = \Gamma_T \left( 1+\epsilon R \right) 
\left(\frac{M^2_\rho}{M^2_\rho+Q^2}\right)^n \sigma^{\gamma p \rightarrow \rho p},
\end{equation}
where $\Gamma_T$ is the transverse photon flux factor, $R=\sigma_L/\sigma_T$ is the ratio of
longitudinal to transverse cross sections, 
$\left(\frac{M^2_\rho}{M^2_\rho+Q^2}\right)^n$ ($n=2$ in PYTHIA)
is an additional factor that accounts for the  suppression of the cross section from virtual photons, 
and $\sigma^{\gamma p \rightarrow \rho p}$ is the photoproduction cross section. The modifications 
to the PYTHIA model implemented
for this analysis mimic those implemented by the HERMES collaboration:
\begin{enumerate}
\item The calculation of $\Gamma_T$ was performed with no high-energy approximations
\item An improved parametrization of $R=\sigma_L/\sigma_T$ 
\item Replacement of the exponent $n=2$ with $n\approx2.6$, more consistent with lower energy
data
\end{enumerate}

The $t$ dependence of the $\rho^\circ$ cross section is parametrized as
\begin{equation}
\frac{d \sigma}{d |t'|} = \sigma^{ep \rightarrow \rho p}(\nu,Q^2) b e^{-b |t'|},
\end{equation}
where $t'=t-t_{min}$ ($<0$ for electroproduction) and $b$ is the slope parameter. Note that at
$t'=0$, $b$ also impacts the overall scale of the forward cross section. The HERMES/PYTHIA model
assumed 
a value of $b\approx7$~GeV$^{-2}$ for all energies. However, CLAS data suggested that this constant
value of $b$ did not adequately describe the $t'$ dependence at JLab energies. The model used
in SIMC fits $b$ as a function of $c \Delta \tau$ (the vector meson formation time). Above 
$c \Delta \tau = 2$~fm, $b$ was taken to be a constant value of 7.0 GeV$^{-2}$, while for 
$c\Delta \tau <2$~fm, $b$ increased from 1.0 GeV$^{-2}$ to 7.0 GeV$^{-2}$ between 
$c\Delta \tau =0.4$~fm and 2~fm. 

Using the above model, the fraction of events due to pions from the decay of produced $\rho$ mesons
was estimated to range from a few percent at low $z$ to about 15$\%$ at $z=0.6$, and was subtracted
on bin by bin basis. The SIMC determination of the exclusive $\rho^\circ$ contribution
to the semi-inclusive yield was also checked independently using a program and model
developed by the CLAS collabration~\cite{harut_pc}. The two calculations were found to
agree to about the 10\% level.

{\bf  Pion decay:}
Pion decay in flight is included in the Monte Carlo simulation. Charged pions predominantly decay 
via $\pi^\pm\rightarrow\mu^\pm\nu_\mu$ with a branching fraction of 99.99$\%$. In SIMC the pion 
can decay at any point along its path in the magnet-free regions and at fixed points in the 
magnetic fields of the HMS. The muon momentum is calculated in the pion center-of-mass frame, 
where the angular distribution is uniform and the muon momentum is fixed. The muon is then 
followed through the spectrometer to the detector hut. Like in the experimental data, the muon is 
treated as if it were a pion in the reconstruction of target variables. In both the experimental 
and simulated data the muons constitute a background that is not removed by means of particle 
identification. However, the distribution of the muons in various reconstructed quantities is much 
broader than that of the pions. The fraction of pions decaying in flight on their way from the 
target can be calculated from the spectrometer central momentum and path length. While at low 
momentum roughly 20$\%$ of all pions decay on their way to the HMS detector hut, only a quarter of 
the muons fall within the acceptance and pass all cuts. More than 85$\%$ of all simulated muons 
that survive all cuts originate in the field free region behind the HMS dipole. We found the muon 
contamination after applying all cuts to be $\sim10\%$ at lowest momentum 1.4 GeV/$c$, and 
$\sim2\%$ in the momentum range $P_\pi=3-4$ GeV/$c$. The uncertainty associated with pion decay is 
estimated to be $\leq$1$\%$.

{\bf Pion Absorption:}
Some pions are lost due to nuclear interactions in the materials that the particles pass through 
on their way from the target to the HMS detector hut. Pions lost in hadronic interactions are 
largely due to absorption and large angle scattering, resulting in pions that do not strike all 
detectors required to form a trigger. The transmission of pions through the spectrometer is 
defined as the fraction of pions that do not interact with any of the materials.
 
The calculation of the pion transmission through the materials is determined by the choice of 
pion-nucleus cross section. In particular, the total cross section, which is defined as the sum of 
all hadronic interactions, represents an underestimate of the transmission. This can be explained 
in terms of the contribution of the individual pieces to the effective loss of pions. Elastic 
scattering is peaked in the forward direction (small angles), so that a large fraction of the 
elastically scattered pions are  expected to still produce a valid pion event. In addition, 
inelastic scattering does not necessarily correspond to an invalid trigger. On the other hand, a 
pion that is truly ``absorbed'' will clearly not result in a trigger. Therefore, the transmission 
is calculated from the reaction cross section which includes all hadronic interactions except for 
elastic scattering (${\sigma_{reac} = \sigma_{absorption}+\sigma_{inelastic}}$). The reaction 
cross section is approximately the average of the total and absorption cross sections and the 
uncertainty on the transmission can be estimated from these two limiting cases. At all kinematic 
settings of E00-108, the pion absorption was estimated to be below $1-2~\%$.

{\bf Kaon contamination:}
For low momentum settings ($P_{HMS}<2.4$ GeV/$c$), which is the case for our $x$-scan and 
$P_t$-scan data, kaon contamination is negligible. At these momenta the real e-$K$ coincidence 
peak will be well outside the e-$\pi$ coincidence peak, so the kaons are eliminated by the 
coincidence timing cut. In addition the HMS aerogel \v{C}erenkov detector can be used for 
separation of pions from kaons. For kinematics with pion momenta above 2.4 GeV/$c$, a correction 
was made to remove kaons from the pion sample. To estimate kaon contamination, the coincidence 
timing distribution was analyzed for $z$-scan data. It was assumed that the ``real'' coincidence 
peak is a sum of $\pi$+$K$, therefore this spectrum was fitted with a sum of two Gaussian 
distributions~\cite{T-Nav07-dis}.

The summary on kaon contamination is presented in Tables \ref{tab:mdual-kaon-z} and 
\ref{tab:mdual-kaon-phms}. The contamination at negative polarity was found to be very small. For 
positive polarity the worst case (about 10$\%$ contamination) was found at $z>0.85$, but a more 
typical contamination is less than 2$\%$

\begin{table}
\caption{\label{tab:mdual-kaon-z} Fraction of $K^+$ relative to $\pi^+$ (in $\%$) for hydrogen 
and deuterium targets, for two different cuts on the number of photoelectrons ($N_{pe}$) in the 
aerogel detector.}
{\centering \begin{tabular}{||c|cc||cc||}
\hline
\hline
      &$(K^+/\pi^+)_H$&            &$(K^+/\pi^+)_D$&             \\
\hline
 $z$  & $N_{pe}>0$    & $N_{pe}>1$ & $N_{pe}>0$    & $N_{pe}>1$  \\
\hline
 0.97 &  9.2          & 8.9       & 11.4         &  10.4      \\
 0.85 &  5.6          & 5.2       &  7.2         &   6.3      \\
 0.74 &  2.7          & 2.0       &  3.5         &   2.5      \\
 0.64 &  0.2          & 0.1       &  0.3         &   0.1      \\
\hline
\hline
\end{tabular}\par}
\end{table}

\begin{table}
\caption{\label{tab:mdual-kaon-phms} Kaon contamination ($\%$) in hydrogen and deuterium data at 
$N_{pe}>1$ cut on aerogel. Estimations are based on linear fits (Eq.~\ref{eq:kaon-corr}) of the 
data from Table~\ref{tab:mdual-kaon-z} }.
{\centering \begin{tabular}{||c|c|cc||cc||}
\hline
\hline
$P_{HMS}$ &  $z$  &  $K^+$(H)     & $K^-$(H)       &  $K^+$(D) & $K^-$(D)  \\
  GeV/$c$   &       &               &                &           &         \\
\hline
3.703     &  0.97 & 9.3$\pm$0.9 &  0.6$\pm$0.2 &  10.9$\pm$0.9
&  0.9$\pm$0.2  \\
3.222     &  0.85 & 7.6$\pm$0.5 &  0.4$\pm$0.3 &   8.9$\pm$0.5
& -0.2$\pm$0.3  \\
2.803     &  0.74 & 4.7$\pm$0.6 &  0.3$\pm$0.4 &   5.6$\pm$0.7
&  0.5$\pm$0.3  \\
2.439     &  0.64 & 3.3$\pm$1.1 & -0.1$\pm$0.2 &   3.6$\pm$1.3
&  0.1$\pm$0.1  \\
\hline
\hline
\end{tabular}\par}
\end{table}

Kaon contamination from the actual data was corrected for as:
\begin{equation}
\label{eq:kaon-corr}
Y_{exp}^{corr} = Y_{exp} \times (1 - R_k),
\end{equation}
where
\begin{eqnarray}
\label{eq:kaon-corr-fact}
\left.
\begin{array}{lll}
R_k^H =  0.265\times (z-0.63) 
&  & \\
R_k^D = 0.294 \times (z-0.63)
\end{array}
\right.
\end{eqnarray}
for hydrogen and deuterium targets.

{\bf Tracking Efficiency and Multiple Tracks:}
The tracking algorithm performs a $\chi^2$ minimization by fitting a straight line through both 
drift chambers. The tracking efficiency is defined as the ratio of events that should have passed 
through the drift chambers and the number of events for which a track was found. The fraction of 
events that should have passed through the drift chambers is defined by a requirement on hits in a 
fiducial area composed of a particular set of scintillator paddles. The efficiency depends on both 
the drift chamber hit efficiency and the tracking algorithm finding a track.

On average, the typical tracking efficiencies for pions in the HMS are better than 95$\%$ in most 
cases and are often better when electrons are detected in coincidence in SOS. At both positive and 
negative polarities of the HMS, and hydrogen and deuterium targets, during data taking the  beam 
current was optimized to keep the HMS rate below 500$-$600 kHz. This helps to minimize the 
well-known negative impact of multiple hits on track reconstruction in HMS. In many cases the 
events with more than 15 hits in both chambers are caused by projectiles scraping the edge of one 
of the magnets and causing a shower of particles, hence multiple hits in the drift chambers. 
Examination of such events and data taken  with high rates in HMS and/or SOS spectrometers 
revealed that sometimes the tracking code picks as the ``best'' among multiple track candidates a 
track that is not what a user looking at the event display would have picked. To eliminate this 
problem a new code was written that uses selection criteria different from the one track 
criteria~\cite{T-Nav07-dis}.

With the new code the resulting tracking efficiencies in HMS and SOS are on the level of 
97$\pm$0.1$\%$ and 98$\pm$0.4$\%$ respectively. (With the improved ``pruning'' algorithm we gain 
about 2-3$\%$ useful tracks). The difference between HMS and SOS mainly reflects the difference in 
incident count rates.

{\bf Coincidence and \v{C}erenkov detector blocking:}
Another source of event loss is connected with coincidence and \v{C}erenkov detector blocking 
effects. The coincidence time is determined by a clock that starts when an HMS signal arrives and 
stops when the SOS signal arrives. Two effects can cause the coincidence timing for good events to 
fail. In the first case, a random SOS single arriving before the coincident particle can stop the 
clock too early, effectively blocking the coincidence. A cut on the coincidence time will largely 
remove these events. The second effect is that a late SOS trigger can confuse the timing logic in 
such a way that the coincidence timing clock, which usually starts with the HMS and stops with the 
SOS, starts and stops with SOS (``retiming''). Wrongly timed events appear at lower (coincidence 
blocking) and higher (retiming) TDC channel numbers. 

The coincidence blocking factor was calculated for a number of runs taken at different trigger 
rates using HMS-SOS raw (not corrected for pathlength) coincidence time (TDC's) spectra.
We found that the coincidence blocking correction, $k_{coin}$, depends nearly linearly on the rate 
of the pretrigger, 5-40 kHz for the case of the electron spectrometer, the SOS. The coincidence 
blocking correction was then parameterized in terms of
\begin{equation}
\label{eq:coin-blk2}
k_{coin} =  1-{\alpha}N_{strig},
\end{equation}
where $\alpha\approx2.218\times 10^{-5}$ and $N_{strig}$ is the SOS trigger rate (in Hz). This 
correction for our coincidence time window of 120 ns was up to 4.5$\%$, with an uncertainty of 
$\sim$0.1$\%$. 

The HMS gas \v{C}erenkov detector is used for electron rejection in the $\pi^-$ production case. 
The effective time window is given by the \v{C}erenkov detector ADC ``gate'' and is approximately 
100 ns wide. The loss of pions due to \v{C}erenkov-detector blocking is due to electrons passing 
through the detector after the first particle (pion), but within the effective ADC gate window. In 
this situation, the signal from the electron will be associated with the original pion trigger and 
the pion event will be mis-identified as an electron. Such mis-identified pions are eliminated due 
to analysis cuts, so that the electron event effectively blocks the HMS \v{C}erenkov detector for 
good pion events. The number of pions lost due to the \v{C}erenkov detector blocking depends only 
on the rate of electrons into the spectrometer and does not depend strongly on variations in run 
to run characteristics. Therefore, we have used a small sample of measurements to determine the 
size of the correction for given kinematics, and for electron rates in the range of 20-500 kHz 
parameterized it in a functional form:
\begin{equation}
\label{eq:cer-blk}
\tau_{cer} = 1 - 1.6\times 10^{-7}\times {\frac {N_{hecl}} {T_{run}} }, \\
\end{equation}
where $N_{hecl}$ is the clean electron trigger (ELCLEAN) counts, defined by high level cuts on 
calorimeter and \v{C}erenkov detector, and $T_{run}$ is the duration of the run (in seconds). 
\v{C}erenkov-detector blocking effect was on the level of $\sim$2$\%$ at 100 kHz, and reached up 
to $\sim$6$\%$ at 400 kHz, with a systematic error less than 1$\%$. The uncertainty in the HMS 
\v{C}erenkov detector blocking correction is largely attributed to the uncertainty in the 
\v{C}erenkov detector timing window. In particular the effective \v{C}erenkov detector gate width 
can be slightly larger than the measured ADC gate ($\approx$100 ns). While the ADC gate is fixed, 
the \v{C}erenkov detector signal itself has some width and the overlap determines an effective 
gate width.

{\bf Computer Dead Time:}
The computer dead time strongly depends on the trigger rate and experimentally is directly 
measured by scalers that record the number of triggers ($N_{trig}$) and pretriggers 
($N_{pretrig}$). Since pretriggers are generated for each particle, and triggers are only read out 
for those events for which the Trigger Supervisor is not busy, the computer live time is 
$N_{trig}/ N_{pretrig}$. 

The computer dead time varied from a few percent at low rates to up $30\%$ at trigger rate $\sim$2 
kHz. The uncertainty in the computer live time measurement is estimated by the deviation of the 
measured value from the value calculated from the total rate. The resulting uncertainty is 
${\sim 0.2 \%}$. The electronic dead time was always $\leq 1\%$ and often negligible.

For the E00-108 experiment, the computer and electronic live time corrections are applied 
run-by-run. More details of the analysis and corrections can be found in 
Ref.~\cite{T-Nav07-dis,H-Mkrt08-dis}.

{\bf Other corrections:}
From a measurement detecting positrons in SOS in coincidence with pions in HMS, we found the 
background originating from $\pi^0$ production and its subsequent decay into two photons and then 
electron-positron pairs, or $e^+e^-\gamma$ directly, negligible. In addition, a small $\sim2\%$ 
correction was made to the deuterium data to account for a small Final-State Interaction effect of 
the pions traversing the deuterium nucleus \cite{Sar05}.

\subsection{Model Cross Section and Monte Carlo Simulations}
We added the possibility of semi-inclusive pion electroproduction to the general Hall C Monte 
Carlo package SIMC \cite{SIMC-SEMI}, using Eq.~(\ref{eq:semi-parton}). 
The CTEQ5 next-to-leading-order (NLO) parton distribution functions were used to parametrize 
$q_i(x,Q^2)$ \cite{CTEQ}, and the fragmentation function parameterization for
$D^+_{q_i \rightarrow \pi}(z,Q^2) + D^-_{q_i \rightarrow \pi}(z,Q^2)$, with $D^+$ ($D^-$) the 
favored (unfavored) fragmentation function, from Binnewies {\sl et al.} \cite{BKK95}. 
The remaining unknowns are the ratio of $D^-/D^+$, the slope $b$ of the $P_t$ dependence, and the 
parameters $A$ and $B$ describing the $\phi$ dependence. Both the $D^-/D^+$ ratio~\cite{Geiger} 
and the $b$-value~\cite{Hommez} are taken from HERMES analysis. The latter is chosen
for consistency with the comparisons shown in our earlier publication~\cite{Nav07}, but
closely coincides with the averaged value for all data. We will study the detailed $P_t$-dependence
of our data later on in Section VIIF.
 
When analyzing our data as a function of $P_t$, we found that the $Q^2$-dependence of the cross 
sections needed to be altered slightly from the factorized high-energy 
expectation~\cite{Mkrt08-q2} to obtain a smooth $P_t$ dependence. This is not too surprising, as 
the (low) energies of our semi-inclusive pion production measurements are beyond the region where 
the BKK fragmentation functions were shown to describe existing data. Hence, we introduced an 
additional $Q^2$-dependent multiplicative term in the model cross section in the form
\begin{equation}
\label{eq:fitfunc1}
F(Q^2) = 1 + {C_1}\cdot \ln(Q^2) + {\frac {C_2}{Q^2}} + {\frac {C_3}{Q^4}. }
\end{equation}
The parameters $C_1$, $C_2$ and $C_3$ were adjusted in such a way that the calculated yields from 
the SIMC simulation match the experimental data. To accomplish this, the ratio of experimental and 
SIMC yields were calculated in a number of $Q^2$ bins, and the model cross section was iterated 
until the ratios approach unity. A variety of fits, with different combinations of data included, 
more complicated fit functions (including $\phi$-dependent terms with additional binning in $\phi$ 
and $P_t$) rendered parameters $C_i$ that remained reasonably stable, within $\pm 10-20 \% $. 
As average ``best values'' for the fit parameters, we adopted~$C_1$ = 0.889, $C_2$ = -2.902 and 
$C_3$ =3.050. Recall that for most of the cross section results (at $P_t \approx$ 0.05 GeV/$c$) we 
neglected the $\phi$-dependence and kept the parameters $A$ and $B$ at 0, in accordance with both 
theoretical expectations (discussed in subsection \ref{sec:pt_phi} ), and our own findings (see 
subsection \ref{sec:z_phi_exp} ).


\section{Systematic Uncertainties}

As part of the analysis, several systematic studies were performed on the data to verify that the 
measured cross sections and ratios are not biased by the detector, event selection and background 
correction effects. The level of corrections applied to the experimental data and related 
systematic uncertainties are listed in Table~\ref{tab:corr_err}. 

\begin{table}
\caption{\label{tab:corr_err} Corrections and systematic uncertainties.}
{\centering \begin{tabular}{||c|c|c||}
\hline
\hline
Source of correction         & Range ($\%$) & Systematics ($\%$) \\
\hline
Detector inefficiencies      &     5-10     &        1-2         \\
Target wall contribution     &     2-3      &        1.0         \\
Accidentals                  &    10-20     &        1-2         \\
Pion absorption              &     1-2      &        1.0         \\
Pion decay                   &     2-10     &       1.0          \\
Kaon contamination           &    0.2-2.0   &       0.5 ($z<0.7$)\\
Radiative corrections        &     5-10     &       1-2          \\
Exclusive tail               &     5-15     &   0.5-2.5 ($z<0.8$)\\
Pions from diffractive $\rho$&     5-15     &       0.5-2.5      \\
Computer Dead Time           &     5-25     &       0.2          \\
Coincidence blocking         &     1-4.5    &       0.1          \\
\v{C}erenkov detector blocking &     2-4      &       $\leq$1      \\
Other corrections            &     1-2      &        1.0         \\
\hline
Total                        &     15-40    &       3.5-7.5      \\
\hline
\hline
\end{tabular}\par}
\end{table}

For absolute cross sections we have added all systematic uncertainties in quadrature. Note, that 
in practice the range of applied corrections and related systematic errors are slightly different 
for $\pi^+$ and $\pi^-$. For example, the \v{C}erenkov blocking is  clearly far larger for 
$\pi^-$. Part of the corrections (such as radiative, pion 
decay, detector inefficiencies) are nearly identical for $\pi^+$ and $\pi^-$ and cancel in the 
ratios, hence related systematic uncertainties are much smaller for the ratios. Below we will 
discuss the most dominant sources of systematic uncertainties related with pions from the
radiative tail from exclusive pion electroproduction and pions from the decay of diffractive
$\rho^o$ mesons.

\subsection{Uncertainties related to the exclusive pion tail}
The model used in SIMC for exclusive pion electroproduction
mainly focused on parallel kinematics (with the outgoing pion 
along the direction of the virtual photon) for the purpose of understanding the $z$-scan data. To 
estimate the possible systematic error that arose from ignoring the $LT$ and $TT$ interference 
terms, and to test the absolute magnitude of the correction, we extracted simulated yields for the 
exclusive radiative tail calculated using our nominal, empirical parameterization as well as the 
MAID model. 

\begin{figure}
\begin{center}
\epsfxsize=3.40in
\epsfysize=3.20in
\epsffile{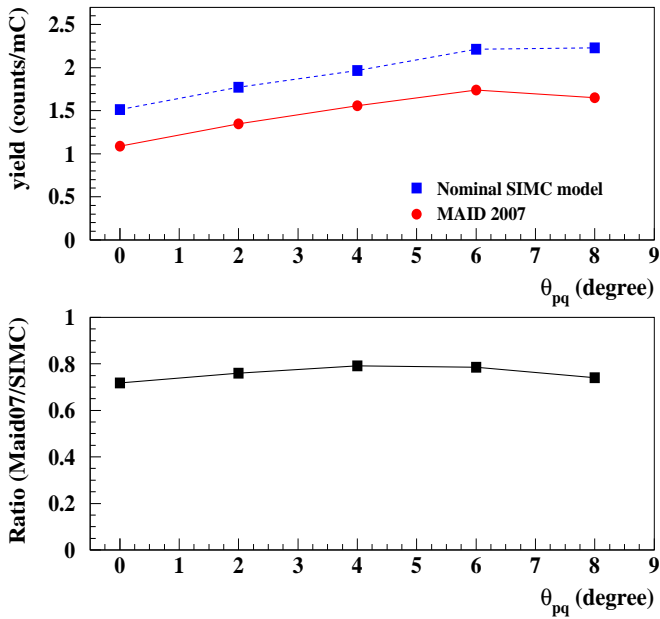}
\caption{\label{fig:excl-dpos} (Color online)
The exclusive radiative tail simulations for positive pions from deuterium using the nominal model 
cross section in SIMC described in the text. The top plot shows the total simulated yield from 
exclusive radiative events vs. the angle between the outgoing pion and the virtual photon. 
All points are at a fixed $z=0.55$. For comparison, the yield using the MAID model (which has 
limited validity for $W>2$~GeV) is also shown. The bottom panel shows the ratio of exclusive 
yields from both models.}
\end{center}
\end{figure}

\begin{figure}
\begin{center}
\epsfxsize=3.40in
\epsfysize=3.20in
\epsffile{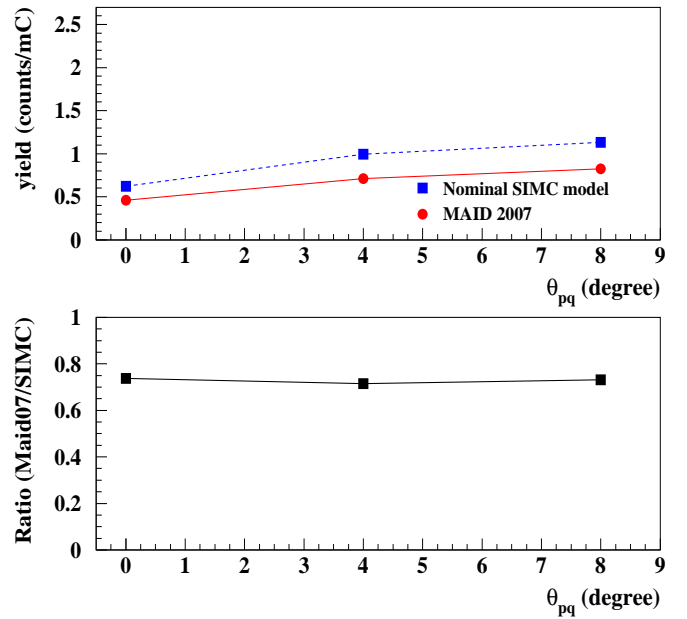}
\caption{\label{fig:excl-dneg} (Color online)
The exclusive radiative tail simulations for negative pions from deuterium 
(as in Fig.~\ref{fig:excl-dpos}). }
\end{center}
\end{figure}

\begin{figure}
\begin{center}
\epsfxsize=3.40in
\epsfysize=3.20in
\epsffile{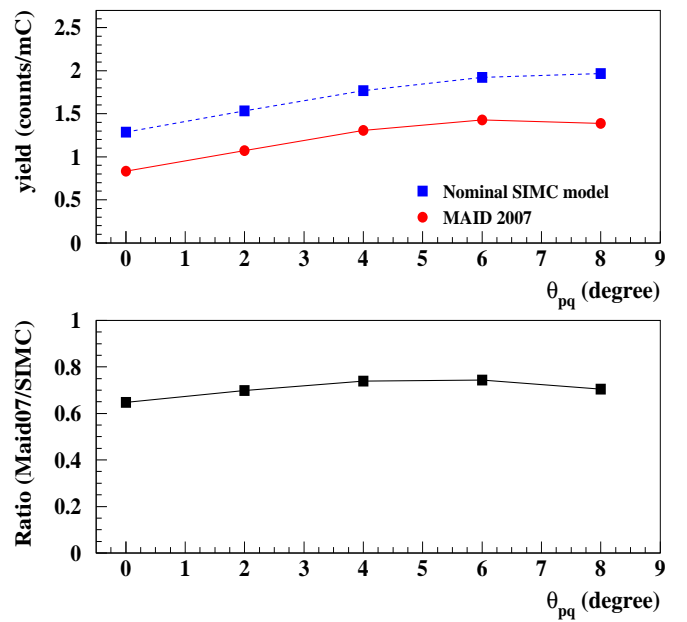}
\caption{\label{fig:excl-hpos} (Color online)
The exclusive radiative tail simulations for positive pions from hydrogen 
(as in Fig.~\ref{fig:excl-dpos}).}
\end{center}
\end{figure}

Figures~\ref{fig:excl-dpos},~\ref{fig:excl-dneg} and \ref{fig:excl-hpos} show the results of these 
simulations (MAID = red circles, SIMC= blue squares). The yields from each model [top panel] as 
well as the ratio between the two [bottom panel] are plotted versus $\theta_{pq}$. While the two 
calculations differ in the absolute magnitude of the exclusive yield, the ratio of yields shows 
little dependence on the outgoing pion angle. This suggests that the choice to ignore the 
interference terms in the SIMC parameterization had minimal effect (the contribution from the 
interference terms should increase at larger pion angles) over the region studied.

The effective $W$ at the vertex for events that contribute to these yields is around 1.9 GeV (on 
average) so the empirical parameterization in SIMC, which agrees well with the JLab data at 
$W=1.95$~GeV, should be more appropriate. Half of the difference between the two results was used 
as an estimate of the systematic uncertainty for the contribution of radiative exclusive events to 
the semi--inclusive yield.

\subsection{Uncertainties related to events from diffractive $\rho$ production}
This uncertainty is related to the choice of the parameterization for the $\rho^o$ cross sections. 
As mentioned, we used cross section based on the PYTHIA~\cite{pythia} generator with modifications 
as implemented by the HERMES collaboration~\cite{Liebing} and additional modifications to 
improve agreement with CLAS data~\cite{Hadjidakis}. To estimate systematic 
uncertainties related to the diffractive $\rho$ subtraction, all calculations were repeated with 
slightly ($\sim$10$\%$) different values of the parameters. Thus it was found that the diffractive 
$\rho$ subtraction contributes a systematic uncertainty of up to $\approx{2.5}\%$ .

\subsection{Other sources of systematic uncertainties}
The systematic uncertainties due to normalization, target thickness, computer and electronic dead 
time, beam charge measurement, beam energy and spectrometer kinematics combine to approximately 
2$\%$. We note that targets and spectrometer polarity were exchanged frequently in this 
experiment, without noticeable effects.

The overall systematic uncertainty due to spectrometer acceptances is estimated to be $\leq{1}\%$. 
This is because the spectrometers have a sufficient wide vertex length acceptance to view a 4 cm 
extended target with limited acceptance losses, given that the SOS spectrometer angle was limited 
to relatively forward angles, and that the particles of interest cover a central region of the SOS 
momentum acceptance only. The HMS spectrometer has a 10 cm uniform vertex length acceptance, and 
was used in the E00-108 experiment to view a 4 cm extended target at angles of 20 degrees or less. 
Hence, the HMS has full acceptance. 

It was verified that the level of changes in our results due to variation of the values of PID 
cuts in the analysis slightly varies from case to case but is less than 1-2$\%$. These variations 
are within the systematic uncertainties assigned to the detector efficiencies, and so are not 
taken as a separate additional source of uncertainty.

\section{Experimental Results}

Some of the results have been described in two previous papers~\cite{Nav07,Mkrt08}. In the first, 
we observed for the first time the quark-hadron duality phenomenon in pion electroproduction, and 
the relation with a precocious low-energy factorization approach in semi-inclusive deep inelastic 
scattering. We quantified the latter by constructing several ratios of pion electroproduction 
cross sections off proton and deuteron targets, and found the ratio of favored to unfavored 
fragmentation functions to closely resemble that of high-energy reactions, up to about $z$ = 0.7 
or missing mass $M_x^2 >$ 2.5 GeV$^2$ or so. In the second, we studied the transverse momentum 
dependence of semi-inclusive pion production, and found the dependence from the deuteron target to 
be slightly shallower than from the proton. In the context of a simple model, we related these 
measurements to the initial transverse momentum widths of down and up quarks, and the transverse 
momentum widths of favored and unfavored fragmentation functions. The results presented here 
supersede those of Ref.~\cite{Mkrt08}, which were for reduced statistics and improper application 
of some corrections.

In this article, we will present our full results in terms of cross sections and ratios for 
semi-inclusive charged-pion electroproduction off proton and deuteron targets, and relate the 
findings to the Quark-Parton Model expectations, further highlighting the onset of a precocious 
high-energy factorized parton model description. However, for completeness we will start with a 
short Section recapitulating two relevant findings of the previous publications.

\subsection{\bf The azimuthal angle $\phi$ and $z$ dependence of the cross sections}
\label{sec:z_phi_exp}
In the E00-108 experiment, the average value of the angle $\phi$ is correlated with $P_t$, see 
Fig.~\ref{fig:phipt} (the effect is tabulated in~\cite{Mkrt08}). However, we will initially show 
results that have only small average value $P_t$ $\approx$ 0.05 GeV/$c$, to later on come back to 
the results as a function of $P_t$, including those for $P_t$ up to 0.45 GeV/$c$, in a separate 
Section.

For the results at low $P_t$, one expects small to negligible contribution from the interference 
terms $A$ and $B$ in Eq.~\ref{eq:semi-parton}. We found no statistically significant difference 
between the results for $\pi^+$ or $\pi^-$, or proton or deuteron targets~\cite{Mkrt08-cos}, and 
therefore combined all four cases together. Taking the systematic uncertainties of approximately 
0.03 into account, we find values of $A$ and $B$ close to zero, with no noticeable $x$ or $z$ 
dependence. When averaged over all data, we find A = 0.02 $\pm$ 0.02 and B = -0.04 $\pm$0.02 at 
$P_t \approx$ 0.05 GeV/$c$. Folding this back into Eq.~\ref{eq:semi-parton}, we can neglect the
azimuthal-angle dependent corrections to the cross section and ratio results presented in the next 
Sections.  The small values of $A$ and $B$ at small $P_t$ for SIDIS kinematics are consistent with 
the expectations from kinematic shifts due to parton motion as described by Cahn~\cite{Cahn} and 
Levelt-Mulders~\cite{Levelt-Mulders}. 

Since the bulk of our data is taken at low $P_t$ = 0.05 GeV/$c$, we will neglect any 
$\phi$-dependence and assume A = B = 0 until we explicitly revisit the $P_t$ dependence of the 
measured cross sections and ratios later on.

In our first publication~\cite{Nav07} we compared the measured $^{1,2}$H(e,e$^\prime \pi^\pm$)$X$ 
cross sections as a function of $z$ (at $x$ = 0.32) with the results of a parton model calculation 
assuming CTEQ5M  Parton Distribution Functions (PDFs) at Next-to-Leading Order~\cite{CTEQ} and the 
parameterized Fragmentation Functions (FFs) of Binnewies, Kniehl and Kramer (BKK)~\cite{BKK95}. 
The ratio of unfavored to favored fragmentation functions $D^-/D^+$, and the slope $b$-values of 
the $P_t$ dependences of the cross sections were taken from HERMES analysis~\cite{Geiger,Hommez}. 
We found excellent agreement between data and Monte Carlo for $z <$ 0.65, but striking deviations 
around $z$ = 0.8. Within our kinematics (at $P_t \sim$ 0), $M_x^2$ is almost directly related to 
$z$, as $M_x^2 = M_p^2 + Q^2(1/x-1)(1-z)$. Hence, we attributed the large ``rise'' in the data 
with respect to the simulation at $z >$ 0.8 to the $N-\Delta(1232)$ region. Indeed, if one 
considers a $^1$H(e,e$^\prime \pi^-$)$X$ spectrum as function of missing mass of the residual 
system $X$, one sees only one prominent resonance region, the $N-\Delta$ region. Apparently, above 
$M_x^2 \approx $ 2.5~GeV$^2$, there are already sufficient resonances to render a spectrum 
mimicking the smooth $z$-dependence as expected according to the factorization ansatz of 
Eq.~(\ref{eq:semi-parton}).

Much of the data shown later on as function of $x$, $Q^2$ and $P_t$, will be centered around 
$z$ = 0.55, so well within the region where effects due to the N-$\Delta$ transition can be 
neglected, and excellent agreement was found between the low-energy pion electroproduction data 
and high-energy parton model expectations. We do note that this $z <$ 0.7, or  
$M_x^2 >$ 2.5 GeV$^2$ cut, corresponds to the prediction of Close and Isgur where duality and 
low-energy factorization would become valid~\cite{CI01,CM09}.

\subsection{\bf The $x$- and $Q^2$-dependence of the cross sections}
At a fixed value $z$ = 0.55, well within the range where we found excellent agreement between our 
cross section data and a naive high-energy ansatz in terms of next-to-leading-order (NLO) parton 
distributions (PDFs) convoluted with the BKK fragmentation functions (FFs), we will now study the 
$x$- and $Q^2$-dependence of the $^{1,2}H(e,e^\prime \pi^\pm)$X cross sections.

We studied the $x$-dependence in the range $0.2\leq x \leq 0.6$ by varying the angle of the 
scattered electron, while keeping the beam energy and the virtual photon energy 
($\nu\approx$ 3.9 GeV) fixed. An additional advantage of this choice of $z$ and $\nu$ is that the 
corresponding outgoing pion momentum is larger than 2 GeV/$c$, well in the region where the 
$\pi-N$ cross sections behave smoothly such that final-state interactions do not overly complicate 
interpretation of the pion yields. Restricting the kinematics to such large pion momenta permits 
to neglect possible differences in $\pi^+$ and $\pi^-$ rescattering. In a simple Glauber 
calculation we estimated the total pion absorption correction due to rescattering to be 2\% for a 
deuterium target, and the difference between $\pi^+$ and $\pi^-$ to be less than 1\%. We apply the 
2\% deuteron correction for all $^2$H data, and assume a constant $b$ = 4.66 (GeV/$c$)$^{-2}$ to 
describe the $P_t$ dependence, somewhat different from what we will derive from the specific 
$P_t$-dependent measurements later on. Even though this does not affect the cross sections 
represented, it does impact the overall agreement with the parton-model calculations, where $b$ 
comes in as an overall normalization. The choice $b$ = 4.66 (GeV/$c$)$^{-2}$ is chosen in these 
figures for consistency with the comparisons shown in our earlier publication~\cite{Nav07}. 
We note that this choice is also consistent with the HERMES findings.

\begin{figure}
\begin{center}
\epsfxsize=3.40in
\epsfysize=3.20in
\epsffile{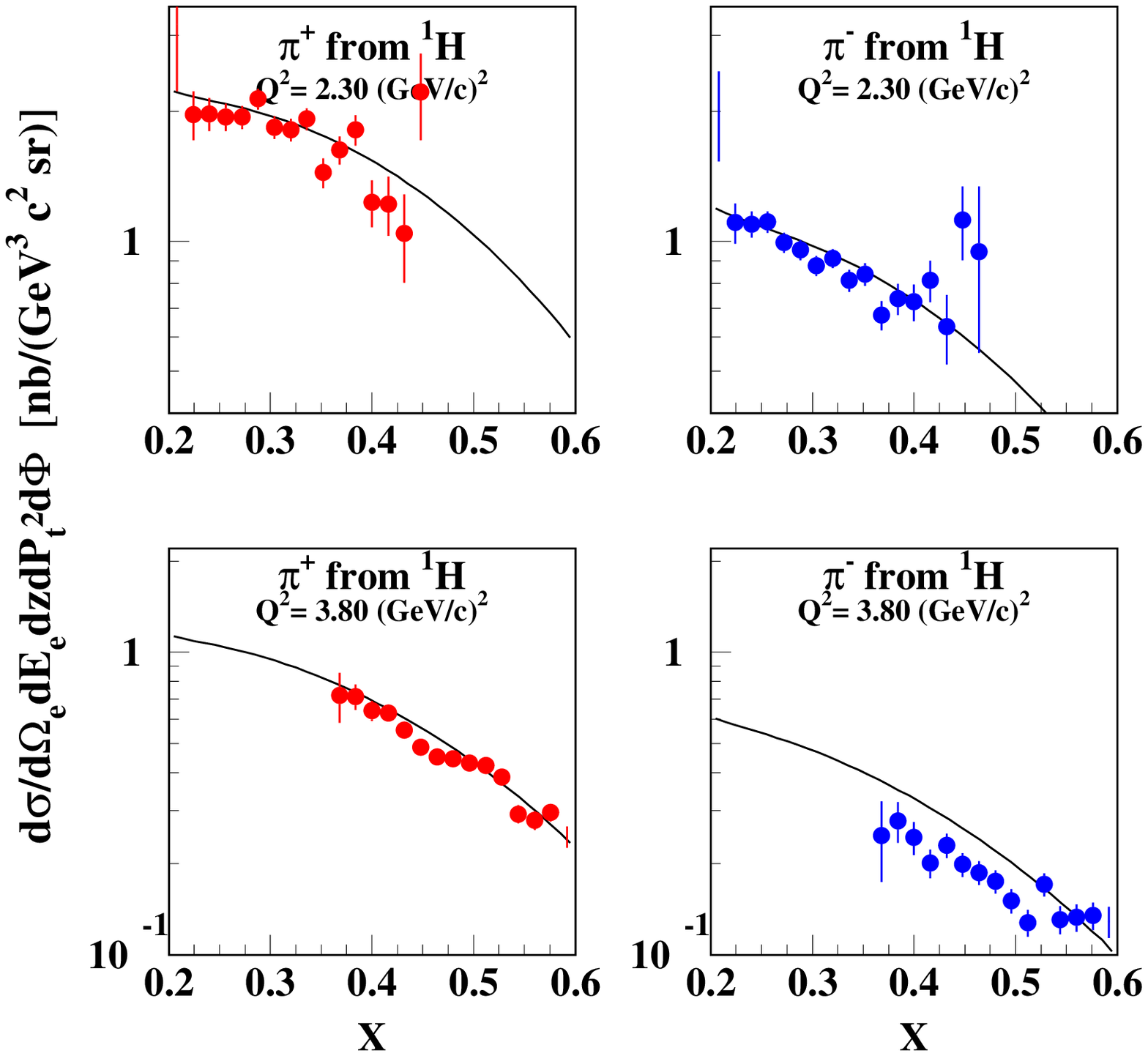}
\caption{\label{fig:xsect_x_h} (Color online)
The $^1$H(e,e$^\prime \pi^+$)X (left) and $^1$H(e,e$^\prime \pi^-$)X (right) 
cross sections  at $z=0.55$ as a function of $x$, at 
Q$^2_{set}$ = 2.30 (GeV/$c$)$^2$) (top) and at 
Q$^2_{set}$ = 3.80 (GeV/$c$)$^2$) (bottom), respectively. 
Solid curves are parton model calculations. Solid symbols are data after 
events from diffractive $\rho$ production are subtracted (see text). }
\end{center}
\end{figure}

\begin{figure}
\begin{center}
\epsfxsize=3.40in
\epsfysize=3.20in
\epsffile{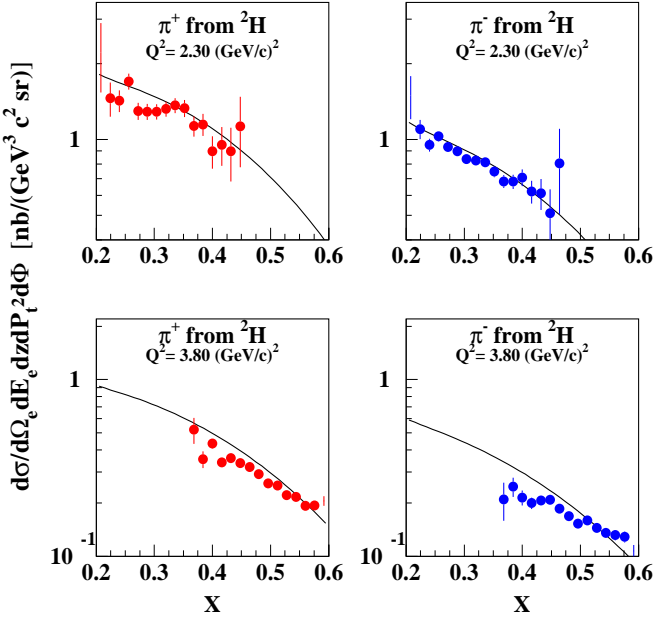}
\caption{\label{fig:xsect_x_d} (Color online)
The $^2$H(e,e$^\prime \pi^+$)X (left) and $^2$H(e,e$^\prime \pi^-$)X (right) 
cross sections  at $z=0.55$ as a function of $x$, at 
Q$^2_{set}$ = 2.30 (GeV/$c$)$^2$) (top) and at 
Q$^2_{set}$ = 3.80 (GeV/$c$)$^2$) (bottom), respectively. 
Solid curves are parton model calculations. Solid symbols are data after 
events from diffractive $\rho$ production are subtracted (see text). }
\end{center}
\end{figure}

We present in Figs.~\ref{fig:xsect_x_h} and ~\ref{fig:xsect_x_d} a selection of differential cross 
sections for the $^1H(e,e^\prime \pi^\pm)$X and $^2H(e,e^\prime \pi^\pm)$X reactions, 
respectively, at low and high ``$x_{set}$'' values of the experiment (by this we mean the $x$ 
value as calculated from the central spectrometer kinematics; using the finite 
spectrometer acceptances we present multiple $x$-bins). Since we vary the scattered electron 
angle, a variation in $x_{set}$ (or $x$) likewise corresponds to a variation of $Q^2_{set}$ 
(or $Q^2$). For $x_{set}$ = 0.32, $Q^2_{set}$ = 2.30 (GeV/$c$)$^2$ and for $x_{set}$ = 0.53, 
$Q^2_{set}$ = 3.8 (GeV/$c$)$^2$ cross sections are shown along with the model calculations. 
For simplicity, we have only considered CTEQ5M parton distributions at NLO~\cite{CTEQ} and the 
BKK~\cite{BKK95} fragmentation functions, allowing for a slightly modified $Q^2$ dependence. The 
scope of this work is to judge how well low-energy pion electroproduction cross sections 
(and ratios) compare with parton model expectations, and comparisons with other possibly more 
sophisticated model calculations is beyond this scope. We conclude that the $x$-dependence agrees 
reasonably well with the model calculations, but differences in the absolute magnitude of the 
cross section are apparent in certain cases.

Next, we want to compare the $Q^2$-dependence of the measured cross sections with the parton model 
expectations. However, as described we varied $x$ by a change in the electron scattering angle, 
which correlates higher $x$ with higher $Q^2$. A similar correlation exists within the finite 
spectrometer acceptances. Hence, we need to remove this correlation to present data as function of 
$Q^2$ only, at a fixed value of $x$. We found that $x$ = 0.40 is the optimal value to choose, 
accessible for each of the five settings of the $x$-scan ($x_{set}$ = 0.26, 0.32, 0.39, 0.46 and 
0.53). The $x$-dependence of the parton model, determined from SIMC simulations over the 
experimental acceptance, was used to scale all data within one $x$-scan setting to $x$ = 0.40. 
This was accomplished by using the ratios of the normalized yields between data and Monte Carlo 
for each $Q_i^2$ bin: $Y_i^{exp}/Y_i^{MC}$. The cross sections of different $x$-scan settings were 
then corrected to $x$ = 0.40 using these ratios and the corresponding model cross sections at  
$x$ = 0.40.

\begin{figure}
\begin{center}
\epsfxsize=3.40in
\epsfysize=3.20in
\epsffile{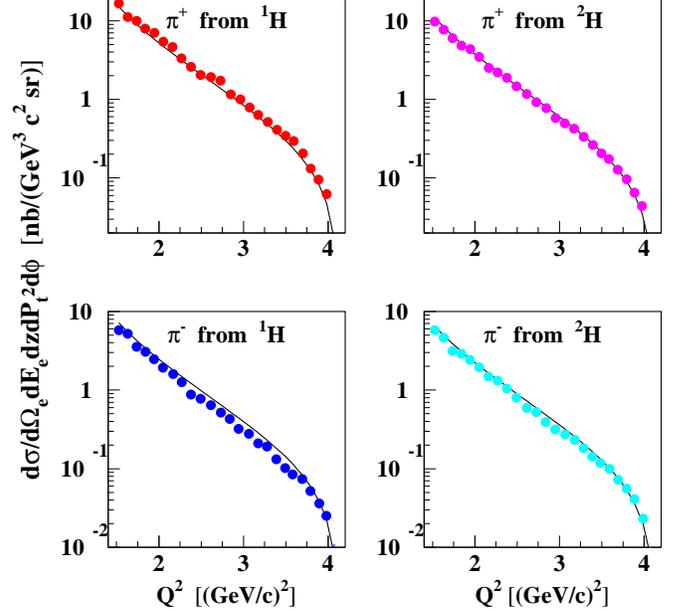}
\caption{\label{fig:xsect-q2} (Color online)
The $^{1,2}$H(e,e$^\prime \pi^\pm$)X cross sections at fixed values of $x$ = 0.40 and $z$ = 0.55, 
as a function of $Q^2$. The solid curves are the simple quark-parton model calculations following 
a high-energy factorized description. Solid symbols are data after events from diffractive $\rho$ 
production are subtracted (see text). }
\end{center}
\end{figure}

The $^{1,2}H(e,e^\prime \pi^\pm)$X cross sections for all five settings of the $x$-scan are shown 
versus $Q^2$ in Fig.~\ref{fig:xsect-q2}, bin-centered to $x$=0.40 with this technique. 
The curves are the parton model calculations, and describe the $Q^2$-dependence of our data 
remarkably well. Note that the $Q^2$-dependence is steeper than naively assumed for a swing in 
$Q^2$ from about 1.5 to 4.0 (GeV/$c$)$^2$, since these are cross sections, ``bin-centered'' to 
fixed $x$ = 0.40, which induces trivial changes in for instance the beam energy and the resulting 
photon flux. Similarly, one can see that the calculated cross section drops very fast around 
$Q^2$ = 4.0 (GeV/$c$)$^2$, which reflects that one has reached the edge of what is kinematically 
possible with our experimental setup. As before, the solid symbols are the data after events from 
coherent diffractive $\rho$ production are subtracted. Such corrections were estimated to be 
$\leq 10 \%$ for 
these cross sections, and do not affect the conclusion that the $Q^2$-dependence of our data 
surprisingly conforms to the high-energy (quark-parton) expectations.

\subsection{\bf The $z$, $x$ and $Q^2$ dependence of the cross sections ratios ($\pi^+/\pi^-$ and 
D/H) }
With the semi-inclusive pion electroproduction data at our relatively low energies closely 
resembling the high-energy parton model expectations, we now turn our attention to various ratios 
constructed from the data, in an effort to quantify the agreement with the quark-parton model. 
Especially the ratio of charged $\pi^+$ and $\pi^-$ semi-inclusive electroproduction cross 
sections (or the ratio of their normalized yields, $\pi^+/\pi^-$) is a quantity relatively easy to 
measure accurately with focusing magnetic spectrometers. In contrast to a large-acceptance 
detector, the acceptance, reconstruction, and detection efficiencies for positively- and 
negatively-charged pions are very similar in a focusing magnetic spectrometer, allowing for 
precision comparisons.
 
With the assumptions of factorization, isospin symmetry and charge conjugation (and neglecting 
heavy quarks in the valence-quark region), the cross sections (or normalized yields) of 
$\pi^\pm$ production on protons and neutrons at fixed $Q^2$ can be presented as:
\begin{widetext}
\begin{eqnarray}
\label{eq:sidis_yields_nucl}
\left.
\begin{array}{lll}
{\sigma_p}^{\pi^+}(x,z) \propto  4u(x)D^+(z) + d(x)D^-(z) + 
 4\bar{u}(x)D^-(z) + \bar{d}(x)D^+(z)  \\
{\sigma_p}^{\pi^-}(x,z) \propto  4u(x)D^-(z) + d(x)D^+(z) + 
 4\bar{u}(x)D^+(z) + \bar{d}(x) D^-(z)  \\
{\sigma_n}^{\pi^+}(x,z) \propto  4d(x)D^+(z) + u(x)D^-(z) +
 4\bar{d}(x)D^-(z) + \bar{u}(x)D^+(z)  \\
{\sigma_n}^{\pi^-}(x,z) \propto  4d(x)D^-(z) + u(x)D^+(z) +
 4\bar{d}(x)D^+(z) + \bar{u}(x) D^-(z),  \\
\end{array}
\right.
\end{eqnarray}
\end{widetext}
with $D^+$ and $D^-$ the favored and unfavored fragmentation functions, respectively.

The ratio of charged pion production on proton and neutron will then be:
\begin{eqnarray}
\label{eq:pirat-form1}
\left.
\begin{array}{lll}
\frac {{\sigma_p}^{\pi^+}}  {{\sigma_p}^{\pi^-}}  = \frac
{ 4u(x) + \bar{d}(x) + (d(x) + 4\bar{u}(x))\cdot \frac {D^-(x)} {D^+(x) } } 
{ (4u(x) + \bar{d}(x))\cdot \frac {D^-(x)} {D^+(x)} + d(x) + 4 \bar{u}(x)}   
&   & \\
\frac {{\sigma_n}^{\pi^+}} {{\sigma_n}^{\pi^-}} = \frac
{ 4d(x) + \bar{u}(x) + (u(x)+4\bar{d}(x))\cdot \frac {D^-(x)} {D^+(x) } }
{ (4d(x) + \bar{u}(x))\cdot \frac {D^-(x)} {D^+(x)} + u(x) + 4 \bar{d}(x)}.
\end{array}
\right. 
\end{eqnarray}
It is obvious that the fragmentation functions do not completely cancel in the $\pi^+/\pi^-$ and 
D/H ratios, and some $z$-dependence remains carried by the term $D^-/D^+$. However, in the ratio 
of charge-combined cross sections, such $z$-dependence will completely cancel. Some of those 
``super-ratios'' were presented in our first publication~\cite{Nav07}, and showed validity of the 
factorization assumption up to $z \sim$ 0.65: no $z$-dependence was found.

\begin{figure}
\begin{center}
\epsfxsize=3.40in
\epsfysize=3.20in
\epsffile{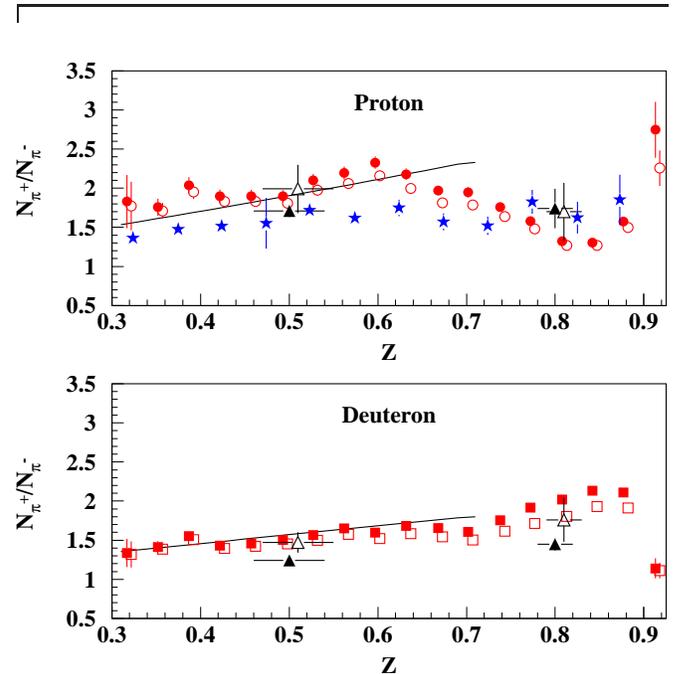}
\caption{\label{fig:pirat-pd-z} (Color online)
The ratio $\pi^+/\pi^-$ for proton (top panel) and deuteron (bottom panel) targets as a function 
of $z$, at $x=0.32$. Solid (open) circles and squares represent the data after (before) events 
from diffractive $\rho$ decay are subtracted. Solid and open triangles represent existing Cornell 
data \protect\cite{Beb77b} at $Q^2$ = 2.0~(GeV/$c$)$^2$ and $x$ = 0.24, and 
$Q^2$ = 4.0 (GeV/$c$)$^2$ and $x$ = 0.50, respectively. Stars represent HERMES 
data~\cite{Airapet01} at average values of $\langle Q^2 \rangle $=2.5 (GeV/$c$)$^2$, 
$\langle W^2 \rangle $=28.6 GeV$^2$, $\langle \nu \rangle$=16.1 GeV and $\langle x \rangle$=0.082.
The solid line is a naive quark-parton model calculation.}
\end{center}
\end{figure}

\begin{figure}
\begin{center}
\epsfxsize=3.40in
\epsfysize=3.20in
\epsffile{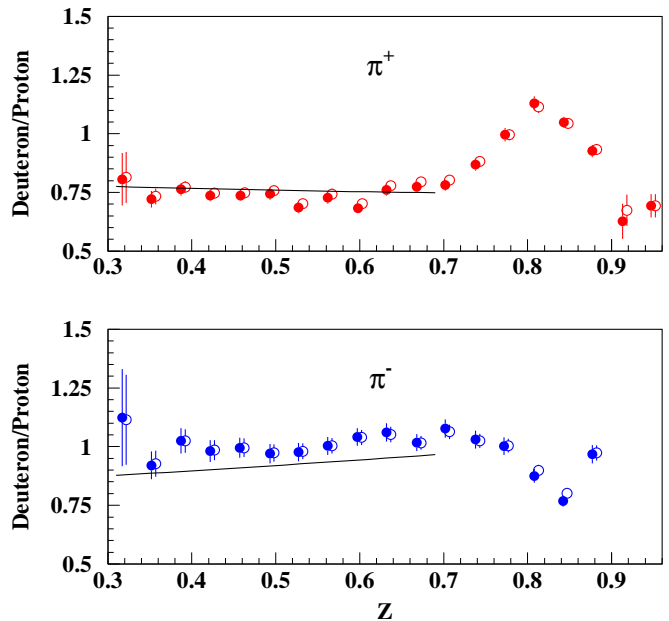}
\caption{\label{fig:dhrat-z} (Color online)
The deuteron over proton (D/H) yield ratio for $\pi^+$ mesons (top panel) and $\pi^-$ mesons 
(bottom panel), as a function of $z$ at fixed $x$ = 0.32. Solid (open) symbols represent the new 
data after (before) events from diffractive $\rho$ decay are subtracted. The solid lines are the 
quark-parton model expectations, plotted up to $z$ = 0.7 where effects from the $N-\Delta$ 
transition may enter the ratios.}
\end{center}
\end{figure}

{\bf The $\pi^+/\pi^-$ and $D/H$ ratios versus $z$:}
Various ratios of cross sections of positively- and negatively-charged pions and proton and 
deuteron targets are shown as a function of $z$ (at $x=0.32$) in Figs.~\ref{fig:pirat-pd-z} and 
~\ref{fig:dhrat-z}. Solid (open) circles and squares again represent the data after (before) 
events from diffractive $\rho$ decay are subtracted. We also added existing data for the 
charged-pion production ratios from Cornell \protect\cite{Beb77b}, with the solid and open 
triangles representing data at $Q^2$ = 2.0~(GeV/$c$)$^2$ and $x$ = 0.24, and 
$Q^2$ = 4.0 (GeV/$c$)$^2$ and $x$ = 0.50, respectively. The solid line is  again the simple 
quark-parton model calculation.

The $\pi^+/\pi^-$ ratio as measured from the proton target is larger than those reported by 
HERMES~\cite{Airapet01}, but agrees well with the older Cornell data~\cite{Beb77b}, and is 
consistent (but not equal) to the rise in $z$ as expected from the quark-parton model calculation 
up to $z\approx$~0.6. At values of 0.65 $<z<$0.85, the ratio decreases because the  
$\pi^-\triangle^{++}$ cross section is larger than the $\pi^+\triangle^o$ one. The sharp rise of 
the ratio at $z>$ 0.85 is due to exclusive $\pi^+$  production. On the other hand, the  
$\pi^+/\pi^-$ ratio measured on the deuteron reproduces the expected rise from the quark-parton 
model calculation very well. The data seem to continue the rising trend for $z >$ 0.7, into the 
region where we noticed effects from the $N-\Delta$ transition before. 

In our previous article~\cite{Nav07} this was explained within the SU(6) symmetric quark model, 
which essentially removes the effect of resonance transitions on this particular ratio, which is 
inversely proportional to the ratio of unfavored to favored fragmentation functions: 
$D^-/D^+ = (4 - r)/(4r - 1)$, with $r$ the ratio of $\pi^+$ over $\pi^-$ yields off a deuteron 
target. The observed $z$-dependence of the resulting $D^-/D^+$ ratio agreed very well with a fit 
by the HERMES collaboration of their data. In ~\cite{Nav07} it was also observed that the 
resulting $D^-/D^+$ ratio was independent of $x$, as it should be, and agreed quite well with 
previous HERMES and EMC data. For completeness and future use, we have added in this manuscript 
the $D^-/D^+$ ratios in tabular format in Table~\ref{tab:ff-ratio-z} and ~\ref{tab:ff-ratio-x}. 
The columns represent the data before and after events from diffractive $\rho$ decay are 
subtracted.

\begin{table}
\caption{\label{tab:ff-ratio-z} The ratio of unfavored to favored fragmentation function $D^-/D^+$ 
as a function of $z$, at $x$=0.32, evaluated at leading order of $\alpha_s$ (neglecting strange 
quarks) from the deuterium data.}
{\centering \begin{tabular}{|c|c|c|}
\hline
z      &  ${{D^-/D^+}(after~\rho)}$ & ${{D^-/D^+}(before~\rho)}$  \\
\hline
0.342  &   0.4620$\pm$0.0710        &  0.4620$\pm$0.0710          \\
0.370  &   0.4196$\pm$0.0475        &  0.4449$\pm$0.0465          \\
0.398  &   0.4838$\pm$0.0453        &  0.5126$\pm$0.0438          \\
0.426  &   0.4764$\pm$0.0429        &  0.5087$\pm$0.0411          \\
0.454  &   0.4575$\pm$0.0414        &  0.4940$\pm$0.0392          \\
0.482  &   0.4425$\pm$0.0413        &  0.4837$\pm$0.0395          \\
0.510  &   0.4059$\pm$0.0318        &  0.4530$\pm$0.0306          \\
0.538  &   0.3635$\pm$0.0270        &  0.4134$\pm$0.0257          \\
0.566  &   0.3699$\pm$0.0266        &  0.4288$\pm$0.0253          \\
0.594  &   0.3638$\pm$0.0274        &  0.4280$\pm$0.0267          \\
0.622  &   0.3448$\pm$0.0298        &  0.4124$\pm$0.0284          \\
0.650  &   0.3157$\pm$0.0289        &  0.3853$\pm$0.0279          \\
0.678  &   0.3587$\pm$0.0314        &  0.4376$\pm$0.0307          \\
0.706  &   0.3934$\pm$0.0327        &  0.4800$\pm$0.0319          \\
0.734  &   0.3137$\pm$0.0273        &  0.3889$\pm$0.0264          \\
0.762  &   0.3164$\pm$0.0254        &  0.3911$\pm$0.0254          \\
0.790  &   0.2738$\pm$0.0223        &  0.3375$\pm$0.0223          \\
0.818  &   0.2625$\pm$0.0198        &  0.3177$\pm$0.0198          \\
0.846  &   0.2380$\pm$0.0177        &  0.2808$\pm$0.0168          \\
0.874  &   0.2294$\pm$0.0161        &  0.2607$\pm$0.0159          \\
0.902  &   0.2423$\pm$0.0243        &  0.2555$\pm$0.0238          \\
0.930  &   0.3025$\pm$0.0739        &  0.2944$\pm$0.0724          \\
\hline
\end{tabular}\par}
\end{table}

\begin{table}
\caption{\label{tab:ff-ratio-x} The ratio of unfavored to favored fragmentation function $D^-/D^+$ 
as a function of $x$, at $z$=0.55, evaluated at leading order of $\alpha_s$ (neglecting strange 
quarks) from the deuterium data.}
{\centering \begin{tabular}{|c|c|c|}
\hline
x      &  ${{D^-/D^+}(after~\rho)}$ & ${{D^-/D^+}(before~\rho)}$  \\
\hline
0.213  &   0.5048$\pm$0.0835        &  0.5682$\pm$0.0800          \\
0.238  &   0.4272$\pm$0.0458        &  0.4789$\pm$0.0435          \\
0.263  &   0.4008$\pm$0.0341        &  0.4472$\pm$0.0322          \\
0.287  &   0.3939$\pm$0.0311        &  0.4361$\pm$0.0294          \\
0.312  &   0.4049$\pm$0.0289        &  0.4446$\pm$0.0277          \\
0.338  &   0.4278$\pm$0.0285        &  0.4660$\pm$0.0274          \\
0.363  &   0.3334$\pm$0.0252        &  0.3631$\pm$0.0242          \\
0.388  &   0.3690$\pm$0.0263        &  0.3987$\pm$0.0253          \\
0.413  &   0.3476$\pm$0.0262        &  0.3732$\pm$0.0249          \\
0.438  &   0.3914$\pm$0.0298        &  0.4177$\pm$0.0287          \\
0.463  &   0.3907$\pm$0.0320        &  0.4142$\pm$0.0310          \\
0.488  &   0.4198$\pm$0.0362        &  0.4420$\pm$0.0349          \\
0.513  &   0.4436$\pm$0.0403        &  0.4546$\pm$0.0395          \\
0.538  &   0.4202$\pm$0.0454        &  0.4385$\pm$0.0440          \\
0.562  &   0.4721$\pm$0.0581        &  0.4890$\pm$0.0572          \\
0.588  &   0.3533$\pm$0.0553        &  0.3668$\pm$0.0539          \\
\hline
\end{tabular}\par}
\end{table}

When expressed as a nuclear (deuteron over proton) ratio of positively-charged $\pi^+$ yields and 
negatively-charged $\pi^-$ yields (see Fig. 12), the data show a relatively flat behavior as a 
function of $z$ up to about $z=0.7$, where the $N-\Delta$ transition comes in again. These ratios 
appear in reasonable agreement with the quark-parton model calculations.

\begin{figure}
\begin{center}
\epsfxsize=3.40in
\epsfysize=3.20in
\epsffile{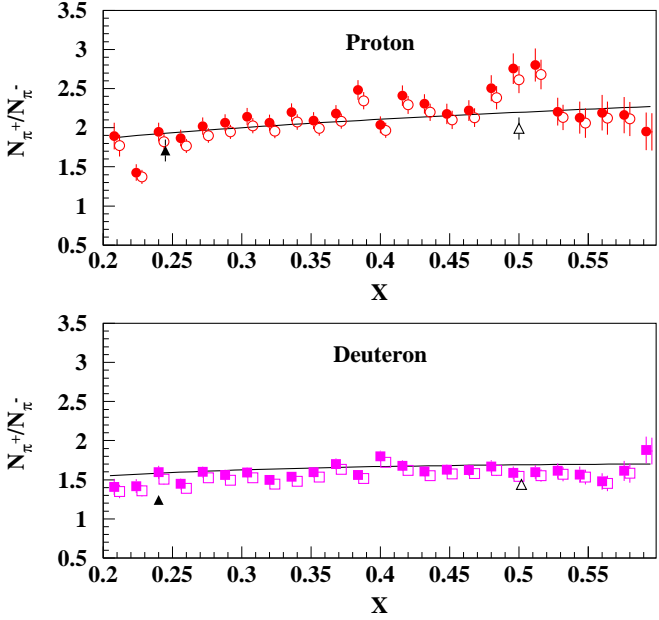}
\caption{\label{fig:pirat-pd-x} (Color online)
Ratio $\pi^+/\pi^-$ for proton (top panel) and deuteron (bottom panel) targets as a function of  
$x$ at $z=0.55$. Solid (open) circles and squares are our data after (before) events from 
diffractive $\rho$ 
decay are subtracted. Solid and open triangle symbols are Cornell data \protect\cite{Beb77b} at 
$Q^2$=2.0~(GeV/$c$)$^2$ and $x=0.24$, and  $Q^2$=4.0 (GeV/$c$)$^2$ and $x=0.50$. The solid lines 
are the quark-parton model expectation.}
\end{center}
\end{figure}

\begin{figure}
\begin{center}
\epsfxsize=3.40in
\epsfysize=3.20in
\epsffile{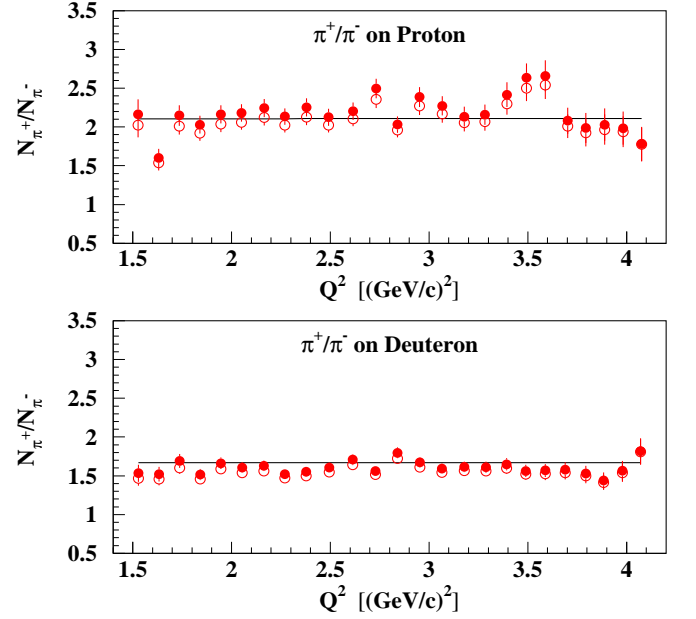}
\caption{\label{fig:pirat-pd-q2} (Color online)
The ratio $\pi^+/\pi^-$ for proton and deuteron targets as a function of $Q^2$ for $x$ = 0.4 and 
$z$ = 0.55. Solid (open) symbols are the ratios after (before) yields from diffractive $\rho$ 
decay are subtracted. The solid lines are the simple quark-parton model expectations.}
\end{center}
\end{figure}

{\bf The $x$- and $Q^2$-dependence of the $\pi^+/\pi^-$ and D/H ratios:}
Given that the $z$-dependence of our low-energy semi-inclusive pion electroproduction data show a 
smooth behavior up to $z$ = 0.7, in reasonable agreement with the quark-parton model expectations, 
we now turn to the $x$ and $Q^2$ dependence of the various ratios. In Figs.~\ref{fig:pirat-pd-x} 
and~\ref{fig:pirat-pd-q2} we show the ratio of positively- to negatively-charged pions versus $x$ 
and $Q^2$, respectively, for both proton and deuteron targets. As before, solid (open) circles and 
squares are the data after (before) corrections are made to subtract pions originating from 
diffractive $\rho$ decay. The $Q^2$ dependence of the deuteron to proton ratio for
$\rho^\circ$ production was studied by the HERMES collaboration~\cite{Osborne-thesis}.
We have also added existing data in Fig.~\ref{fig:pirat-pd-x} from 
Cornell~\cite{Beb77b}. Solid and open triangles represent data at $x$ = 0.24 and 
$Q^2$ = 2.0~(GeV/$c$)$^2$, and at $x$ = 0.50 and $Q^2$ = 4.0~(GeV/$c$)$^2$. As can be seen, the 
Cornell data are in good agreement with our data. The positively- to negatively-charged pion 
ratios are also in surprisingly good agreement with the quark-parton model prediction.

The $Q^2$ dependence of the ratios was extracted by using the $\pi^+$ and $\pi^-$ production cross 
sections at $x$ = 0.26, 0.32, 0.39, 0.46 and 0.53, and ``bin-centering'' these cross sections, and 
thus their ratios, as before, to one common value of $x$ = 0.4. This was done by using the hadron 
part of the model cross section in SIMC. The correction was checked by running SIMC and taking 
cross section ratios for proton targets at the five $x$-scan central settings. The size of the 
applied corrections amounts to $\sim$ 15\% maximum for the proton target, and is always below 
10$\%$ for the deuteron target.

The results, again at a value of $z$ = 0.55, are shown in Fig.~\ref{fig:pirat-pd-q2}. The $Q^2$ 
dependence of these ratios is in very good agreement with the quark-parton model expectations, 
indicated by the solid curve. This teaches that whereas the $Q^2$-dependence of the measured pion 
electroprodution cross sections is in reasonable, but not excellent, agreement with the 
quark-parton model expectations, as shown in Fig.~\ref{fig:xsect-q2}, any spurious or 
higher-twist-related $Q^2$-dependence get completely absorbed in ratios (or have an origin in the 
$x$-dependence of such ratios, as there is a strong kinematical correlation between $x$ and $Q^2$ 
within the E00-108 experimental setup). This is good news for low-energy access to quark-parton 
model physics in semi-inclusive meson electroproduction.

\begin{figure}
\begin{center}
\epsfxsize=3.40in
\epsfysize=3.20in
\epsffile{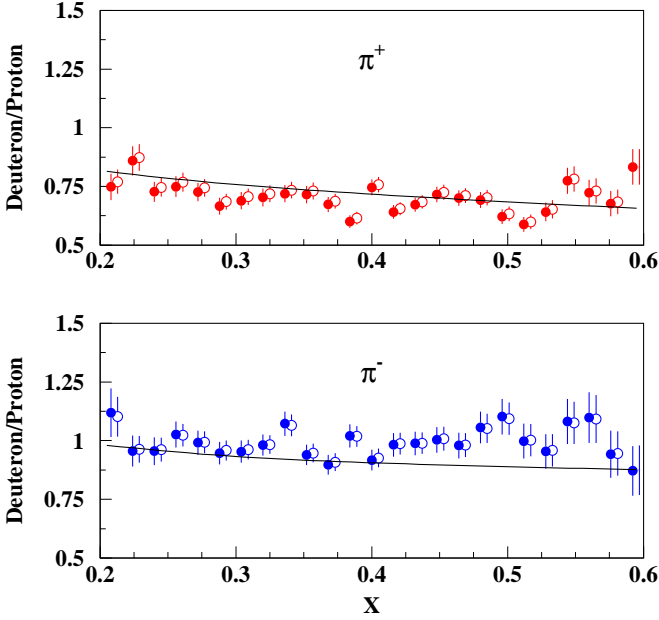}
\caption{\label{fig:dhrat-x} (Color online)
The ratio deuteron over proton for $\pi^+$ (top panel) and $\pi^-$ (bottom) as a function of $x$ 
at $z=0.55$. The solid (open) symbols are the ratios after (before) yields from diffractive $\rho$ 
decay are subtracted. The solid lines indicate the simple quark-parton model calculations.}
\end{center}
\end{figure}

\begin{figure}
\begin{center}
\epsfxsize=3.40in
\epsfysize=3.20in
\epsffile{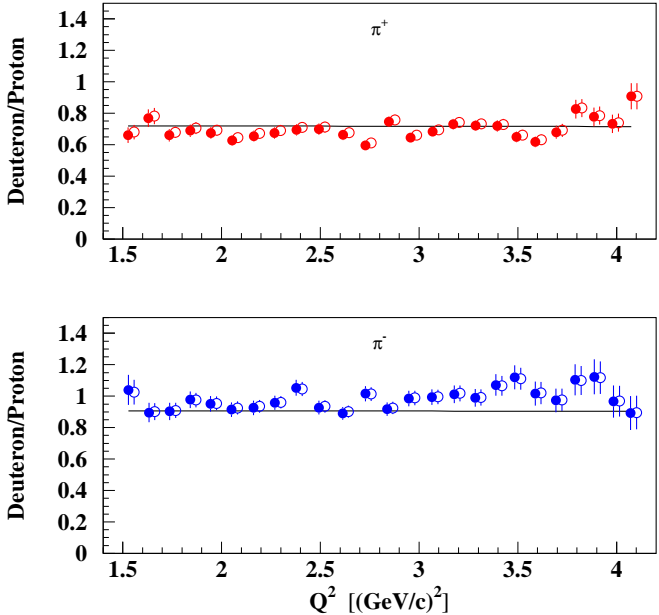}
\caption{\label{fig:dhrat-q2} (Color online)
The ratio deuteron over proton for $\pi^+$ (top panel) and $\pi^-$ (bottom) as a function of 
$Q^2$, at $x$ = 0.40 and $z$ = 0.55. The solid (open) symbols are the ratios after (before) yields 
from diffractive $\rho$ decay are subtracted. 
The solid lines indicate the simple quark-parton model calculations.}
\end{center}
\end{figure}

Lastly, we show in Figs.~\ref{fig:dhrat-x} and ~\ref{fig:dhrat-q2} the deuteron over proton ratios 
for $\pi^+$ and $\pi^-$  electroproduction, as a function of $x$ and $Q^2$, respectively. These 
nuclear D/H ratios are at a common $z$ = 0.55 again, and the solid curves shown correspond as 
before to the quark-parton model expectations. 

The conclusions from these nuclear D/H ratios are not unexpected. The dependences on $x$ and $Q^2$ 
from the quark-parton model is remarkably close to the data, again confirming for these ratios 
that higher-twist effects are small or nearly cancel in ratios. The absolute magnitudes of the 
ratios slightly differs from the quark-parton model estimates, which reflects the similar 
difference noted in Fig.~\ref{fig:dhrat-z}. Most obvious is the nuclear D/H ratio for the $\pi^-$ 
electroproduction case, where the data are some 10\% higher than the calculated quark-parton model 
ratio. The origin of the discrepancy is not yet clear, but on the other hand the data are at a 
relatively large $x$ of 0.4, where the parton distributions themselves start having noticeable 
uncertainties. The latter is investigated in more detail in the next section, by constructing 
direct ratios of the $d$ and $u$ valence quark ratios from the data.

\subsection{\bf The ratio of $d$/$u$ valence quarks constructed from
charged-pion yields}
The cross section for $\pi^\pm$ production on a deuteron at fixed $Q^2$ can be presented, in 
similar format and under identical assumptions, as the sum of the pion cross sections of 
Eq.~(\ref{eq:sidis_yields_nucl}):
\begin{eqnarray}
\label{eq:sigma_deutr}
\left.
\begin{array}{lll}
 {\sigma_d}^{\pi^+}(x,z) \propto (4D^+(z) + D^-(z))(u(x) + d(x)) + \\
  \hspace*{0.8in}  (4D^-(z) + D^+(z))(\bar{u}(x) + \bar{d}(x))  \\
 {\sigma_d}^{\pi^-}(x,z) \propto (4D^-(z) + D^+(z))(u(x) + d(x)) + \\
  \hspace*{0.8in}  (4D^+(z) + D^-(z))(\bar{u}(x) + \bar{d}(x))  \\
\end{array}
\right.
\end{eqnarray}
The measured cross sections or yields for $\pi^\pm$ production on the proton and deuteron can in 
the quark-parton model be directly used to form relations in terms of the $u_v$ and $d_v$ valence 
quark distributions:
\begin{eqnarray}
\label{eq:sigma_diff}
\left.
\begin{array}{lll}
{\sigma_p}^{\pi^+} - {\sigma_p}^{\pi^-} \propto (D^+ - D^-)(4u_v - d_v) \\
{\sigma_d}^{\pi^+} - {\sigma_d}^{\pi^-} \approx 
({\sigma_p}^{\pi^+} - {\sigma_p}^{\pi^-}) + 
({\sigma_n}^{\pi^+} - {\sigma_n}^{\pi^-}) \\
\hspace*{0.8in}    \propto (D^+ - D^-)(3u_v + 3d_v),\\
\end{array}
\right.
\end{eqnarray}
where $u_v =u-\bar u$, and $d_v=d-\bar d$. Of course, only the full parton distribution $u$ 
(and $d$) is physical, but at intermediate to large $x$, $x >$ 0.3, sea quark contributions are 
small and it is common to consider valence quark distributions only in this region.

The $d_v/u_v$ ratio can be directly extracted from a specific combination of the measured proton 
and deuteron $\pi^\pm$ cross sections as follows:
\begin{equation}
\label{eq:ratios}
R_{pd}^-(x)  = \frac { \sigma_p^{\pi^+}(x,z) - \sigma_p^{\pi^-}(x,z) } 
{ \sigma_d^{\pi^+}(x,z) - \sigma_d^{\pi^-}(x,z)}
  = \frac { 4u_v(x) - d_v(x) } {3[u_v(x) + d_v(x)] },
\end{equation}
from which one finds
\begin{equation}
d_v/ u_v = {(4 - 3R_{pd}^-) / (3R_{pd}^- + 1)}. \\
\end{equation}
Studying the $x$ and $z$ (and $P_t$) dependences of $R_{pd}^-$ and $d_v/ u_v $ thus provides an 
excellent test of the validity of the high-energy factorized view of the SIDIS process, and the 
various assumptions made.

\begin{figure}
\begin{center}
\epsfxsize=3.40in
\epsfysize=3.20in
\epsffile{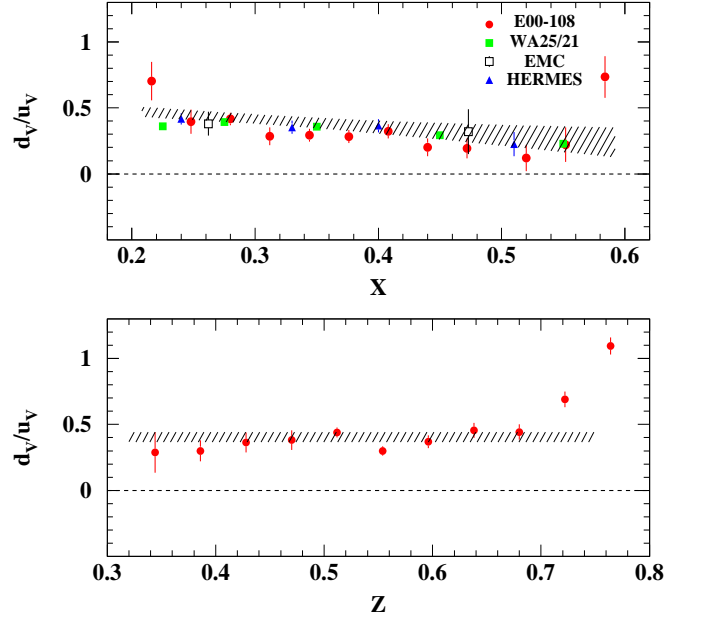}
\caption{\label{fig:du-rat-xz} (Color online)
Top panel: The ratio of valence quarks $d_v/u_v$ as a function of $x$ at $z$=0.55. Solid circles 
are our data from E00-108 experiment (at $P_t \approx 0$) after events from $\rho$ decay are 
subtracted. Solid and open squares represent data from WA-21/25 \protect\cite{WA-21-25} and 
EMC \protect\cite{EMC}. Solid triangle symbols are HERMES data \protect\cite{HERMES} integrated 
over the $0.2<z<0.7$ range. Bottom panel: The ratio of valence quarks $d_v/u_v$ as a function of 
$z$ at $x$=0.32. Solid circles are our data from E00-108 after events from diffractive $\rho$ decay
are subtracted. The shaded bands on both panels reflect the values of and uncertainties in this 
ratio using CTEQ parton distribution functions, based on Eq.~\ref{eq:ratios}~\cite{CTEQ}.}
\end{center}
\end{figure}

The ratio $d_v/u_v$ is shown in Fig.~\ref{fig:du-rat-xz}, both as a function of $x$ at $z$=0.55 
(top panel), and as a function of $z$ at $x$=0.32 (bottom panel).
The ratios extracted from our SIDIS data are also compared to WA-21/25 data from neutrino and 
anti-neutrino deep inelastic scattering off proton targets (solid squares)~\cite{WA-21-25}, and to 
ratios extracted from forward hadron production data from the European Muon Collaboration (open 
squares)~\cite{EMC}. The shaded bands on both panels represent the values (including their present 
uncertainties) as calculated from Eq.~\ref{eq:ratios} using CTEQ parton distribution 
functions~\cite{CTEQ}.

The experimentally extracted ratios appear somewhat low as compared to the quark-parton model 
expectations using the CTEQ parton distributions, but possibly within uncertainties. For the 
results of the present experiment, one should not only take into account experimental systematic 
uncertainties, but also possible biases due to various assumptions in low-energy factorization and 
symmetry in fragmentation functions, etc. Nonetheless, the E00-108 data (at $P_t \approx 0$) are 
in good agreement with previous extractions of WA21/25 and EMC, with vastly different techniques. 

The undershoot as compared to CTEQ parton distribution function expectations can be further 
investigated by investigating the dependence on $z$ of the measured ratios at a fixed value of 
$x$ (= 0.32). If isospin symmetry between favored ($D^+$) and unfavored ($D^-$) fragmentation 
functions of light quarks ($u$ and $d$) and  anti-quarks ($\bar u$ and $\bar d$) breaks down 
($D_u^{\pi^+}\neq D_{\bar u}^{\pi^-} \neq D_d^{\pi^-}\neq D_{\bar d}^{\pi^+}$ and 
$D_u^{\pi^-}\neq D_{\bar u}^{\pi^+} \neq D_d^{\pi^+}\neq D_{\bar d}^{\pi^-}$), the ratios of 
Eq.~\ref{eq:ratios} may contain additional $z$-dependent factors, related to asymmetries between 
the fragmentation functions. Thus, a dependence of the extracted ``$d_v/u_v$ ratio'' on $z$ will 
be a good indication for a breakdown of the symmetry assumptions, or of the factorized formalism. 
Indeed, one can witness in the bottom panel of Fig.~\ref{fig:du-rat-xz} a sharp increase of the 
extracted $d_v/u_v$ ratio at $z >$ 0.7. This is likely not surprising as $z >$ 0.7 corresponds in 
E00-108 kinematics to missing mass $M_x^2<2.5$~GeV$^2$), where {\sl e.g.} the $\Delta$- and 
higher-resonance contributions become dominant.

Below $z \approx 0.7$, the extracted $d_v/u_v$ ratio is found to be reasonably independent of $z$, 
within the uncertainties of the data. On average, the data is somewhat low as compared to the 
quark-parton model expectations based upon CTEQ parton distribution functions, similar as was 
found in the $x$-dependence of this ratio. As a reminder, the data presented in 
Fig.~\ref{fig:du-rat-xz} are at an average low $P_t \sim$ 0.05 GeV/$c$; we will revisit any 
possible $P_t$ dependence of the extracted ratio later on.

Even though the extracted $d_v/u_v$ ratios from the E00-108 experiment tend to undershoot the 
expectations based upon CTEQ6 parton distributions, the agreement with the existing WA21/25 and 
EMC data is good, and possibly points to the applicability of the assumed factorization and access 
to the quark-parton model in relatively low-energy SIDIS data. This is consistent with our earlier 
findings in Ref.~\cite{Nav07}.

\subsection{\bf Nuclear Al/D ratios }

We have also analyzed the pion production ratio from aluminum to deuterium targets, Al/D, by using 
the data from the ``dummy'' target cells. The nuclear EMC effect, the modification of the 
(inclusive) nuclear structure functions as compared to those of the free nucleon,  was originally 
a revelation and firmly injected the subject of quarks into nuclear physics. In the valence quark 
region, a linear decrease in the nuclear ratio of structure functions (typically A/D) of about 
unity at $x = 0.3$ to a maximum depletion of 10-20\% around $x = 0.7$ has been found. For medium 
to heavy nuclei, $A > 12$,  the effect can be well described by either an atomic mass number 
$A^{-1/3}$ or nuclear density $\rho$ dependence.

In semi-inclusive pion production, nuclear effects are more complicated, because in addition to 
influencing the electron-quark scattering part, they can affect the quark-hadron fragmentation 
process. For the purpose of the present discussion, we assume that the nuclear effects on parton 
distributions and fragmentation functions simply factorize. This has by no way been based on 
rigorous experimental verification.

Experimental results on semi-inclusive leptoproduction of hadrons from nuclei are usually 
presented in terms of multiplicity ratios between nuclear (A) and deuteron (D) targets as a 
function of $z$ and $\nu$:
\begin{equation}
\label{eq:nucl_atten1}
{ R_A^h =  {1\over{N_A^{DIS}}}{dN_A^h\over dz}/
{1\over{N_D^{DIS}}}{dN_D^h\over dz} \approx 
{dN_A^h\over dz}/{dN_D^h\over dz}  }, 
\end{equation}
where the latter applies, if one can ignore EMC-type effects, true for $x\sim$0.3. Using the 
factorized assumption and neglecting the nuclear EMC effect for now (we will return to the 
$x$-dependence in the nuclear ratios later on), we first constructed a nuclear attenuation from 
the ratios of the normalized yields,
\begin{equation}
\label{eq:nucl_atten2}
{ R_A^h \approx {dN_A^h\over dz}/{dN_D^h\over dz} =
{Y_A^h\over Y_D^h} }, \\
\end{equation}
where $Y_A^h$ and $Y_D^h$ are the normalized yields of the electro-produced pions from aluminum 
nuclei and deuterium, respectively.

\begin{figure} 
\begin{center}
\epsfxsize=3.40in
\epsfysize=1.80in
\epsffile{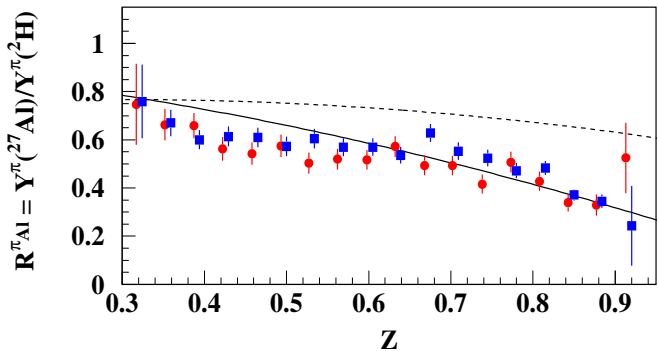}
\caption{\label{fig:atten_z} (Color online)
The ratio of aluminum over deuteron for both $\pi^+$ (circles) and $\pi^-$ (squares) data as a 
function of $z$ at $x$ = 0.32. The dashed curve is a calculation based on the model for hadron 
formation of Bialas and Chmaj~\protect\cite{Bialas83}, with the hadron formation time inserted 
from a string tension model~\cite{Gal-Mos}. The solid curve is a prediction based on gluon 
radiation theory~\cite{Accardi}. For the latter, we scaled between data for both $^{14}N$ and 
$^{64}Cu$, assuming $(1-R_A^h)\sim A^{1/3}$, and took the average value.}
\end{center}
\end{figure}

In Fig.~\ref{fig:atten_z} we present the ratio of the normalized pion electroproduction yields,  
Al/D, for both $\pi^+$ (solid circles) and $\pi^-$ (solid squares) versus $z$, at fixed $x$=0.32. 
The general features of the data, a value of $R$ below unity and decreasing with $z$, are similar 
to what has been observed in other experiments and which globally have been explained within 
various models (see, e.g., ~\cite{Air07} and references therein). This applies even in the region 
of $z >$ 0.7 where for both the deuteron target and the nearly-isoscalar aluminum target nucleon 
resonances come into play (which within the symmetric SU(6) quark model cancel out).

\begin{figure}
\begin{center}
\epsfxsize=3.40in
\epsfysize=3.20in
\epsffile{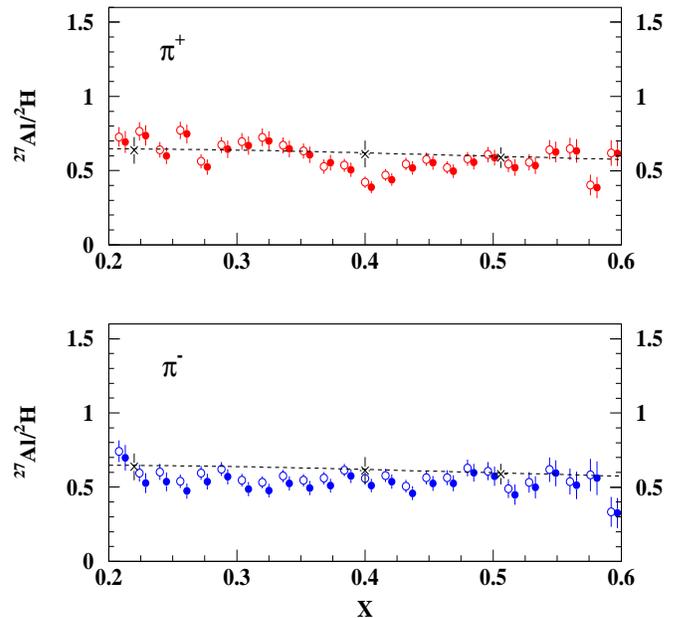}
\caption{\label{fig:ald-rat-x} (Color online)
The ratio of aluminum over deuteron for $\pi^+$ (top panel) and $\pi^-$ (bottom) as a function of 
$x$ at $z=0.55$. 
Solid (open) symbols are our data after (before) events from diffractive $\rho$ production are 
subtracted. The cross symbols are Al/D data from the SLAC collaboration~\cite{Gomez}, whereas the 
dashed lines are the results of their A-dependent global fit to nuclear EMC effect. The data and 
parameterization are both normalized to take into account any hadron attenuation effects.}
\end{center}
\end{figure}

The $x$-dependences of the Al/D cross section ratio for both $\pi^+$ and $\pi^-$, at $z=0.55$, are 
shown in Fig.~\ref{fig:ald-rat-x}. We show data both before and after the events from diffractive 
$\rho$ production are removed to demonstrate a slightly larger impact on the $\pi^-$ data. 
The dashed line represents the A-dependent EMC effect fit from the SLAC collaboration~\cite{Gomez} 
fit. The parameterization is normalized to take into account hadron attenuation effects. 
Overall, the agreement is quite good, even if our data scatter somewhat and the $x$-dependence of 
the cross section ratio Al/D is weak in this region. This confirms that nuclear ratios already 
behave like a high-energy parton model  expectation at relatively low energies and $W^2$.

\begin{figure}
\begin{center}
\epsfxsize=3.40in
\epsfysize=3.20in
\epsffile{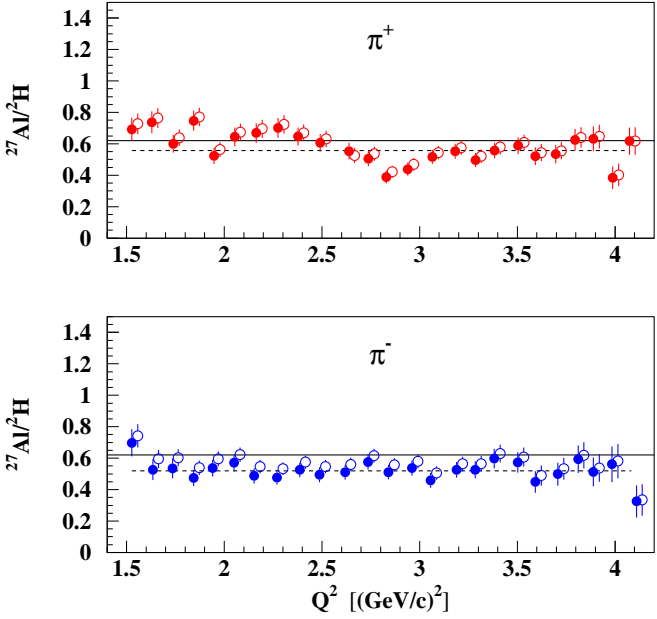}
\caption{\label{fig:ald-rat-q2} (Color online)
The ratio aluminum over deuteron for $\pi^+$ and $\pi^-$ as a function of $Q^2$ at $z=0.55$ and 
$x=0.40$. Solid (open) symbols are data after (before) events from diffractive $\rho$ production are 
subtracted. The solid curves represent a constant value of 0.62, as expected from the gluon 
radiation calculation at $z=0.55$. The dashed lines represent constant fits to the data, with 
value for $\pi^+$ ($\pi^-$) of 0.556$\pm$0.011 (0.520$\pm$0.011).}
\end{center}
\end{figure}

The Al/D cross section ratios versus $Q^2$ are extracted using again our Monte Carlo simulations 
to ``bin-center'' the data to a fixed $x$ = 0.40. We show the extracted ratios for both $\pi^+$ 
and $\pi^-$ data at $z$ = 0.55 and $x$ = 0.40 as a function of $Q^2$ in Fig.~\ref{fig:ald-rat-q2}. 
The multiplicity ratio for Al/D is nearly flat with $Q^2$, comparable to what was also observed 
for the D/H ratio in Fig.\ref{fig:dhrat-q2}. Similarly, only a very weak $Q^2$ dependence for this 
ratio was observed by the HERMES experiment~\cite{{Air07},{Hunen},{Haarlem}}. The dashed lines 
represent constant fits to the data, with a best-fit value for $\pi^+$ ($\pi^-$) of 
0.556 $\pm$ 0.011 (0.520 $\pm$ 0.011). 

Recently, there has been discussion on a possible flavor dependence of the EMC 
effect~\cite{CBT-09}. This would result in a possible different depletion of up quarks as compared 
to down quarks. Predictions indicate a somewhat larger depletion for up quarks than for down 
quarks, or equivalently larger $\pi^+$ attenuation than $\pi^-$ attenuation. We find the opposite, 
although the uncertainties are large and many complicated nuclear effects may contribute, 
including effects that do not obey a factorized form. Further study of this requires precision 
measurement of the $z$- and $x$-dependences of these ratios, and their 
differences~\cite{{DGH-09},{DPCG-10}}.

\subsection{\bf The $P_t$ dependence of the cross sections}
\begin{figure}
\begin{center}
\epsfxsize=3.40in
\epsfysize=3.20in
\epsffile{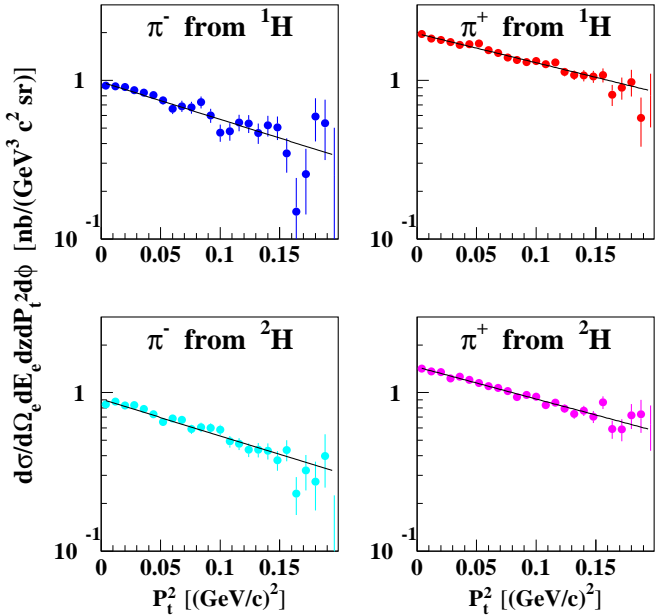}
\caption{\label{fig:slope-pt2} (Color online)
The $P^2_t$ dependence of differential cross sections per nucleon for $\pi^\pm$ production on 
hydrogen (H) and deuterium (D) targets at $\langle z\rangle $=0.55 and $\langle x\rangle $=0.32. 
The solid lines are exponential fits. The error bars are statistical only. }
\end{center}
\end{figure}

The extracted cross sections as a function of the pion transverse momentum squared $P^2_t$ are 
shown in Fig.~\ref{fig:slope-pt2} and listed in Table~\ref{tab:xsect-vrs-pt2}. 

The solid lines are exponential fits. The acceptance-averaged values of $\cos\phi$ range from -0.3 
at low $P_t$ to nearly -1 at high $P_t$, while the average values of $\cos2\phi$ range from 0.03 
at $P_t\leq$ 0.1 to 1 at high $P_t$.

A recent study~\cite{S-T-M-2010} analyzed these data in combination with the CLAS 
data~\cite{Osip09}, and concluded that in the kinematics similar to the CLAS data, the Hall C 
data could be relatively well described by a Gaussian model with average transverse momentum 
width of 0.24 (GeV/c)$^2$. The good description of the $\pi^\pm$ cross sections from different 
targets was argued to indicate that the assumption of flavor-independent Gaussian width for both 
the transverse widths of quark and fragmentation functions was reasonable, in the valence-$x$ 
region for $z$=0.55.

If taken as standalone data, a careful examination of Fig.~\ref{fig:slope-pt2} shows that the 
$P^2_t$-dependences for the four cases are 
similar, but not identical within statistical uncertainties. For a more quantitative understanding 
of the possible implications, we study the data in the context of a simple model in which the 
$P_t$ dependence is described in terms of two Gaussian distributions for each case. 

The probability of producing a pion with a transverse momentum $P_t$ relative to the virtual 
photon ($\vec{q}$) direction is described by a convolution of the quark distribution functions  
and $p_t$-dependent fragmentation functions $D^+(z,p_t)$ and $D^-(z,p_t)$, where $p_t$ is the 
transverse momentum of the pion relative to the quark direction, with the condition 
$P_t = z k_t + p_t$ assumed.

\begin{figure}
\begin{center}
\epsfxsize=3.4in
\epsfysize=3.2in
\epsffile{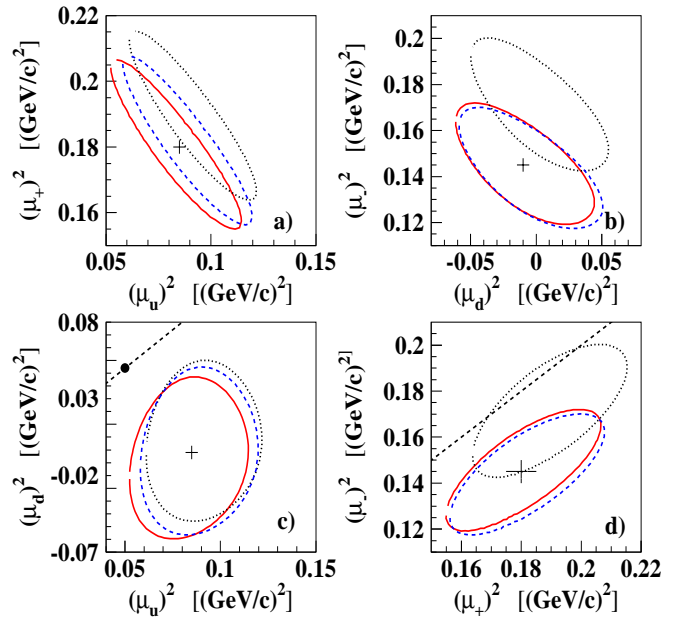}
\caption{\label{fig:fit1} (Color online) Fit parameters (crosses) and one-standard-deviation 
contours (continuous ellipses) from the seven-parameter fit to the data shown in 
Fig.~\ref{fig:slope-pt2}: 
a) $u$ quark width squared $\mu_u^2$ versus favored fragmentation width squared $\mu_+^2$; 
b) $\mu_d^2$ versus $\mu_-^2$; 
c) $\mu_u^2$ versus $\mu_d^2$; 
d) $\mu_-^2$ vs $\mu_+^2$. 
The dashed and dotted contours are for the case of no diffractive $\rho$ subtraction and a 30\% 
reduction in the size of the exclusive radiative tail subtraction, respectively. The large dot 
near the middle of panel c is from a di-quark model~\protect\cite{Jakob}. The dashed straight 
lines in panels c and d indicate $\mu_u^2=\mu_d^2$ and $\mu_-^2=\mu_+^2$, respectively.}
\end{center}
\end{figure}

Following Ref.~\cite{Anselmino}, we assume that the widths of the quark and fragmentation 
functions are Gaussian and that the convolution of these distributions combines quadratically. 
The main difference from Ref.~\cite{Anselmino} is that we allow separate widths for up and down 
quarks, and separate widths for favored and unfavored fragmentation functions. The widths of the 
up and down distributions are denoted by $\mu_u$ and $\mu_d$, respectively, and the favored 
(unfavored) fragmentation widths are given by $\mu_+$ ($\mu_-$). Following Cahn~\cite{Cahn} and 
more recent studies~\cite{Anselmino}, we assume that only the fraction $z$ of the quark transverse 
momentum contributes to the pion transverse momentum. We assume further that sea quarks are 
negligible (typical global fits show less than 10$\%$ contributions at $x=0.3$). To make the 
problem tractable, in the $\phi$-dependence we take only the leading-order terms in $(P_t/Q)$, 
which was shown to be a reasonable approximation up to moderate $P_t$ in Ref.~\cite{Anselmino}. 
This simple model then gives:
\begin{eqnarray}
\label{eq:sidis_xsects}
\left.
\begin{array}{lll}
\sigma_p^{\pi^+}(P_t) = C [4 c_1(P_t) e^{-b_u^+ P_t^2} + 
(\frac d u) (\frac {D^-} {D^+}) c_2(P_t) e^{-b_d^- P_t^2}] \\
\sigma_p^{\pi^-}(P_t) = C [4 (\frac {D^-} {D^+}) c_3(P_t) e^{-b_u^- P_t^2} + 
(\frac d u) c_4(P_t)e^{-b_d^+ P_t^2}] \\
\sigma_n^{\pi^+}(P_t) = C [4 (\frac d u) c_4(P_t)e^{-b_d^+ P_t^2} + 
(\frac {D^-} {D^+}) c_3(P_t)e^{-b_u^- P_t^2}] \\
        &        &  \\
\sigma_n^{\pi^-}(P_t) = C [4 (\frac d u)(\frac {D^-} {D^+})c_2(P_t)e^{-b_d^- P_t^2} 
                   +  c_1(P_t)e^{-b_u^+ P_t^2}], \\
\end{array}
\right.
\end{eqnarray}
\noindent where $C$ is an arbitrary normalization factor, and the inverse of the total widths for 
each combination of quark flavor and fragmentation function are given by
\begin{eqnarray}
\label{eq:b}
\left.
\begin{array}{lll}
b_u^\pm=(z^2\mu_u^2 + \mu_\pm^2)^{-1} \\
b_d^\pm=(z^2\mu_d^2 + \mu_\pm^2)^{-1} \\
\end{array}
\right.
\end{eqnarray}
and we assume $\sigma_d = (\sigma_p + \sigma_n$)/2. The $\phi$-dependence is taken into account 
through the terms:
\begin{eqnarray}
\label{eq:c}
\left.
\begin{array}{lll}
c_1(P_t) = 1 + c_0(P_t,\langle\cos(\phi)\rangle) \mu_u^2 b_u^+ \\
c_2(P_t) = 1 + c_0(P_t,\langle\cos(\phi)\rangle) \mu_d^2 b_d^- \\
c_3(P_t) = 1 + c_0(P_t,\langle\cos(\phi)\rangle) \mu_u^2 b_u^- \\
c_4(P_t) = 1 + c_0(P_t,\langle\cos(\phi)\rangle) \mu_d^2 b_d^+ \\
c_0(P_t,\langle\cos(\phi)\rangle) = \frac { 4 z (2-y) \sqrt{1-y}}
{ \sqrt{Q^2} [1 + (1-y)^2 ] } \sqrt{P_t^2} \langle\cos(\phi)\rangle. \\ 
\end{array}
\right.
\end{eqnarray}
We fit the $P_t$-dependence of the four cross sections of Eq.~\ref{eq:sidis_xsects} for the four 
widths ($\mu_u$, $\mu_d$, $\mu_+$, and $\mu_-$), $C$, and the ratios $D^-/D^+$ and $d/u$, where 
the fragmentation ratio is understood to represent the data-averaged value at $z=0.55$, and the 
quark distribution ratio is understood to represent the average value at $x=0.3$. The fit 
describes the data reasonably well ($\chi^2=68$ for 73 degrees of freedom), and finds the ratio  
$d/u=0.39\pm 0.03$, in good agreement with the LO GRV98 fit~\cite{GRV} for valence quarks 
(about 0.40). The fit also gives a reasonable value for the  ratio $D^-/D^+=0.43\pm 0.01$ 
(a fit to HERMES results~\cite{Geiger}, $D^-/D^+=1/(1+z)^2$, predicts 0.42 at $z=0.55$). 
Both $d/u$ and $D^-/D^+$ are largely uncorrelated with the other fit parameters and their values 
are largely determined by the magnitude of the cross sections. To estimate the effect of 
experimental systematic uncertainties on our fit results, we repeated the fits with: no 
diffractive $\rho$ subtraction; 30\% smaller exclusive radiative tail subtraction; relative target 
thickness changed by 1\%; and difference in $\pi^+$ and $\pi^-$ absorptions changed by 1\%. 
The last three changes had a negligible effect compared to statistical errors. The first change 
mainly affected $\mu_-^2$, shifting it to a more positive value by almost the size of the 
statistical error, as shown in Fig.~\ref{fig:fit1}. 
We found no significant change to the fit parameters upon adding to $\mu_u^2$ and $\mu_d^2$ an 
average nucleon transverse momentum squared of 0.001 (GeV/$c$)$^2$ (evaluated using  the Paris 
wave function~\cite{Paris}) for the deuteron model.

Since the data are at fixed $z$, the main terms that distinguish large fragmentation widths from 
large quark widths are the $\phi$-dependent $c_i$ terms. While  there is a significant inverse 
correlation between the two most important quark and fragmentation widths, ($\mu_u$ and $\mu_+$, 
respectively), the fit indicates a preference for $\mu_u$ to be smaller than $\mu_+$ as shown in 
Fig.~\ref{fig:fit1}a.  The fit also indicates a preference for $\mu_d$ to be smaller than $\mu_-$ 
as shown in Fig.~\ref{fig:fit1}b. So in both cases, fragmentation widths appear to somewhat 
dominate over quark widths, within our simple model.

The fit parameters indicate a non-zero $k_t$ width squared for $u$ quarks ($\mu_u^2=0.07\pm 0.03$ 
(GeV/$c$)$^2$), but a $d$-quark width squared that is consistent with zero 
($\mu_d^2=-0.01\pm 0.05$ (GeV/$c$)$^2$), as illustrated in Fig.~\ref{fig:fit1}c. We do note that 
intrinsic transverse momentum width for $u$ and $d$ quarks presented here are far different from 
the earlier published~\cite{Mkrt08}. The previous analysis results used a limited and not required 
cut in the reconstructed vertex coordinate reducing statistics, and also has improper corrections 
for contributions of pions from both the decay of diffractive $\rho$ production and the exclusive 
radiative tail. Still, the difference in the two results calls for a future careful measurement 
over a large of kinematics ($Q^2$, $P_t$ and $cos(\phi)$).

The results are consistent with a di-quark model~\cite{Jakob} in which the $d$ quarks are only 
found in an axial di-quark, while the $u$ quarks are predominantly found in a scalar di-quark. 
We plotted the results with equal axial and scalar di-quarks masses ($M_a$ and $M_s$) of 0.6 GeV; 
picking $M_a<M_s$ results in $\mu_d^2<\mu_u^2$, and visa verse, with the average remaining near 
0.06 (GeV/$c$)$^2$. 

Using the fit parameters, we find the magnitude of the $\cos(\phi)$ term $A$ at $P_t=0.4$ GeV/$c$ 
to be about $-0.15\pm0.05$ for all four cases. These results are similar in sign and magnitude to 
those found in the HERMES experiment~\cite{HermesPhi}.
 
We find that the fragmentation widths $\mu_+$ and $\mu_-$ are correlated, as illustrated in 
Fig.~\ref{fig:fit1}d, although the allowed range is not large, and the central values 
($\mu_+^2=0.18\pm 0.02$ (GeV/$c$)$^2$ and $\mu_-^2=0.14\pm 0.02$ (GeV/$c$)$^2$) are in reasonable 
agreement with each other and with the flavor-averaged value of 0.20 (GeV/$c$)$^2$ found in 
Ref.~\cite{Anselmino}. While there is a slight tendency for the favored width to be larger than 
the unfavored one, a reasonable fit can be obtained setting the widths equal to each other 
($\chi^2=71$ for 74 d.f., $\mu_+^2=\mu_-^2=0.17\pm0.03$ (GeV/$c$)$^2$). Taking into account the 
systematic uncertainties, the favored and unfavored widths are consistent with each other.

\subsection{The $P_t$ dependence of the ratios}

\begin{figure}
\begin{center}
\epsfxsize=3.40in
\epsfysize=3.20in
\epsffile{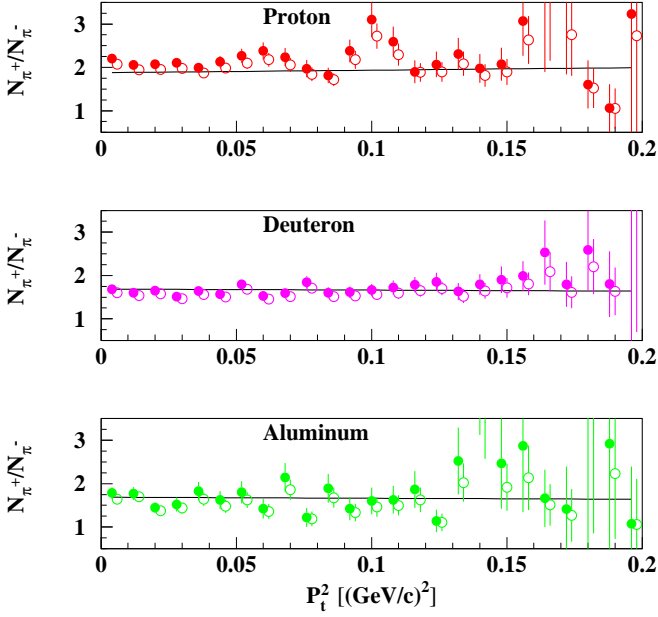}
\caption{\label{fig:pirat-pt2} (Color online)
The ratio $\pi^+/\pi^-$ for proton, deuteron and aluminum as a function of $P^2_t$ at $z=0.55$ and 
$x=0.32$. Solid (open) symbols are our data after (before) events from diffractive $\rho$ decay are
subtracted. The solid lines are simple quark-parton model expectations.}
\end{center}
\end{figure}

{\bf $\pi^+/\pi^-$ ratios versus $P^2_t$:}
The ratios of charged pions for proton, deuteron and aluminum targets as a function of $P^2_t$ at 
$z=0.55$ and $x=0.32$ are shown in Fig.~\ref{fig:pirat-pt2}. Solid (open) symbols are our data 
after (before) events from $\rho$ decay are subtracted. The solid lines represent the expectations 
from the simple quark-parton model.

The average values of the pion ratios for deuteron and aluminum are smaller than that for the 
proton, but they are nearly flat with $P^2_t$ for all three targets.

\begin{figure}
\begin{center}
\epsfxsize=3.40in
\epsfysize=3.20in
\epsffile{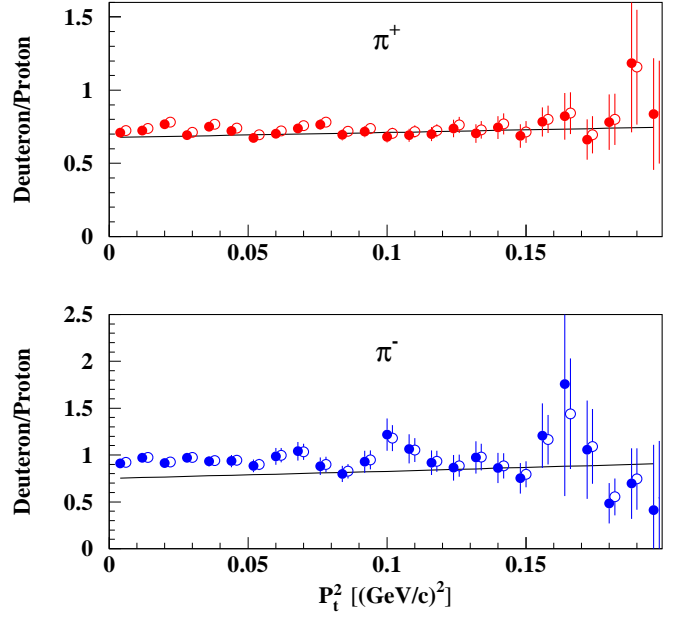}
\caption{\label{fig:dhrat-pt2} (Color online)
The ratio deuteron over proton for $\pi^+$ (top panel) and $\pi^-$ (bottom) as a function of 
$P^2_t$ at $z=0.55$ and $x=0.32$. Solid (open) symbols are our data after (before) events from 
diffractive $\rho$ decay are subtracted. The solid lines are our simple quark-parton model
expectations.}
\end{center}
\end{figure}

{\bf D/H ratios versus $P^2_t$:}
The deuteron over proton ratios for $\pi^+$ (top panel) and $\pi^-$ (bottom) as a function of 
$P^2_t$ at $z=0.55$ and $x=0.32$ are shown in Fig.~\ref{fig:dhrat-pt2}. 
Solid (open) symbols are our data after (before) events from $\rho$ decay are subtracted. 
The solid lines shown correspond as before to the simple quark-parton model calculation. As can be 
seen, the D/H ratios for $\pi^+$ are in good agreement with the quark-parton model prediction. 
For $\pi^-$ the experimental data on average are $\sim$10-15$\%$ higher relative to the model 
expectation.

\begin{figure}
\begin{center}
\epsfxsize=3.40in
\epsfysize=2.20in
\epsffile{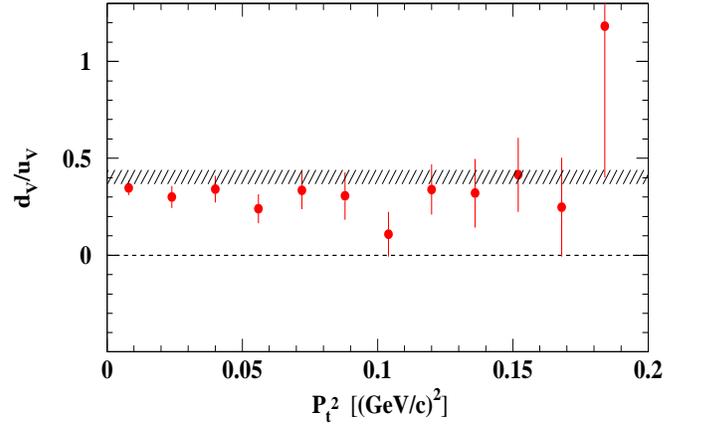}
\caption{\label{fig:du-rat-pt2} (Color online)
The extracted ratio $d_v/u_v$ as a function of $P_t^2$ at $x$=0.32 and $z$=0.55. The solid circles 
are the E00-108 data after events from diffractive $\rho$ decay are subtracted. 
The dashed band is a quark-parton model expectation using CTEQ parton distribution function 
parameterizations~\protect\cite{CTEQ}. }
\end{center}
\end{figure}

{\bf $P^2_t$ dependence of the $d_v/u_v$ ratios:}
For fixed $x$ = 0.32 ($Q^2$ = 2.30 (GeV/$c$)$^2$) and $z$ = 0.55, we show in 
Fig.~\ref{fig:du-rat-pt2} the extracted ratios of the down to up valence quark distributions 
$d_v/u_v$ as a function of $P_t^2$. The extracted ratios shown before, as function of $x$ and $z$ 
in Fig.~\ref{fig:du-rat-xz}, were from the lowest $P_t^2$ bin only. As before, the extracted 
ratios are on average below the quark-parton model expectations, here based upon the GRV parton 
distributions but also consistent with the earlier comparisons with CTEQ parton distributions. 
Given the statistical precision of the E00-108 data, it can not be ruled out that the $d_v/u_v$ 
valence quark distribution ratio may have a dependence on $P_t$ (or intrinsic quark momentum 
$k_t$). Such a dependence is in principle possible within a transverse-momentum dependent 
framework~\cite{Mkrt08}. It has been calculated to be small for up and down spin-averaged parton 
distributions in Lattice QCD, with far larger dependences found in spin-dependent parton 
distributions~\cite{Hag09}. 

\begin{figure}
\begin{center}
\epsfxsize=3.40in
\epsfysize=3.20in
\epsffile{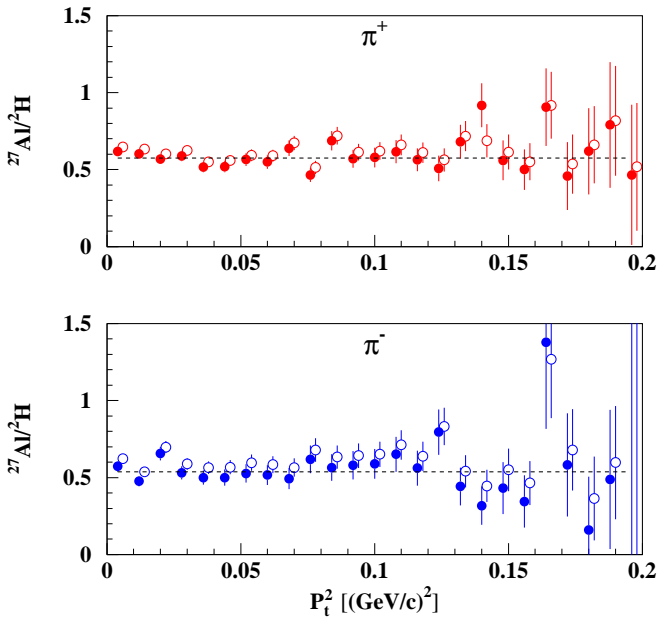}
\caption{\label{fig:ald-rat-pt2} (Color online)
The cross section ratio aluminum over deuteron for $\pi^+$ (top) and $\pi^-$ (bottom) as a 
function of $P^2_t$ at $x=0.32$ and $z=0.55$. Solid (open) symbols are data after (before) events 
from diffractive $\rho$ production are subtracted. The dashed lines are constant fits to the data.}
\end{center}
\end{figure}

{\bf $P^2_t$ dependence of the Al/D ratios:}
In Fig.~\ref{fig:ald-rat-pt2} the ratio aluminum over deuteron for $\pi^+$  and $\pi^-$  as a 
function of $P^2_t$ at $x=0.32$ and $z=0.55$ is shown.
Solid (open) symbols are data after (before) events from coherent $\rho$ production are 
subtracted. The data show that the Al/D ratio is 
reduced at high $z$, as was observed by HERMES group~\cite{Haarlem}. Our data show slight 
differences in attenuation of $\pi^+$ and $\pi^-$. The reduction seems to be stronger for $\pi^+$.

In our kinematic range the multiplicity ratio for Al/D as a function of $P_t^2$ is nearly flat. 
The dashed lines in Fig.~\ref{fig:ald-rat-pt2} represent constant fits to the data, with a best-fit
value for $\pi^+$ and $\pi^-$ of 0.575$\pm$0.010 and 0.538$\pm$0.014, and a $\chi^2/ndf$ of 1.23 
and 1.16, respectively. Similar flat behavior for the region $P_t^2\leq$0.2~(GeV/$c$)$^2$ was 
observed by HERMES group for variety of nuclei.

For future use, we have presented in this manuscript the various ratios versus $z$, $x$ and 
$P_t^2$ in tabular format in Tables~\ref{tab:pirat-vrs-z}-\ref{tab:nuclrat-vrs-pt2}.

\section{SUMMARY AND CONCLUSIONS}

In summary, we have measured semi-inclusive electroproduction of charged pions ($\pi^{\pm}$) from 
both proton and deuteron targets, using a 5.479 GeV energy electron beam in Hall C 
at Jefferson Lab. We have 
observed, for the first time, the quark-hadron duality phenomenon in pion electroproduction 
reactions. This has important consequences for a viable access to a quark-parton model description 
in semi-inclusive deep inelastic scattering experiments at relatively low energies. Several ratios 
constructed from the data exhibit, provided that $W^2>$ 4.0 GeV$^2$ and $z<$ 0.7 (or beyond the 
$\Delta$-resonance region in missing mass), the features of factorization in a sequential 
electron-quark scattering and a quark-pion fragmentation process. We find the azimuthal dependence 
of the data to be small, as compared to the typically larger azimuthal dependences found in 
exclusive pion electroproduction data, but consistent with data from other groups and  theoretical 
expectations~\cite{Cahn,Anselmino} based on a semi-inclusive deep inelastic scattering approach.

Examination of the $P_t$ dependence of the cross section shows a possible flavor dependence of the 
transverse-momentum dependence of the quark distribution and/or fragmentation functions. In the 
context of a simple model with only valence quarks and only two fragmentation functions, we find 
the transverse momentum $k_t$ width of $u$ quarks to be larger than that for $d$ quarks, for which 
the width is consistent with zero within the statistical uncertainties. 
We find that the transverse momentum $p_t$ widths of the favored and unfavored fragmentation 
functions are similar to each other, and both larger than the two quark widths. This is consistent 
with theoretical expectations based on fits to the world data. We have shown the sensitivity of 
our results to be small to possible corrections due to both radiative events from exclusive pion 
production channels and pions originating from diffractive $\rho$ scattering (and decay). In many 
cases, the corrections are negligible, although they can become large at large values of $z$. 
We believe our work will provide a fruitful basis for future studies of the quark-parton model and 
more sophisticated model calculations at relatively low energies.

\vspace{0.2in}
\centerline{ACKNOWLEDGMENTS}

The authors wish to thank H. Avakian, A. Afanasev, A Bruell, C.E. Carlson, W. Melnitchouk and 
M. Schlegel for discussions, and for many useful suggestions. 
This work is supported in part by research grants from the U.S. Department of Energy 
DE-FG02-99ER-41065 (Florida International University), DE-AC02-06CH11357 (Argonne National 
Laboratory), DE-FG02-04ER41330 (Mississippi State University), and the U.S. National Science 
Foundation $\#$0072466 and 0347438 (North Carolina A$\&$T State University), and $\#$0400332 and 
0653508 (Hampton University). 
We acknowledge support from the Natural Sciences and Engineering Research Council of Canada 
(University of Regina), and the South African National Research Foundation (University of 
Johannesburg). 
The Southeastern Universities Research Association operates the Thomas Jefferson National 
Accelerator Facility under the U.S. Department of Energy contract DEAC05-84ER40150.

\begin{widetext}
\begin{center}
\begin{table}
\caption{\label{tab:xsect-vrs-pt2} Experimental differential cross sections per nucleus (in nb/GeV$^3/c^2/sr$) versus $P^2_t$ (in GeV$^2/c^2$) for 
$\pi^+$ and $\pi^-$ production on Hydrogen and Deuterium targets. Left (right) part of the table represents the values before (after) events from 
diffractive $\rho$ decay have been subtracted. Error bars are statistical only.}
{\centering  \begin{tabular}{||c|c c c c ||c c c c ||}
\hline
\hline
 $P^2_t$ & $\sigma^+_{H}\pm d\sigma$ & $\sigma^-_{H}\pm d\sigma$ & $\sigma^+_{D}\pm d\sigma$ & $\sigma^-_{D}\pm d\sigma$  
 & $\sigma^+_{H}\pm d\sigma$ & $\sigma^-_{H}\pm d\sigma$ & $\sigma^+_{D}\pm d
\sigma$ & $\sigma^-_{D}\pm d\sigma$      \\ 
\hline
  0.004 & 2.068$\pm$0.029 & 1.039$\pm$0.022 & 3.052$\pm$0.040 & 1.912$\pm$0.034 & 1.959$\pm$0.029 & 0.927$\pm$0.022 & 2.831$\pm$0.039 & 1.688$\pm$0.034\\
  0.012 & 1.943$\pm$0.031 & 1.033$\pm$0.026 & 2.951$\pm$0.041 & 1.982$\pm$0.040 & 1.832$\pm$0.031 & 0.916$\pm$0.026 & 2.727$\pm$0.041 & 1.754$\pm$0.039\\
  0.020 & 1.910$\pm$0.035 & 1.023$\pm$0.031 & 2.925$\pm$0.046 & 1.884$\pm$0.048 & 1.798$\pm$0.035 & 0.908$\pm$0.031 & 2.700$\pm$0.046 & 1.655$\pm$0.047\\
  0.028 & 1.857$\pm$0.040 & 0.983$\pm$0.038 & 2.682$\pm$0.051 & 1.899$\pm$0.058 & 1.745$\pm$0.039 & 0.866$\pm$0.037 & 2.457$\pm$0.051 & 1.668$\pm$0.057\\
  0.036 & 1.789$\pm$0.043 & 0.954$\pm$0.042 & 2.754$\pm$0.059 & 1.811$\pm$0.064 & 1.674$\pm$0.042 & 0.838$\pm$0.041 & 2.521$\pm$0.058 & 1.579$\pm$0.063\\
  0.044 & 1.810$\pm$0.046 & 0.925$\pm$0.045 & 2.630$\pm$0.063 & 1.685$\pm$0.065 & 1.695$\pm$0.045 & 0.807$\pm$0.044 & 2.405$\pm$0.062 & 1.458$\pm$0.065\\
  0.052 & 1.820$\pm$0.047 & 0.862$\pm$0.046 & 2.523$\pm$0.066 & 1.531$\pm$0.068 & 1.711$\pm$0.047 & 0.747$\pm$0.046 & 2.303$\pm$0.066 & 1.305$\pm$0.067\\
  0.060 & 1.656$\pm$0.047 & 0.775$\pm$0.049 & 2.415$\pm$0.069 & 1.597$\pm$0.073 & 1.549$\pm$0.046 & 0.662$\pm$0.048 & 2.197$\pm$0.068 & 1.377$\pm$0.072\\
  0.068 & 1.605$\pm$0.049 & 0.806$\pm$0.056 & 2.367$\pm$0.073 & 1.571$\pm$0.079 & 1.491$\pm$0.049 & 0.687$\pm$0.056 & 2.145$\pm$0.072 & 1.347$\pm$0.078\\
  0.076 & 1.507$\pm$0.049 & 0.788$\pm$0.060 & 2.263$\pm$0.074 & 1.397$\pm$0.080 & 1.397$\pm$0.049 & 0.676$\pm$0.059 & 2.050$\pm$0.073 & 1.178$\pm$0.079\\
  0.084 & 1.455$\pm$0.051 & 0.843$\pm$0.064 & 2.094$\pm$0.077 & 1.433$\pm$0.084 & 1.344$\pm$0.051 & 0.730$\pm$0.063 & 1.874$\pm$0.076 & 1.210$\pm$0.084\\
  0.092 & 1.414$\pm$0.053 & 0.710$\pm$0.061 & 2.144$\pm$0.081 & 1.410$\pm$0.086 & 1.305$\pm$0.053 & 0.602$\pm$0.061 & 1.931$\pm$0.080 & 1.197$\pm$0.085\\
  0.100 & 1.430$\pm$0.056 & 0.572$\pm$0.058 & 2.101$\pm$0.085 & 1.380$\pm$0.087 & 1.323$\pm$0.056 & 0.468$\pm$0.058 & 1.890$\pm$0.084 & 1.171$\pm$0.086\\
  0.108 & 1.383$\pm$0.061 & 0.587$\pm$0.060 & 1.886$\pm$0.084 & 1.196$\pm$0.082 & 1.268$\pm$0.060 & 0.477$\pm$0.060 & 1.671$\pm$0.083 & 0.985$\pm$0.081\\
  0.116 & 1.412$\pm$0.067 & 0.648$\pm$0.064 & 1.935$\pm$0.090 & 1.152$\pm$0.084 & 1.300$\pm$0.066 & 0.542$\pm$0.064 & 1.730$\pm$0.089 & 0.951$\pm$0.083\\
  0.124 & 1.237$\pm$0.069 & 0.638$\pm$0.069 & 1.787$\pm$0.095 & 1.068$\pm$0.086 & 1.129$\pm$0.068 & 0.535$\pm$0.069 & 1.587$\pm$0.093 & 0.872$\pm$0.085\\
  0.132 & 1.182$\pm$0.074 & 0.565$\pm$0.069 & 1.663$\pm$0.102 & 1.062$\pm$0.091 & 1.081$\pm$0.074 & 0.466$\pm$0.068 & 1.470$\pm$0.101 & 0.873$\pm$0.090\\
  0.140 & 1.180$\pm$0.088 & 0.625$\pm$0.079 & 1.735$\pm$0.116 & 1.058$\pm$0.100 & 1.074$\pm$0.087 & 0.521$\pm$0.078 & 1.537$\pm$0.115 & 0.858$\pm$0.099\\
  0.148 & 1.157$\pm$0.093 & 0.605$\pm$0.086 & 1.601$\pm$0.125 & 0.947$\pm$0.108 & 1.057$\pm$0.092 & 0.506$\pm$0.085 & 1.407$\pm$0.123 & 0.751$\pm$0.107\\
  0.156 & 1.177$\pm$0.109 & 0.440$\pm$0.086 & 1.933$\pm$0.164 & 1.065$\pm$0.135 & 1.081$\pm$0.108 & 0.347$\pm$0.085 & 1.737$\pm$0.162 & 0.870$\pm$0.134\\
  0.164 & 0.922$\pm$0.125 & 0.255$\pm$0.094 & 1.368$\pm$0.160 & 0.650$\pm$0.124 & 0.814$\pm$0.124 & 0.149$\pm$0.093 & 1.178$\pm$0.157 & 0.461$\pm$0.123\\
  0.172 & 1.001$\pm$0.143 & 0.357$\pm$0.114 & 1.370$\pm$0.189 & 0.840$\pm$0.164 & 0.899$\pm$0.142 & 0.257$\pm$0.114 & 1.173$\pm$0.186 & 0.646$\pm$0.163\\
  0.180 & 1.073$\pm$0.189 & 0.686$\pm$0.181 & 1.614$\pm$0.259 & 0.725$\pm$0.188 & 0.978$\pm$0.187 & 0.593$\pm$0.180 & 1.436$\pm$0.255 & 0.548$\pm$0.186\\
  0.188 & 0.684$\pm$0.201 & 0.637$\pm$0.223 & 1.677$\pm$0.350 & 1.013$\pm$0.297 & 0.580$\pm$0.199 & 0.536$\pm$0.222 & 1.458$\pm$0.345 & 0.797$\pm$0.293\\
  0.196 & 0.878$\pm$0.302 & 0.323$\pm$0.252 & 1.394$\pm$0.404 & 0.324$\pm$0.260 & 0.805$\pm$0.299 & 0.250$\pm$0.252 & 1.257$\pm$0.399 & 0.189$\pm$0.259\\
\hline
\hline
\end{tabular}\par}
\end{table}
\end{center}
\end{widetext}

\begin{widetext}
\begin{center}
\begin{table}
\caption{\label{tab:pirat-vrs-z} The $R=\pi^+/\pi^-$ ratios versus Z for H, D and Al targets. 
Left (right) part of the table represents the ratios before (after) events from diffractive 
$\rho$ decay have been subtracted. Error bars are statistical only.}
{\centering  \begin{tabular}{||c|c c c||c c c||}
\hline
\hline
   z  &   $R_H\pm$dR    &   $R_D\pm$dR    &  $R_{Al}\pm$dR  &   $R_H\pm$dR    &   $R_D\pm$dR    &  $R_{Al}\pm$dR \\
\hline
0.317 & 1.773$\pm$0.311 & 1.317$\pm$0.167 & 1.330$\pm$0.335 & 1.828$\pm$0.340 & 1.337$\pm$0.179 & 1.358$\pm$0.368\\
0.352 & 1.710$\pm$0.100 & 1.384$\pm$0.065 & 1.356$\pm$0.141 & 1.756$\pm$0.110 & 1.415$\pm$0.072 & 1.401$\pm$0.162\\
0.387 & 1.953$\pm$0.091 & 1.508$\pm$0.053 & 1.343$\pm$0.117 & 2.038$\pm$0.102 & 1.553$\pm$0.059 & 1.389$\pm$0.136\\
0.422 & 1.832$\pm$0.076 & 1.396$\pm$0.045 & 1.483$\pm$0.136 & 1.898$\pm$0.084 & 1.429$\pm$0.050 & 1.556$\pm$0.161\\
0.457 & 1.832$\pm$0.071 & 1.421$\pm$0.045 & 1.554$\pm$0.141 & 1.898$\pm$0.079 & 1.456$\pm$0.049 & 1.638$\pm$0.169\\
0.493 & 1.809$\pm$0.068 & 1.455$\pm$0.046 & 1.437$\pm$0.124 & 1.898$\pm$0.077 & 1.501$\pm$0.052 & 1.517$\pm$0.153\\
0.527 & 1.974$\pm$0.072 & 1.499$\pm$0.047 & 1.672$\pm$0.145 & 2.098$\pm$0.084 & 1.565$\pm$0.054 & 1.848$\pm$0.195\\
0.562 & 2.061$\pm$0.071 & 1.573$\pm$0.047 & 1.622$\pm$0.134 & 2.194$\pm$0.083 & 1.649$\pm$0.054 & 1.778$\pm$0.178\\
0.597 & 2.159$\pm$0.066 & 1.519$\pm$0.043 & 1.608$\pm$0.123 & 2.327$\pm$0.079 & 1.597$\pm$0.050 & 1.764$\pm$0.166\\
0.632 & 1.997$\pm$0.061 & 1.584$\pm$0.045 & 1.445$\pm$0.109 & 2.178$\pm$0.075 & 1.681$\pm$0.054 & 1.577$\pm$0.149\\
0.668 & 1.814$\pm$0.051 & 1.542$\pm$0.040 & 1.773$\pm$0.128 & 1.970$\pm$0.064 & 1.657$\pm$0.051 & 2.101$\pm$0.203\\
0.702 & 1.787$\pm$0.052 & 1.501$\pm$0.039 & 1.555$\pm$0.116 & 1.947$\pm$0.066 & 1.608$\pm$0.049 & 1.792$\pm$0.182\\
0.738 & 1.637$\pm$0.047 & 1.613$\pm$0.040 & 1.742$\pm$0.138 & 1.760$\pm$0.060 & 1.755$\pm$0.051 & 2.111$\pm$0.243\\
0.772 & 1.479$\pm$0.041 & 1.713$\pm$0.039 & 1.506$\pm$0.108 & 1.580$\pm$0.053 & 1.916$\pm$0.053 & 1.784$\pm$0.185\\
0.808 & 1.269$\pm$0.033 & 1.806$\pm$0.038 & 1.776$\pm$0.129 & 1.323$\pm$0.040 & 2.018$\pm$0.052 & 2.061$\pm$0.228\\
0.842 & 1.267$\pm$0.029 & 1.929$\pm$0.039 & 1.718$\pm$0.140 & 1.306$\pm$0.034 & 2.133$\pm$0.050 & 1.936$\pm$0.256\\
0.877 & 1.499$\pm$0.041 & 1.912$\pm$0.046 & 1.735$\pm$0.173 & 1.574$\pm$0.049 & 2.109$\pm$0.060 & 2.184$\pm$0.325\\
0.913 & 2.257$\pm$0.227 & 1.111$\pm$0.100 & 0.683$\pm$0.268 & 2.746$\pm$0.359 & 1.140$\pm$0.127 & 0.528$\pm$0.382\\
\hline
\hline
\end{tabular}\par}
\end{table}
\end{center}
\end{widetext}

\begin{widetext}
\begin{center}
\begin{table}
\caption{\label{tab:pirat-vrs-x} The $R=\pi^+/\pi^-$ ratios versus X for H, D and Al targets. 
Left (right) part of the table represents the ratios before (after) events from diffractive $\rho$ decay 
have been subtracted. Error bars are statistical only.}
{\centering  \begin{tabular}{||c|c c c||c c c||}
\hline
\hline
   x  &   $R_H\pm$dR    &   $R_D\pm$dR    &  $R_{Al}\pm$dR   &  $R_H\pm$dR    &   $R_D\pm$dR    &  $R_{Al}\pm$dR  \\
\hline
0.208 & 1.773$\pm$0.139 & 1.349$\pm$0.086 & 1.338$\pm$0.157 & 1.894$\pm$0.172 & 1.410$\pm$0.103 & 1.426$\pm$0.204 \\
0.224 & 1.371$\pm$0.089 & 1.358$\pm$0.073 & 1.607$\pm$0.186 & 1.427$\pm$0.105 & 1.418$\pm$0.087 & 1.742$\pm$0.260 \\
0.240 & 1.823$\pm$0.099 & 1.509$\pm$0.071 & 1.474$\pm$0.158 & 1.946$\pm$0.119 & 1.595$\pm$0.085 & 1.612$\pm$0.218 \\
0.256 & 1.765$\pm$0.090 & 1.392$\pm$0.059 & 1.692$\pm$0.174 & 1.867$\pm$0.106 & 1.447$\pm$0.068 & 1.886$\pm$0.242 \\
0.272 & 1.899$\pm$0.093 & 1.526$\pm$0.061 & 1.344$\pm$0.137 & 2.021$\pm$0.110 & 1.600$\pm$0.071 & 1.413$\pm$0.178 \\
0.288 & 1.948$\pm$0.091 & 1.494$\pm$0.059 & 1.507$\pm$0.141 & 2.064$\pm$0.106 & 1.559$\pm$0.068 & 1.594$\pm$0.178 \\
0.304 & 2.025$\pm$0.096 & 1.527$\pm$0.058 & 1.524$\pm$0.153 & 2.142$\pm$0.111 & 1.593$\pm$0.067 & 1.596$\pm$0.195 \\
0.320 & 1.958$\pm$0.091 & 1.446$\pm$0.055 & 1.993$\pm$0.193 & 2.063$\pm$0.104 & 1.496$\pm$0.062 & 2.213$\pm$0.258 \\
0.336 & 2.073$\pm$0.100 & 1.481$\pm$0.055 & 1.763$\pm$0.168 & 2.197$\pm$0.116 & 1.538$\pm$0.063 & 1.914$\pm$0.215 \\
0.352 & 1.991$\pm$0.095 & 1.536$\pm$0.058 & 1.711$\pm$0.175 & 2.091$\pm$0.108 & 1.596$\pm$0.065 & 1.843$\pm$0.223 \\
0.368 & 2.083$\pm$0.097 & 1.632$\pm$0.061 & 1.641$\pm$0.168 & 2.179$\pm$0.109 & 1.702$\pm$0.069 & 1.784$\pm$0.212 \\
0.384 & 2.344$\pm$0.110 & 1.516$\pm$0.056 & 1.318$\pm$0.132 & 2.481$\pm$0.126 & 1.560$\pm$0.063 & 1.372$\pm$0.158 \\
0.400 & 1.966$\pm$0.096 & 1.725$\pm$0.063 & 1.249$\pm$0.144 & 2.038$\pm$0.107 & 1.801$\pm$0.071 & 1.290$\pm$0.176 \\
0.416 & 2.292$\pm$0.115 & 1.618$\pm$0.061 & 1.262$\pm$0.139 & 2.409$\pm$0.130 & 1.680$\pm$0.068 & 1.294$\pm$0.165 \\
0.432 & 2.198$\pm$0.111 & 1.553$\pm$0.059 & 1.746$\pm$0.191 & 2.306$\pm$0.124 & 1.604$\pm$0.066 & 1.887$\pm$0.240 \\
0.448 & 2.101$\pm$0.117 & 1.576$\pm$0.061 & 1.497$\pm$0.162 & 2.178$\pm$0.131 & 1.627$\pm$0.068 & 1.570$\pm$0.192 \\
0.464 & 2.129$\pm$0.120 & 1.577$\pm$0.065 & 1.292$\pm$0.154 & 2.220$\pm$0.134 & 1.626$\pm$0.072 & 1.326$\pm$0.179 \\
0.480 & 2.385$\pm$0.148 & 1.618$\pm$0.072 & 1.521$\pm$0.180 & 2.505$\pm$0.166 & 1.668$\pm$0.080 & 1.595$\pm$0.211 \\
0.496 & 2.615$\pm$0.173 & 1.545$\pm$0.073 & 1.431$\pm$0.180 & 2.755$\pm$0.195 & 1.587$\pm$0.080 & 1.487$\pm$0.208 \\
0.512 & 2.680$\pm$0.189 & 1.554$\pm$0.079 & 1.671$\pm$0.265 & 2.803$\pm$0.211 & 1.598$\pm$0.087 & 1.782$\pm$0.322 \\
0.528 & 2.131$\pm$0.162 & 1.571$\pm$0.089 & 1.554$\pm$0.242 & 2.205$\pm$0.177 & 1.613$\pm$0.096 & 1.630$\pm$0.284 \\
0.544 & 2.061$\pm$0.189 & 1.535$\pm$0.095 & 1.444$\pm$0.233 & 2.129$\pm$0.206 & 1.566$\pm$0.102 & 1.483$\pm$0.259 \\
0.560 & 2.122$\pm$0.210 & 1.454$\pm$0.099 & 1.352$\pm$0.247 & 2.191$\pm$0.228 & 1.479$\pm$0.106 & 1.377$\pm$0.270 \\
0.576 & 2.112$\pm$0.217 & 1.584$\pm$0.122 & 1.061$\pm$0.272 & 2.161$\pm$0.230 & 1.613$\pm$0.130 & 1.059$\pm$0.298 \\
0.592 & 1.946$\pm$0.238 & 1.867$\pm$0.169 & 2.257$\pm$0.642 & 1.953$\pm$0.240 & 1.879$\pm$0.171 & 2.242$\pm$0.643 \\
\hline
\hline
\end{tabular}\par}
\end{table}
\end{center}
\end{widetext}

\begin{widetext}
\begin{center}
\begin{table}
\caption{\label{tab:pirat-vrs-pt2} The $R=\pi^+/\pi^-$ ratios versus $P^2_t$ (in GeV$^2/c^2$) for H, D and Al 
targets. Left (right) part of the table represents the ratios before (after) events from diffractive $\rho$ 
decay have been subtracted. Error bars are statistical only.}
{\centering  \begin{tabular}{||c|c c c||c c c||}
\hline
\hline
$P^2_t$ &   $R_H\pm$dR    &   $R_D\pm$dR    &  $R_{Al}\pm$dR  &   $R_H\pm$dR    &   $R_D\pm$dR    &  $R_{Al}\pm$dR  \\
\hline
0.004   & 2.072$\pm$0.059 & 1.601$\pm$0.039 & 1.638$\pm$0.103 & 2.201$\pm$0.069 & 1.679$\pm$0.045 & 1.784$\pm$0.133 \\
0.012   & 1.940$\pm$0.065 & 1.535$\pm$0.041 & 1.695$\pm$0.119 & 2.058$\pm$0.075 & 1.603$\pm$0.047 & 1.770$\pm$0.146 \\
0.020   & 1.951$\pm$0.077 & 1.575$\pm$0.050 & 1.372$\pm$0.101 & 2.071$\pm$0.090 & 1.653$\pm$0.059 & 1.444$\pm$0.126 \\
0.028   & 1.978$\pm$0.093 & 1.464$\pm$0.054 & 1.436$\pm$0.125 & 2.109$\pm$0.110 & 1.516$\pm$0.063 & 1.523$\pm$0.162 \\
0.036   & 1.873$\pm$0.099 & 1.564$\pm$0.064 & 1.642$\pm$0.154 & 1.991$\pm$0.117 & 1.645$\pm$0.076 & 1.826$\pm$0.210 \\
0.044   & 1.982$\pm$0.113 & 1.508$\pm$0.068 & 1.479$\pm$0.154 & 2.126$\pm$0.136 & 1.573$\pm$0.080 & 1.621$\pm$0.209 \\
0.052   & 2.099$\pm$0.131 & 1.682$\pm$0.085 & 1.621$\pm$0.181 & 2.266$\pm$0.160 & 1.793$\pm$0.104 & 1.801$\pm$0.254 \\
0.060   & 2.176$\pm$0.157 & 1.460$\pm$0.078 & 1.361$\pm$0.170 & 2.378$\pm$0.196 & 1.529$\pm$0.093 & 1.425$\pm$0.226 \\
0.068   & 2.056$\pm$0.164 & 1.512$\pm$0.089 & 1.859$\pm$0.232 & 2.239$\pm$0.205 & 1.595$\pm$0.108 & 2.141$\pm$0.337 \\
0.076   & 1.842$\pm$0.161 & 1.710$\pm$0.111 & 1.186$\pm$0.168 & 1.970$\pm$0.198 & 1.844$\pm$0.140 & 1.220$\pm$0.217 \\
0.084   & 1.714$\pm$0.146 & 1.512$\pm$0.105 & 1.677$\pm$0.232 & 1.815$\pm$0.175 & 1.605$\pm$0.129 & 1.891$\pm$0.331 \\
0.092   & 2.176$\pm$0.207 & 1.528$\pm$0.108 & 1.333$\pm$0.199 & 2.379$\pm$0.262 & 1.617$\pm$0.133 & 1.424$\pm$0.266 \\
0.100   & 2.722$\pm$0.294 & 1.563$\pm$0.115 & 1.459$\pm$0.223 & 3.098$\pm$0.402 & 1.663$\pm$0.142 & 1.599$\pm$0.307 \\
0.108   & 2.294$\pm$0.256 & 1.595$\pm$0.127 & 1.497$\pm$0.232 & 2.591$\pm$0.349 & 1.723$\pm$0.164 & 1.628$\pm$0.320 \\
0.116   & 1.884$\pm$0.215 & 1.649$\pm$0.138 & 1.627$\pm$0.283 & 1.896$\pm$0.263 & 1.786$\pm$0.178 & 1.864$\pm$0.425 \\
0.124   & 1.897$\pm$0.228 & 1.701$\pm$0.157 & 1.108$\pm$0.200 & 2.069$\pm$0.292 & 1.857$\pm$0.206 & 1.138$\pm$0.259 \\
0.132   & 2.083$\pm$0.279 & 1.525$\pm$0.154 & 2.023$\pm$0.441 & 2.312$\pm$0.369 & 1.631$\pm$0.197 & 2.527$\pm$0.769 \\
0.140   & 1.813$\pm$0.258 & 1.645$\pm$0.182 & 3.418$\pm$0.843 & 1.976$\pm$0.332 & 1.795$\pm$0.239 & 5.186$\pm$2.065 \\
0.148   & 1.896$\pm$0.298 & 1.716$\pm$0.225 & 1.913$\pm$0.543 & 2.073$\pm$0.384 & 1.904$\pm$0.307 & 2.467$\pm$1.041 \\
0.156   & 2.631$\pm$0.553 & 1.808$\pm$0.260 & 2.135$\pm$0.746 & 3.071$\pm$0.803 & 1.989$\pm$0.343 & 2.871$\pm$1.527 \\
0.164   & 3.509$\pm$1.355 & 2.088$\pm$0.451 & 1.511$\pm$0.480 & 5.313$\pm$3.421 & 2.533$\pm$0.741 & 1.663$\pm$0.660 \\
0.172   & 2.754$\pm$0.945 & 1.612$\pm$0.368 & 1.271$\pm$0.607 & 3.439$\pm$1.600 & 1.797$\pm$0.520 & 1.412$\pm$0.979 \\
0.180   & 1.525$\pm$0.462 & 2.205$\pm$0.639 & 4.001$\pm$3.144 & 1.607$\pm$0.556 & 2.593$\pm$0.962 & 10.05$\pm$21.82 \\
0.188   & 1.049$\pm$0.463 & 1.631$\pm$0.551 & 2.233$\pm$1.506 & 1.058$\pm$0.555 & 1.802$\pm$0.756 &  2.917$\pm$2.849\\
0.196   & 2.727$\pm$2.279 & 4.250$\pm$3.553 & 1.060$\pm$1.050 & 3.235$\pm$3.431 & 6.571$\pm$9.187 &  1.075$\pm$1.323\\
\hline
\hline
\end{tabular}\par}
\end{table}
\end{center}
\end{widetext}

\begin{widetext}
\begin{center}
\begin{table}
\caption{\label{tab:nuclrat-vrs-z} The deuteron over proton ($R_{D/H}$= D/H) and aluminum over deuteron ($R_{Al/D}$=Al/D) ratios for $\pi^+$ and $\pi^-$ versus z. 
Left (right) part of the table represents the ratios before (after) events from diffractive $\rho$ decay have been subtracted. Error bars are statistical only.}
{\centering  \begin{tabular}{||c|c c c c||c c c c||}
\hline
\hline
   z  & R$^+_{D/H}\pm$dR & R$^-_{D/H}\pm$dR & R$^+_{Al/D}\pm$dR & R$^-_{Al/D}\pm$dR & R$^+_{D/H}\pm$dR & R$^-_{D/H}\pm$dR & R$^+_{Al/D}\pm$dR & R$^-_{Al/D}\pm$dR\\
\hline
0.317 & 0.814$\pm$0.109  & 1.115$\pm$0.192  & 0.770$\pm$0.146   & 0.763$\pm$0.159   & 0.806$\pm$0.112  & 1.123$\pm$0.207  & 0.759$\pm$0.153   & 0.747$\pm$0.168  \\
0.352 & 0.733$\pm$0.034  & 0.927$\pm$0.055  & 0.689$\pm$0.052   & 0.691$\pm$0.060   & 0.721$\pm$0.035  & 0.920$\pm$0.059  & 0.670$\pm$0.055   & 0.663$\pm$0.065  \\
0.387 & 0.773$\pm$0.025  & 1.024$\pm$0.050  & 0.622$\pm$0.038   & 0.688$\pm$0.049   & 0.763$\pm$0.026  & 1.025$\pm$0.054  & 0.600$\pm$0.040   & 0.659$\pm$0.053  \\
0.422 & 0.747$\pm$0.023  & 0.985$\pm$0.043  & 0.636$\pm$0.039   & 0.597$\pm$0.045   & 0.736$\pm$0.024  & 0.981$\pm$0.046  & 0.614$\pm$0.041   & 0.563$\pm$0.049  \\
0.457 & 0.748$\pm$0.023  & 0.995$\pm$0.040  & 0.631$\pm$0.039   & 0.579$\pm$0.043   & 0.737$\pm$0.023  & 0.995$\pm$0.043  & 0.610$\pm$0.041   & 0.543$\pm$0.046  \\
0.493 & 0.758$\pm$0.024  & 0.973$\pm$0.037  & 0.602$\pm$0.038   & 0.618$\pm$0.041   & 0.744$\pm$0.025  & 0.970$\pm$0.041  & 0.573$\pm$0.041   & 0.575$\pm$0.046  \\
0.527 & 0.702$\pm$0.023  & 0.980$\pm$0.034  & 0.635$\pm$0.038   & 0.562$\pm$0.039   & 0.685$\pm$0.024  & 0.977$\pm$0.039  & 0.605$\pm$0.041   & 0.504$\pm$0.044  \\
0.562 & 0.742$\pm$0.024  & 1.004$\pm$0.033  & 0.601$\pm$0.035   & 0.576$\pm$0.038   & 0.727$\pm$0.025  & 1.003$\pm$0.038  & 0.569$\pm$0.037   & 0.520$\pm$0.043  \\
0.597 & 0.702$\pm$0.020  & 1.039$\pm$0.032  & 0.606$\pm$0.033   & 0.577$\pm$0.035   & 0.683$\pm$0.021  & 1.041$\pm$0.037  & 0.570$\pm$0.036   & 0.517$\pm$0.040  \\
0.632 & 0.778$\pm$0.022  & 1.052$\pm$0.033  & 0.579$\pm$0.032   & 0.635$\pm$0.037   & 0.760$\pm$0.024  & 1.060$\pm$0.039  & 0.536$\pm$0.035   & 0.572$\pm$0.043  \\
0.668 & 0.795$\pm$0.021  & 1.016$\pm$0.029  & 0.670$\pm$0.033   & 0.582$\pm$0.034   & 0.774$\pm$0.023  & 1.017$\pm$0.036  & 0.628$\pm$0.037   & 0.494$\pm$0.041  \\
0.702 & 0.802$\pm$0.021  & 1.063$\pm$0.031  & 0.607$\pm$0.033   & 0.584$\pm$0.034   & 0.781$\pm$0.023  & 1.077$\pm$0.039  & 0.553$\pm$0.037   & 0.493$\pm$0.040  \\
0.738 & 0.882$\pm$0.023  & 1.025$\pm$0.030  & 0.580$\pm$0.031   & 0.526$\pm$0.033   & 0.868$\pm$0.025  & 1.030$\pm$0.038  & 0.524$\pm$0.035   & 0.416$\pm$0.040  \\
0.772 & 0.996$\pm$0.024  & 1.004$\pm$0.029  & 0.540$\pm$0.027   & 0.612$\pm$0.034   & 0.996$\pm$0.028  & 1.002$\pm$0.037  & 0.472$\pm$0.031   & 0.507$\pm$0.043  \\
0.808 & 1.115$\pm$0.026  & 0.898$\pm$0.024  & 0.543$\pm$0.025   & 0.542$\pm$0.032   & 1.130$\pm$0.030  & 0.875$\pm$0.029  & 0.483$\pm$0.028   & 0.428$\pm$0.039  \\
0.842 & 1.044$\pm$0.022  & 0.802$\pm$0.020  & 0.430$\pm$0.021   & 0.457$\pm$0.030   & 1.048$\pm$0.024  & 0.768$\pm$0.023  & 0.371$\pm$0.023   & 0.339$\pm$0.036  \\
0.877 & 0.933$\pm$0.023  & 0.973$\pm$0.032  & 0.406$\pm$0.026   & 0.448$\pm$0.037   & 0.927$\pm$0.025  & 0.967$\pm$0.039  & 0.345$\pm$0.028   & 0.329$\pm$0.044  \\
0.913 & 0.674$\pm$0.067  & 0.000$\pm$0.000  & 0.382$\pm$0.137   & 0.622$\pm$0.117   & 0.627$\pm$0.075  & 0.000$\pm$0.000  & 0.243$\pm$0.165   & 0.525$\pm$0.146  \\
0.947 & 0.693$\pm$0.049  & 0.049$\pm$0.012  & 1.971$\pm$0.119   & 5.758$\pm$1.535   & 0.693$\pm$0.049  & 0.049$\pm$0.012  & 1.971$\pm$0.119   & 5.758$\pm$1.535  \\
0.983 & 0.330$\pm$0.025  & 0.211$\pm$0.019  & 1.758$\pm$0.130   & 1.748$\pm$0.214   & 0.330$\pm$0.025  & 0.211$\pm$0.019  & 1.758$\pm$0.130   & 1.748$\pm$0.214  \\
\hline
\hline
\end{tabular}\par}
\end{table}
\end{center}
\end{widetext}

\begin{widetext}
\begin{center}
\begin{table}
\caption{\label{tab:nuclrat-vrs-x} The deuteron over proton ($R_{D/H}$= D/H) and aluminum over deuteron ($R_{Al/D}$=Al/D) ratios for $\pi^+$ and $\pi^-$ versus x. 
Left (right) part of the table represents the ratios before (after) events from diffractive $\rho$ decay have been subtracted. Error bars are statistical only.}
{\centering  \begin{tabular}{||c|c c c c||c c c c||}
\hline
\hline
   x  & R$^+_{D/H}\pm$dR & R$^-_{D/H}\pm$dR & R$^+_{Al/D}\pm$dR & R$^-_{Al/D}\pm$dR & R$^+_{D/H}\pm$dR & R$^-_{D/H}\pm$dR & R$^+_{Al/D}\pm$dR & R$^-_{Al/D}\pm$dR\\
\hline
0.208 & 0.770$\pm$0.052  & 1.102$\pm$0.085  & 0.727$\pm$0.066   & 0.742$\pm$0.074   & 0.748$\pm$0.056  & 1.119$\pm$0.103  & 0.693$\pm$0.074   & 0.698$\pm$0.086  \\
0.224 & 0.873$\pm$0.057  & 0.963$\pm$0.057  & 0.765$\pm$0.063   & 0.595$\pm$0.058   & 0.859$\pm$0.062  & 0.956$\pm$0.066  & 0.737$\pm$0.069   & 0.527$\pm$0.067  \\
0.240 & 0.746$\pm$0.040  & 0.962$\pm$0.051  & 0.639$\pm$0.053   & 0.603$\pm$0.054   & 0.727$\pm$0.043  & 0.955$\pm$0.059  & 0.600$\pm$0.057   & 0.536$\pm$0.063  \\
0.256 & 0.768$\pm$0.041  & 1.023$\pm$0.048  & 0.771$\pm$0.058   & 0.539$\pm$0.046   & 0.749$\pm$0.044  & 1.026$\pm$0.055  & 0.747$\pm$0.064   & 0.474$\pm$0.052  \\
0.272 & 0.744$\pm$0.037  & 0.993$\pm$0.046  & 0.564$\pm$0.048   & 0.595$\pm$0.047   & 0.726$\pm$0.039  & 0.991$\pm$0.052  & 0.524$\pm$0.052   & 0.537$\pm$0.053  \\
0.288 & 0.685$\pm$0.034  & 0.958$\pm$0.043  & 0.673$\pm$0.054   & 0.622$\pm$0.046   & 0.666$\pm$0.036  & 0.947$\pm$0.048  & 0.644$\pm$0.059   & 0.570$\pm$0.052  \\
0.304 & 0.706$\pm$0.035  & 0.962$\pm$0.042  & 0.696$\pm$0.057   & 0.547$\pm$0.043   & 0.689$\pm$0.037  & 0.953$\pm$0.047  & 0.669$\pm$0.062   & 0.487$\pm$0.048  \\
0.320 & 0.719$\pm$0.036  & 0.983$\pm$0.040  & 0.724$\pm$0.059   & 0.533$\pm$0.041   & 0.703$\pm$0.038  & 0.981$\pm$0.045  & 0.701$\pm$0.064   & 0.477$\pm$0.046  \\
0.336 & 0.734$\pm$0.035  & 1.065$\pm$0.046  & 0.670$\pm$0.053   & 0.575$\pm$0.042   & 0.719$\pm$0.037  & 1.072$\pm$0.052  & 0.646$\pm$0.057   & 0.525$\pm$0.047  \\
0.352 & 0.730$\pm$0.036  & 0.947$\pm$0.040  & 0.632$\pm$0.052   & 0.546$\pm$0.042   & 0.715$\pm$0.037  & 0.939$\pm$0.044  & 0.606$\pm$0.056   & 0.494$\pm$0.047  \\
0.368 & 0.687$\pm$0.031  & 0.908$\pm$0.038  & 0.526$\pm$0.048   & 0.560$\pm$0.042   & 0.673$\pm$0.032  & 0.897$\pm$0.042  & 0.553$\pm$0.053   & 0.510$\pm$0.047  \\
0.384 & 0.615$\pm$0.027  & 1.018$\pm$0.044  & 0.536$\pm$0.045   & 0.615$\pm$0.043   & 0.600$\pm$0.027  & 1.020$\pm$0.049  & 0.506$\pm$0.048   & 0.575$\pm$0.047  \\
0.400 & 0.757$\pm$0.033  & 0.926$\pm$0.040  & 0.422$\pm$0.039   & 0.558$\pm$0.042   & 0.746$\pm$0.034  & 0.917$\pm$0.044  & 0.388$\pm$0.041   & 0.511$\pm$0.046  \\
0.416 & 0.655$\pm$0.028  & 0.987$\pm$0.046  & 0.470$\pm$0.040   & 0.579$\pm$0.044   & 0.641$\pm$0.029  & 0.982$\pm$0.050  & 0.439$\pm$0.042   & 0.537$\pm$0.048  \\
0.432 & 0.684$\pm$0.029  & 0.989$\pm$0.046  & 0.544$\pm$0.042   & 0.506$\pm$0.043   & 0.672$\pm$0.030  & 0.988$\pm$0.050  & 0.517$\pm$0.045   & 0.459$\pm$0.047  \\
0.448 & 0.725$\pm$0.032  & 1.008$\pm$0.051  & 0.576$\pm$0.045   & 0.564$\pm$0.046   & 0.715$\pm$0.033  & 1.003$\pm$0.055  & 0.553$\pm$0.047   & 0.526$\pm$0.050  \\
0.464 & 0.711$\pm$0.032  & 0.981$\pm$0.051  & 0.521$\pm$0.043   & 0.564$\pm$0.050   & 0.700$\pm$0.033  & 0.979$\pm$0.055  & 0.496$\pm$0.045   & 0.526$\pm$0.054  \\
0.480 & 0.702$\pm$0.033  & 1.052$\pm$0.062  & 0.579$\pm$0.047   & 0.629$\pm$0.056   & 0.692$\pm$0.034  & 1.056$\pm$0.068  & 0.557$\pm$0.049   & 0.597$\pm$0.061  \\
0.496 & 0.633$\pm$0.031  & 1.094$\pm$0.069  & 0.608$\pm$0.051   & 0.607$\pm$0.061   & 0.621$\pm$0.032  & 1.102$\pm$0.076  & 0.588$\pm$0.053   & 0.574$\pm$0.065  \\
0.512 & 0.599$\pm$0.031  & 1.002$\pm$0.069  & 0.544$\pm$0.055   & 0.490$\pm$0.064   & 0.588$\pm$0.032  & 0.998$\pm$0.074  & 0.521$\pm$0.057   & 0.449$\pm$0.068  \\
0.528 & 0.652$\pm$0.039  & 0.958$\pm$0.069  & 0.555$\pm$0.055   & 0.533$\pm$0.069   & 0.641$\pm$0.040  & 0.954$\pm$0.074  & 0.534$\pm$0.058   & 0.498$\pm$0.074  \\
0.544 & 0.781$\pm$0.054  & 1.076$\pm$0.090  & 0.640$\pm$0.066   & 0.618$\pm$0.083   & 0.774$\pm$0.055  & 1.081$\pm$0.096  & 0.625$\pm$0.069   & 0.594$\pm$0.088  \\
0.560 & 0.731$\pm$0.054  & 1.092$\pm$0.102  & 0.646$\pm$0.076   & 0.538$\pm$0.088   & 0.723$\pm$0.056  & 1.098$\pm$0.108  & 0.632$\pm$0.079   & 0.512$\pm$0.092  \\
0.576 & 0.684$\pm$0.053  & 0.945$\pm$0.095  & 0.403$\pm$0.070   & 0.582$\pm$0.109   & 0.677$\pm$0.054  & 0.942$\pm$0.100  & 0.385$\pm$0.072   & 0.561$\pm$0.114  \\
0.592 & 0.833$\pm$0.076  & 0.874$\pm$0.105  & 0.618$\pm$0.087   & 0.334$\pm$0.101   & 0.833$\pm$0.076  & 0.871$\pm$0.105  & 0.617$\pm$0.087   & 0.325$\pm$0.102  \\
\hline
\hline
\end{tabular}\par}
\end{table}
\end{center}
\end{widetext}

\begin{widetext}
\begin{center}
\begin{table}
\caption{\label{tab:nuclrat-vrs-pt2} The deuteron over proton ($R_{D/H}$= D/H) and aluminum over deuteron ($R_{Al/D}$=Al/D) ratios for $\pi^+$ and $\pi^-$ versus 
$P^2_t$ (in GeV$^2/c^2$). Left (right) part of the table represents the ratios before (after) events from diffractive $\rho$ decay have been subtracted. 
Error bars are statistical only.}
{\centering  \begin{tabular}{||c|c c c c||c c c c||}
\hline
\hline
$P^2_t$ & R$^+_{D/H}\pm$dR & R$^-_{D/H}\pm$dR & R$^+_{Al/D}\pm$dR & R$^-_{Al/D}\pm$dR & R$^+_{D/H}\pm$dR & R$^-_{D/H}\pm$dR & R$^+_{Al/D}\pm$dR & R$^-_{Al/D}\pm$dR\\
\hline
0.004   & 0.724$\pm$0.017  & 0.922$\pm$0.024  & 0.646$\pm$0.027   & 0.623$\pm$0.031   & 0.709$\pm$0.018  & 0.912$\pm$0.027  & 0.618$\pm$0.029   & 0.573$\pm$0.035  \\
0.012   & 0.739$\pm$0.018  & 0.975$\pm$0.029  & 0.633$\pm$0.027   & 0.538$\pm$0.030   & 0.723$\pm$0.018  & 0.972$\pm$0.033  & 0.603$\pm$0.029   & 0.477$\pm$0.034  \\
0.020   & 0.781$\pm$0.020  & 0.925$\pm$0.035  & 0.602$\pm$0.027   & 0.698$\pm$0.039   & 0.767$\pm$0.021  & 0.915$\pm$0.039  & 0.569$\pm$0.029   & 0.656$\pm$0.044  \\
0.028   & 0.711$\pm$0.020  & 0.973$\pm$0.045  & 0.624$\pm$0.031   & 0.588$\pm$0.038   & 0.692$\pm$0.021  & 0.969$\pm$0.051  & 0.589$\pm$0.034   & 0.530$\pm$0.043  \\
0.036   & 0.766$\pm$0.024  & 0.942$\pm$0.050  & 0.550$\pm$0.031   & 0.564$\pm$0.041   & 0.750$\pm$0.025  & 0.934$\pm$0.057  & 0.516$\pm$0.034   & 0.500$\pm$0.046  \\
0.044   & 0.741$\pm$0.024  & 0.944$\pm$0.056  & 0.560$\pm$0.034   & 0.567$\pm$0.046   & 0.722$\pm$0.026  & 0.935$\pm$0.064  & 0.519$\pm$0.037   & 0.500$\pm$0.052  \\
0.052   & 0.694$\pm$0.024  & 0.898$\pm$0.060  & 0.593$\pm$0.038   & 0.596$\pm$0.053   & 0.674$\pm$0.026  & 0.883$\pm$0.068  & 0.565$\pm$0.041   & 0.527$\pm$0.060  \\
0.060   & 0.722$\pm$0.027  & 0.998$\pm$0.076  & 0.590$\pm$0.040   & 0.584$\pm$0.055   & 0.702$\pm$0.029  & 0.986$\pm$0.089  & 0.550$\pm$0.044   & 0.516$\pm$0.063  \\
0.068   & 0.757$\pm$0.031  & 1.034$\pm$0.086  & 0.673$\pm$0.046   & 0.565$\pm$0.060   & 0.739$\pm$0.033  & 1.040$\pm$0.101  & 0.639$\pm$0.050   & 0.493$\pm$0.068  \\
0.076   & 0.782$\pm$0.034  & 0.899$\pm$0.082  & 0.515$\pm$0.043   & 0.679$\pm$0.077   & 0.765$\pm$0.037  & 0.881$\pm$0.095  & 0.466$\pm$0.047   & 0.619$\pm$0.090  \\
0.084   & 0.718$\pm$0.035  & 0.827$\pm$0.077  & 0.720$\pm$0.057   & 0.633$\pm$0.075   & 0.694$\pm$0.037  & 0.799$\pm$0.087  & 0.687$\pm$0.063   & 0.565$\pm$0.086  \\
0.092   & 0.738$\pm$0.037  & 0.948$\pm$0.100  & 0.614$\pm$0.054   & 0.642$\pm$0.080   & 0.717$\pm$0.039  & 0.928$\pm$0.117  & 0.571$\pm$0.059   & 0.580$\pm$0.092  \\
0.100   & 0.705$\pm$0.037  & 1.181$\pm$0.138  & 0.621$\pm$0.058   & 0.651$\pm$0.083   & 0.681$\pm$0.039  & 1.217$\pm$0.172  & 0.579$\pm$0.064   & 0.589$\pm$0.097  \\
0.108   & 0.717$\pm$0.042  & 1.054$\pm$0.127  & 0.660$\pm$0.067   & 0.712$\pm$0.095   & 0.692$\pm$0.045  & 1.066$\pm$0.157  & 0.616$\pm$0.075   & 0.651$\pm$0.113  \\
0.116   & 0.723$\pm$0.045  & 0.932$\pm$0.112  & 0.610$\pm$0.067   & 0.638$\pm$0.096   & 0.699$\pm$0.048  & 0.919$\pm$0.132  & 0.563$\pm$0.074   & 0.561$\pm$0.114  \\
0.124   & 0.762$\pm$0.055  & 0.886$\pm$0.117  & 0.563$\pm$0.075   & 0.832$\pm$0.121   & 0.739$\pm$0.060  & 0.864$\pm$0.137  & 0.507$\pm$0.083   & 0.795$\pm$0.147  \\
0.132   & 0.729$\pm$0.060  & 0.978$\pm$0.142  & 0.717$\pm$0.098   & 0.541$\pm$0.104   & 0.704$\pm$0.064  & 0.973$\pm$0.172  & 0.680$\pm$0.110   & 0.442$\pm$0.123  \\
0.140   & 0.769$\pm$0.072  & 0.885$\pm$0.135  & 0.688$\pm$0.108   & 0.446$\pm$0.104   & 0.745$\pm$0.078  & 0.862$\pm$0.160  & 0.918$\pm$0.143   & 0.318$\pm$0.124  \\
0.148   & 0.714$\pm$0.074  & 0.794$\pm$0.140  & 0.614$\pm$0.114   & 0.550$\pm$0.138   & 0.687$\pm$0.080  & 0.753$\pm$0.163  & 0.560$\pm$0.129   & 0.432$\pm$0.168  \\
0.156   & 0.800$\pm$0.093  & 1.164$\pm$0.264  & 0.551$\pm$0.120   & 0.465$\pm$0.143   & 0.783$\pm$0.100  & 1.208$\pm$0.345  & 0.500$\pm$0.132   & 0.345$\pm$0.170  \\
0.164   & 0.843$\pm$0.142  & 1.441$\pm$0.589  & 0.918$\pm$0.217   & 1.269$\pm$0.384   & 0.821$\pm$0.159  & 1.758$\pm$1.195  & 0.905$\pm$0.252   & 1.378$\pm$0.561  \\
0.172   & 0.695$\pm$0.127  & 1.092$\pm$0.398  & 0.536$\pm$0.192   & 0.680$\pm$0.265   & 0.662$\pm$0.138  & 1.057$\pm$0.525  & 0.458$\pm$0.221   & 0.583$\pm$0.335  \\
0.180   & 0.800$\pm$0.175  & 0.553$\pm$0.198  & 0.661$\pm$0.251   & 0.364$\pm$0.272   & 0.781$\pm$0.190  & 0.484$\pm$0.215  & 0.619$\pm$0.281   & 0.160$\pm$0.345  \\
0.188   & 1.157$\pm$0.392  & 0.744$\pm$0.328  & 0.817$\pm$0.357   & 0.597$\pm$0.367   & 1.185$\pm$0.472  & 0.696$\pm$0.376  & 0.790$\pm$0.409   & 0.488$\pm$0.453  \\
0.196   & 0.850$\pm$0.352  & 0.545$\pm$0.604  & 0.518$\pm$0.415   & 2.079$\pm$2.119   & 0.836$\pm$0.382  & 0.412$\pm$0.698  & 0.466$\pm$0.456   & 2.849$\pm$4.515  \\
\hline
\hline
\end{tabular}\par}
\end{table}
\end{center}
\end{widetext}



\end{document}